\documentclass[twocolumn]{aastex61}

\usepackage{rotating}

\usepackage{amsmath}

\newcommand{\masr}{mas~yr$^{-1}$}
\newcommand{\uas}{$\mu$as}
\newcommand{\jb}{Jy~beam$^{-1}$}
\newcommand{\mjb}{mJy~beam$^{-1}$}
\newcommand{\ujb}{$\mu$Jy~beam$^{-1}$}
\newcommand{\unsim}{{\sim}}    % for use as a unary command, like ~2.

\submitjournal{the Astrophysical Journal}
\accepted{2018 February 13}
\shortauthors{Walker \& Hardee et al.}
\shorttitle{The Sub-Parsec Scale Jet in M\,87}

\begin{document}

\title{THE STRUCTURE AND DYNAMICS OF THE SUB-PARSEC SCALE JET IN M\,87 BASED ON 50
VLBA OBSERVATIONS OVER 17 YEARS AT 43 GHZ}

\correspondingauthor{R. Craig Walker}
\email{cwalker@nrao.edu}

\author[0000-0002-6710-6411]{R. Craig Walker} 
\altaffiliation{The National Radio Astronomy Observatory (NRAO) is a facility of the 
National Science Foundation (NSF), operated under cooperative agreement by Associated
Universities, Inc. (AUI).}
\affiliation{National Radio Astronomy Observatory, Socorro, NM 87801}

\author{Philip E. Hardee}
\affiliation{Department of Physics \& Astronomy, The University of Alabama,
Tuscaloosa, AL 35487}

\author{Frederick B. Davies}
\affiliation{Department of Physics, University of California, Santa Barbara, CA 93106-9530, USA}

\author[0000-0002-4245-2318]{Chun Ly}
\affiliation{Steward Observatory, 933 N Cherry Ave, Tucson, AZ 85721}
\affiliation{MMT Observatory, 933 N Cherry Ave, Tucson, AZ 85721}

\author[0000-0003-4740-8321]{ William Junor}
\affiliation{ISR-2, MS-B244, Los Alamos National Laboratory, P.O. Box 1663, Los Alamos, NM 87545}

\defcitealias{MLWH2016}{MLWH}

%   Ap. J. has a 250 word limit on the abstract.

\begin{abstract}

The central radio source in M\,87 provides the best opportunity to study jet formation because 
it has a large angular size for the gravitational radius of the black hole and has a bright jet that is well resolved by VLBI observations.
We present intensive monitoring observations from 2007 and 2008, plus roughly annual observations that span 17 years, all made with the the Very Long Baseline Array at 43 GHz with a resolution of about 30 by 60 $R_{\rm S}$.
Our high-dynamic-range images clearly show the wide-opening-angle structure and the counter-jet.
The jet and counter-jet are nearly symmetric in the inner 1.5 milli-arcseconds (mas; 0.12 pc in projection) with both being edge brightened.
Both show deviations from parabolic shape in the form of an initial rapid expansion and subsequent contraction followed by further rapid expansion and, beyond the visible counter-jet, subsequent collimation.
Proper motions and counter-jet/jet intensity ratios both indicate acceleration from apparent speeds of $\lesssim 0.5c$ to $\gtrsim 2c$ in the inner $\unsim 2$ mas and suggest a helical flow.
The jet displays a sideways shift with an approximately 8 to 10 year quasi-periodicity.
The shift propagates outwards non-ballistically and significantly more slowly than the flow speed revealed by the fastest moving components.
Polarization data show a systematic structure with magnetic field vectors that suggest a toroidal field close to the core.

\end{abstract}


\section{INTRODUCTION}
\label{Sec:Intro}

M\,87 (Virgo A, NGC\,4486, 3C\,274) is a giant elliptical galaxy near the center of the Virgo Cluster that contains a very massive black hole and a prominent jet.
The scale, in gravitational units per unit angle, is the best for any jet with observed structure making M\,87 a prime target for studies of the jet launch region.
At a distance to M\,87 of $16.7$~Mpc \citep{Mei07}, $1~{\rm mas} \approx 0.081$ pc and $1$~\masr $\approx 0.264c$.
At this assumed distance, the black hole in the center of M\,87 has a mass, determined from stellar dynamics, of $(6.1 \pm 0.4) \times 10^9$ M$_{\odot}$ \citep[][adjusted for a different assumed distance]{G2011}.   
This gives a scale of $1~{\rm mas} \approx 140$~R$_{\rm S}$ where the Schwarzschild radius $R_{\rm S} \equiv 2GM/c^2 \approx 1.8\times10^{15}$~cm.\footnote{The M\,87 black hole mass is still controversial, with results from gas dynamics giving masses of about half that given above \citep{WBHS2013}.
If the smaller mass proves to be correct, all distances given in this paper in $R_{\rm S}$ will need to be revised upward accordingly.}
With this scale, Very Long Baseline Array \citep[VLBA;][]{VLBA1993} observations at 43~GHz (7~mm), with a resolution of 0.2 to 0.4 mas, probe structures with sizes on the order of 100~$R_{\rm S}$.

The well-known, kpc-scale jet in this galaxy is a prominent source of radio, optical, near-ultraviolet (NUV) and X-ray emission.
There are remarkable similarities between the radio, optical and NUV emission on projected scales of $0.1\!-\!1$ kpc.
The bright knots, the filamentary emission between the knots, and the bends and twists in the jet can easily be identified in the optical and NUV as well as the radio \citep{Perlman2001}.
While X-ray images from {\it Chandra} \citep{Marshall2002, Wilson2002} are not of comparable resolution to the radio or optical, the same bright knots and overall jet structure can easily be seen in the X-rays.
 
The M\,87 jet has long been known to have a wide-opening-angle base on scales well below a mas \citep{Junor1999}.
On average the jet expands parabolically out to about the first bright knot (known as HST-1) at $\unsim 0.9\arcsec$ (projected distance $\unsim 70$ pc) from the core \citep{AN2012,Hada2013}, but with evidence for faster expansion at the smallest scales \citep[see][]{Hada2013, Hada2016}.
Older VLBI observations at 1.6~GHz indicate a full width half-maximum (FWHM) apparent opening angle of $\Theta^{\rm app} \approx 6.9\arcdeg$ out to $\unsim 50$~mas (projected distance $\unsim 4$ pc) from the radio core  \citep{Reid1989}.
Beyond HST-1 the jet expands conically \citep{AN2012}, and Very Large Array (VLA) observations at 15~GHz indicate a FWHM apparent opening angle of $\Theta^{\rm app} \approx 6.5\arcdeg$ from HST-1 to knot A located at $\unsim 12\arcsec$ (projected distance $\unsim 930$ pc) from the radio core \citep{OHC89}.
The radio core in M\,87 is thought to be nearly coincident with the position of the black hole, unlike the situation in many blazars \citep[e.g.,][]{Mar08}.
Multi-frequency VLBA astrometric observations indicate an offset of only 41 \uas~at 43 GHz \citep{Hada2011}, equivalent to about 6~$R_{\rm S}$.
 
Proper motions of jet components in the radio and optical range from subluminal to superluminal.
A component located at $\unsim 18$~mas (projected distance $\unsim 1.4$ pc), shows an angular motion of $1.1 \pm 0.3$~ mas yr$^{-1}$ that corresponds to a subluminal motion of $\beta^{\rm app} \equiv v^{\rm app}/c \approx 0.3$  \citep{Reid1989}.
\citet{Kov2007} find motions of less than a few percent of the speed of light at 15 GHz for several components within 25 mas of the core with the VLBA.
That result is based partly on the MOJAVE program \citep{MOJAVE}, a recent reanalysis of which gives hints of acceleration and mild superluminal motion \citep{Britzen2017}.
At higher frequencies, both subluminal and mildly superluminal (up to $\beta^{\rm app}\approx 2.5$) speeds, with evidence for acceleration, have been seen based on the 43 GHz VLBA data to be presented in this paper \citep[][hereafter MLWH]{MLWH2016} and based on KaVA (KVN and VERA Array (Korean VLBI Network and VLBI Exploration of Radio Astronomy)) data at 22 GHz \citep{Hada2017}.
Slow motions were seen by \citet{Hada2016} with 4 epochs, two at 86 GHz and two at 43 GHz.
With one exception, those components were at core distances of about 1 mas or less.
The fastest observed proper motion, found in the optical band at HST-1, is superluminal with $\beta^{\rm app}  = 6.1 \pm 0.6$ \citep{BSM99}.
The fastest observed radio proper motions, also at HST-1, are superluminal with $\beta^{\rm app} = 4.3 \pm 0.7$ \citep{CHS07}, and $4.17 \pm 0.07$ and $4.08 \pm 0.08$ \citep[both][]{Giroletti2012}.
\citet{Giroletti2012} also find less reliable evidence for a component moving at $\beta^{\rm app} = 6.4 \pm 0.8$, similar to the optical result.
\citet{Asada2014} find evidence from their 1.6 GHz global VLBI data plus other data from the literature for acceleration from subluminal to superluminal speeds between about 160 mas and HST-1.
In general, at and beyond HST-1 the radio proper motions are less than optical proper motions 
in the same region, and proper motions suggest jet deceleration \citep[see][]{HE2011}.

An important parameter for interpretation of the data is the jet angle to the line-of-sight.  
There is no one, simple way to determine that angle so the reasoning leading to the choice of the angle used in this paper is explained here at some length.
The fastest observed proper motions and modeling can be used to constrain the viewing angle.
The highest observed optical superluminal speed requires a jet viewing angle of $\theta < {18.6\arcdeg}^{+2.0}_{-1.6}$.
Recent model dependent values for the viewing angle are  $\theta = 14\arcdeg$ used by \citet{NM2014} to provide a self-consistent shock model for optical relativistic motions and particle acceleration at HST-1, and a value of $\theta = 15\arcdeg$  found by \citet{WZ2009} using synchrotron spectrum model fitting to the kpc jet.
\citetalias{MLWH2016} found that a viewing angle of $\theta = 19\arcdeg \pm 4\arcdeg$  (see Table 6 in \citetalias{MLWH2016}) provided a fit to the observed acceleration and collimation of the jet assuming a Poynting flux dominated approximation for the jet dynamics.

Less model dependent values for the viewing angle can be found in \citetalias{MLWH2016}.
A value of  $\theta = 17.2\arcdeg \pm 3.3\arcdeg$ [see eq.\ 4 in \citetalias{MLWH2016}] is obtained by \citetalias{MLWH2016} between 0.5 and 1~mas from the core using apparent component speeds in the jet, $\beta^{\rm app}_{\rm j} = 0.21 \pm 0.04$, and the counter-jet, $\beta^{\rm app}_{\rm cj} = 0.14 \pm 0.02$ along with an intensity ratio, $I_{\rm j}/I_{\rm cj} = 9.5 \pm 1.5$.
An additional analysis in \citetalias{MLWH2016}, based on the assumption that the significantly different speeds in northern and southern limbs of the jet are the result of jet rotation, leads to a viewing angle value of $\theta = 19.2\arcdeg \pm 3.7\arcdeg$ (see eq.\ 9 in \citetalias{MLWH2016}).
Thus, \citetalias{MLWH2016} concluded that a viewing angle of $\theta \approx 18\arcdeg$ represented a reasonable average based on these two techniques and Poynting flux dominated jet dynamics.
We note here that the jet rotation and Poynting flux model viewing angle estimate lies above the maximum value allowed by the optically observed $\beta_j^{\rm app}=6.1$ motion at HST-1, and the average viewing angle estimate lies uncomfortably close to the maximum viewing angle.
Thus, we adopt the smaller viewing angle from \citetalias{MLWH2016} of $\theta \approx 17\arcdeg$ in order to relate angular scales to intrinsic scales, and point out that all the recent model dependent viewing angles lie within $3\arcdeg$ of this choice.

At a viewing angle of $\theta = 17\arcdeg$, intrinsic lengths are $\unsim 3.42$ times longer than apparent (projected) lengths, and the intrinsic jet opening angle between HST-1 (intrinsic distance $\unsim 240$~pc) and knot A (intrinsic distance $\unsim 3180$~pc) is $\Theta \unsim 2\arcdeg$.
At that viewing angle, the observed optical and radio superluminal motions of $\beta^{\rm app} = 6.1$ and $\beta^{\rm app} = 4.3$ imply  intrinsic Lorentz factors of $\gamma = 10.9$ ($\beta = 0.9958$) and $\gamma = 4.6$ ($\beta = 0.9763$) respectively.

The jet is observed to be edge-brightened from about 50 to 200~mas and also beyond $1\arcsec$.
At angular distances out to $\unsim 80$~mas, the edge-brightened radio jet exhibits quasi-periodic deviation of the brightness centroid \citep{Reid1989} that likely indicates brightness alternation from one side to the other.
Beyond HST-1 the jet is similarly edge-brightened at 15~GHz and side-to-side alternation in the radio brightness profile is apparent from HST-1 through knot D located at $\unsim 3.5\arcsec$ (intrinsic distance $\unsim 906$~pc) from the 15~GHz radio core \citep{OHC89}.
The jet also appears to contain twisted filaments from HST-1 to knot A \citep{LHE2003} that are related to side-to-side variation in the brightness profile \citep{HE2011}.

The jet structure seen in the optical and NUV  is quite similar to that seen at 15~GHz, but there are important differences in detail.
The optical/NUV emission is more concentrated in the knots  and towards the jet axis, and the outer edges of the jet are less  well defined than in the radio \citep[e.g.,][]{SBM1996,Madrid2007}.
The major bright radio or optical knots can also be identified in the  lower-resolution X-ray images, e.g., \citet{PW2005}, but they can  differ in position or structure from their radio or optical  counterparts.
At 15~GHz the projected magnetic field lies more or less along the edge of the jet flow \citep{OHC89}, which may be indicative of a shear layer \citep{L1981}.
The optical/NUV knots also differ somewhat in polarization from the radio knots in that they tend to indicate magnetic field perpendicular to the jet flow, and suggest the presence of shocks \citep{P1999}.
In general, the broader radio profile, the more centrally concentrated optical/NUV structure, differences in the polarization structure, and the slower observed motions in the radio suggest velocity shear across the jet with the limb brightening observed in the radio being associated with the shear layer.

In this paper, we present results from monitoring observations of M\,87 at 43 GHz using the VLBA.
Images from 23 epochs in 2007 and early 2008 are used to study the structure and dynamics of the inner jet.
An image showing the average structure was made by stacking the 23 images and is used for the structure analysis.
In that image, the limb brightened jet is observed to 20~mas beyond the core.
A faint structure, that extends slightly more than 1~mas beyond the core in the opposite direction, is speculated to be a counter-jet \citep{Ly2007, Kov2007}.
We use transverse intensity slices to obtain the maximum intensity as a function of distance from the core and the width as a function of distance from the core.
The individual images, and movies made with them, are used to study the rapid and complex changes near the core in an effort to extract velocities and possible acceleration.
The 2008 observations showed a significant increase in flux density of the VLBA core that coincided with a flare seen at TeV energies as reported in \citet{Acc09}.
An early version of the 23-epoch average image, along with images from four individual epochs, was presented in that initial flare report.
The flare led to an effort to catch additional cases of correlated radio and TeV flares,
a result of which is that we have an ongoing series of roughly annual 43 GHz VLBA 
observations of M\,87.  We report results, based on those observations to 2016, along with legacy observations going back to 1999, of a study of changes to the jet envelope
on time scales of years.
Preliminary presentations of some of our data, including an initial movie, have appeared in \citet{Walker2008} and \citet{Walker2016} while an extensive analysis of the apparent velocity field in 2007 is given in \citetalias{MLWH2016}.

The paper is organized as follows.
In Section~\ref{Sec:Obs} we describe the observations.
In Section~\ref{Sec:Results}, we present the observational results with subsections on the average structure (\ref{SSec:Avg}), on the rapid structural evolution (\ref{SSec:RSE}), on the polarized structure (\ref{SSec:Pol}), on the long term evolution between 1999 and 2016 (\ref{SSec:LongTerm}), and on the counter-jet structure based on an average image and on recent, high-sensitivity observations (\ref{SSec:CJ}).
In Section~\ref{Sec:Implications} we examine some of the possible implications of our observations for jet structure and collimation properties that will provide a framework for future theoretical and numerical modeling of jet  acceleration and collimation regions.
Finally, we summarize and further discuss our results in Section~\ref{Sec:SD}.

For a listing of the symbols used in this paper, see Table~\ref{SymDef} in Appendix~\ref{AppSymbols}.

\section{THE OBSERVATIONS}
\label{Sec:Obs}

\subsection{Data Aquisition}

The primary goals of the 43 GHz VLBA M\,87 project were to determine the apparent jet velocity and as much as possible about the dynamics of the jet close to the black hole.
M\,87 had been observed before, but with some combination of inadequate resolution to see very close to the core and time sampling intervals that were too long unless the motions are very slow \citep[e.g.,][]{Reid1989,Junor1995}.
For example, if the apparent motions are at the speed of light, a component would be moving at 4 mas per year which is about 20 beam widths for the VLBA at 43 GHz.
Some other sources have apparent speeds of more than ten times that.
If fast motions exist in M\,87 near the core as they do on larger scales,
they probably had not been detected because they were seriously undersampled.
So a pilot project was conducted in 2006 during which time intervals of between three days and three months were sampled.
The results were somewhat ambiguous, but it was concluded that the changes that were seen could be followed with a three week sampling interval.
Such observations were made throughout 2007.
During that period it became apparent that the sampling interval was too long --- apparently related features moved two or more beam widths between observations and relating features was difficult.
In early 2008, the observations were continued at about 1 week intervals.
Despite the use of dynamic scheduling, the short interval forced the use of less optimal observing conditions, so the data quality in 2008 was not as high as in 2007. Many sessions had missing antennas which is not tolerated well when imaging a complex source with the sparse UV coverage of the VLBA.
The primary data set for this paper is the first 11 epochs from 2007 and 12 epochs from 2008.  One additional epoch, 2007 Nov.\ 02, is shown, but the remaining five
2007 epochs have technical problems so they are not used in this paper.

The correlation between a TeV flare and a radio flare noted in Section~\ref{Sec:Intro} led to an ongoing effort to catch other such flares in the hopes of further constraining the TeV emission location and mechanism.
The TeV instruments could only observe at night which constrains the coordinated observations of M\,87 to happen during the early part of each year.
Observations were made with the VLBA once or twice each year to monitor the ambient condition of the radio jet.
If a high energy flare was seen, additional observations were triggered.
There were triggers in 2010 and 2016 but no clear correlated radio flare was found in our data.
With our data plus an additional observation of their own that had high flux density and was almost coincident with the TeV flare, \citet{Hada2012} reported a weak radio flare associated with the 2010 TeV event.
In 2012, there was a period of enhanced TeV emission that failed to reach the trigger level for our project, but for which observations on the VERA and the European VLBI Network detected a significant radio flare \citep{Hada2014}.
A by-product of our 2010 trigger response is that we have a six-image stack for 2010.
In 2016, most of the follow-up observations were too short to make good images, so only three full-track observations are available and one of those had poor observing conditions.

All of the observations reported here were made on the 10-station VLBA.
Only the first two archival observations in 1999 and 2000 used additional antennas as noted in Table~\ref{ObsTable} in the Appendix.
That table gives the dates, image noise, and peak and total flux densities from the individual observations that contributed to this paper (not for all of the observations that were made).
The long-term analysis also used four average observations, listed in the Appendix in Table~\ref{StackTable}, one from each year with six or more individual observations.

The observations from 2004 and earlier are from the VLBA archive.
They provide roughly annual information on M\,87 at 43 GHz back to 1999.
Several of those observations involved co-authors on this paper and several involved the use of M\,87 as a phase reference source for astrometric studies or for observations of nearby weak Virgo Cluster sources.
Literature references for the projects involved can be found in Table~\ref{ObsTable}.
Most of those observations involved less, sometimes significantly less, time on source and less bandwidth than the post-2006 observations so the image quality is generally not up to the same standard.

Starting with the pilot project in 2006, the observations involved a full track with data spanning most of the time that M\,87 was visible from at least a few stations of the VLBA.
The total time on M\,87 at 43 GHz varied from about 5 to about 7 hours.
The sources OJ\,287, 3C\,279, and occasionally 3C\,84 or J1635+3808 were observed to use as polarization calibrators, fringe finders, and bandpass calibrators.
Beginning in 2007, short phase-referencing scans on M\,84 were observed in an attempt to measure the relative proper motions of M\,84 and M\,87 and to constrain any core
wander.
Such data are also available from 2001 October 12 and 2004 April 4 \citep{Ly2007}.
To improve the quality of that astrometry, two to four roughly half-hour bursts of scans on sources around the sky were added to calibrate the atmosphere starting with 2008 March 12.
Beginning in 2009, roughly 20\% of the M\,87 on-source time used 24 GHz to allow study of the spectral index and, possibly, Faraday rotation.
The results from most of the polarization data, from the 24 GHz imaging, and the M\,87/M\,84 astrometry will be reported elsewhere.

Various changes have occurred to the VLBA hardware over the years spanned by this project.
The most important is that the recording bandwidth has been increased both by increasing availability of recording media and, eventually, by a significant hardware upgrade.
The bandwidths used are shown in Table~\ref{ObsTable}.
Between the pilot project, recorded at 128 Mbps (16 MHz per polarization) and the most recent observations, recorded at 2048 Mbps (256 MHz per polarization), the bandwidth improvements alone gave a sensitivity increase of a factor of four.
The data were correlated at the Pete V. Domenici Science Operations Center in Socorro, New Mexico, on the original hardware correlator up through 2009.
A new Distributed FX (DiFX) software correlator that provides significantly enhanced capabilities \citep{DiFX2011} began operation in time to be used for our 2010 and later epochs.
A description of the status of the VLBA hardware at any given time, at least since 2009, can be found in the Observational Status Summaries such as \citet{OSS} and others that can be found at the same site.

\subsection{Data Reduction}

The data were reduced  with the Astronomical Image Processing System \citep[AIPS;][]{Greisen95} following the usual procedures for VLBI data reduction including correction for instrumental offsets using the autocorrelations, bandpass calibration based on strong calibrator observations, and correction for atmospheric opacity based on the system temperature data.
The a-priori amplitude calibration depended on the gains provided by VLBA operations, which are based on results from regular single-dish pointing observations of Jupiter, Saturn, and Venus averaged over many months.
Additional calibrations were done to enhance the M\,87/M\,84 astrometry and are included here for completeness.
The Earth Orientation Parameters were updated to values more accurate than those available at the time of correlation.
For many of the epochs, corrections were made for the ionospheric delays based on global ionospheric maps produced by the Jet Propulsion Laboratory which are available at the NASA's Crustal Dynamics Data Information System (CDDIS).
Atmospheric delay corrections were made using AIPS task DELZN based on either delays measured on sources around the sky in the aforementioned 2 to 4 bursts of about a half hour duration or, for observations on or before 2008 March 06, on the residual fringe rates of the project sources and calibrators.
These additional calibrations do not affect the images of M\,87.

The images of M\,87 were made using data that are both amplitude and phase self-calibrated.
The flux scale for each epoch was set by normalizing the self-calibration gain adjustments to the a-priori gains for observations above $30\arcdeg$ elevation on only those antennas with good weather and instrumental conditions for that epoch.
The flux densities are believed to be accurate to within about 10\%, although the errors on the first few observations (especially 1999) may be higher.
The phase self-calibration does not preserve source position so, for analysis of the dynamics, the images were aligned on the image peaks.
During the radio flare of 2008, the astrometric observations of the M\,87/M\,84 relative positions suggest that there is a small core shift as new components appear ( Davies, F. et al., in preparation).  
For the 2008 event, the brightest we have seen by far, the effect was under 100 \uas\  while at other times it is less than a few tens of \uas, so the effect is small enough to ignore for purposes of this paper.
For the imaging of this complex and fairly extended source, we found it important to use the robustness parameter \citep{Briggs} to gain some of the advantages of both natural and uniform weighting together, and to be able to significantly vary the dirty beam, which seems to help convergence during the self-calibration and imaging iterations.
We also found that significantly better images could be obtained using the multi-resolution capability of the imaging task IMAGR \citep{Greisen09}.
To aid comparison between epochs, nearly all images in this paper have had the point CLEAN components restored with a common beam of $0.43 \times 0.21$~mas, elongated in position angle $-16\arcdeg$.  
That resolution amounts to about $60 \times 30\,R_{\rm S}$ with the adopted distance and black hole mass.

The observations were scheduled and correlated in a manner consistent with obtaining images of the polarized structure of M\,87.
The primary calibrators for this purpose were OJ\,287 and 3C\,279.
These sources were used to determine the instrumental polarization using the standard procedures in AIPS meant for application to resolved calibrators of unknown structure.
For the electric vector position angle (EVPA) calibration, we have utilized results from the Boston University Blazar project \citep[][and references therein]{JM2016}, which includes the two calibrators.
That project has many more sources which aids the calibration significantly.
We do not have simultaneous epochs, but past experience suggests that the absolute position angles will be the same when the images have the same total intensity structure, the same polarized intensity structure, and the same run of polarization angles along the source.
When an epoch of the Blazar Project can be found with such a match for the  total intensity and polarization structure for a calibrator, the EVPA calibration can be set to give the same polarization position angles for that calibrator.
This method was used to calibrate the EVPA for both epochs for which we present polarization images later.
Confidence in the method was gained from the fact that the two calibrators gave independent, but consistent calibrations for each epoch and the polarization angles for M\,87 were consistent between the epochs.
To our knowledge, this method of EVPA calibration has not been used before but should be useful for other projects.

For this project, obtaining a reliable polarization calibration proved to be possible, but very time consuming.  It has only been completed to our satisfaction for two epochs, the results from which will be shown in Section~\ref{SSec:Pol}.
Those two epochs give similar results, so they likely reveal the nature of the polarization structure within about 0.5 mas of the core.
To avoid significant additional delay in the publication of the total intensity
results, we have chosen to defer a full polarization analysis.
Once the rest of the polarization results are available, we should be able to study time dependencies and to detect the polarization farther down the jet.
Also the 24 GHz data should allow study of Faraday rotation along the jet.
Detection of polarization farther along the jet depends on higher sensitivity which can be obtained by stacking the 2007/2008 data and/or by using the recent, higher-sensitivity observations.  
Those recent observations are subject to frequency-dependent instrumental effects within our observing bands which, because of software limitations (since fixed), hindered our initial calibration attempts.

\section{OBSERVATIONAL RESULTS}
\label{Sec:Results}

The results of our M\,87 VLBA observational program are presented in this section.  
There are five sub-sections.  
Each sub-section begins by presenting images that address a different aspect of the information available from the observations.  
Then the images are analyzed to extract more quantitative information that should help understand the nature of the inner jet.
Further discussion of what is learned from the data is deferred to Section~\ref{Sec:Implications}.

Section~\ref{SSec:Avg} presents an average intensity image based on the 2007 and 2008 data, showing the persistent structure.
An analysis based on slices is used to extract some of the detailed features of the jet shape which can shed light on the jet propagation near the core.  

Section~\ref{SSec:RSE} shows the individual images from the 2007 and 2008 observations.
These are the main movie observations, and the movies are available in the online version of the paper.  
An effort is made to measure component motions using traditional, component-tracking methods.  
This is not as sophisticated as what was done in \citetalias{MLWH2016}, but provides added confidence in the detection of superluminal motion and of acceleration very near the core.  

Section~\ref{SSec:Pol} presents the polarization results from the two epochs from 2007 for which we already have a successful polarization calibration.
There is a systematic structure to the polarization near the core which suggests a toroidal field geometry.  

Section~\ref{SSec:LongTerm} gathers the early observations, from before 2007, plus the results from ongoing efforts to catch more cases of activity related to TeV flares since 2009, 
to provide roughly annual images between 1999 and 2016.
A movie based on these results is available in the online version of the paper.
The jet is shown to move transversely over a period of several years.
An analysis of that transverse motion, including derivation of a propagation speed of the pattern that is that sideways motion, is presented.

Section~\ref{SSec:CJ} uses the inner few mas of an average of nearly all our data over the full 17 years, along with images from two of the recent, high-sensitivity observations, processed specifically for high resolution, to delineate the structure of the counter-jet.

\subsection{Average Intensity}
\label{SSec:Avg}
\begin{figure*}   % two column       %======================================
\epsscale{1.17}
% \plotone{Collimation6.eps}  %  Old version.
% \plotone{M87_SUM_A_STACK_MATHSC_1_NewF1.eps}
\plotone{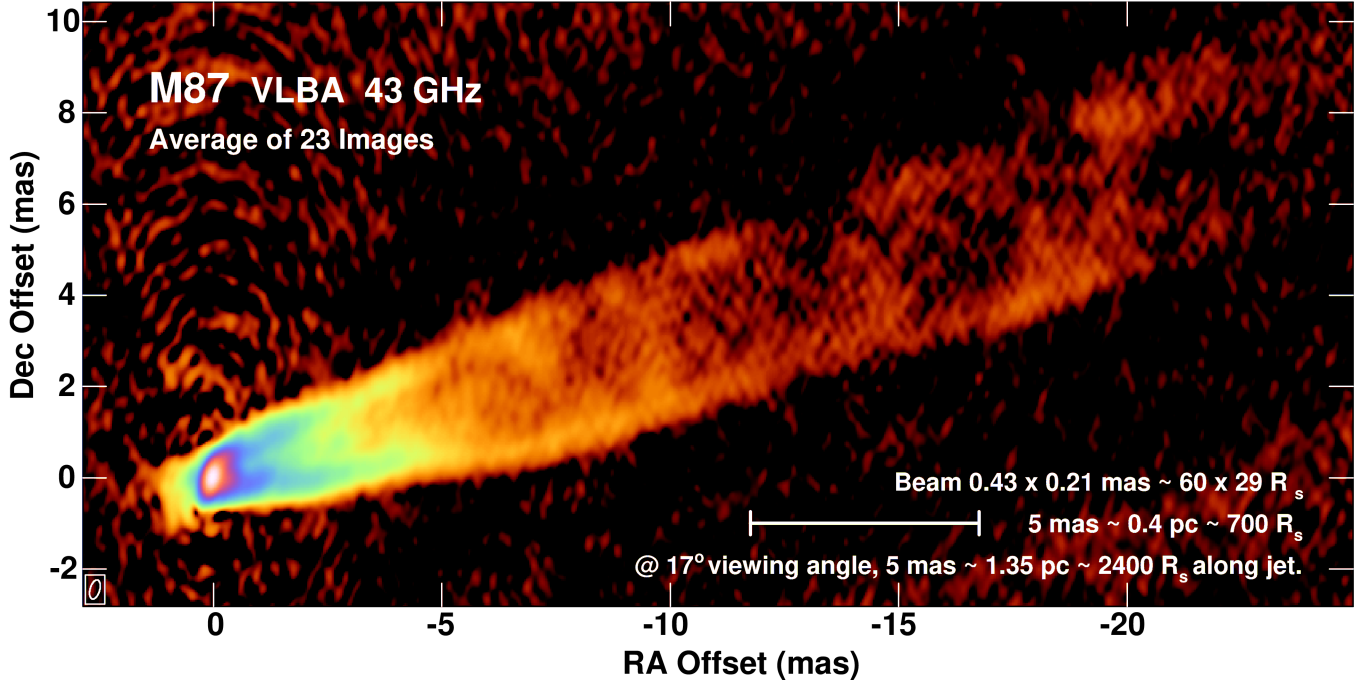}
\caption{A 23-epoch average radio image of the jet and counter-jet in M\,87 based on data from 2007 and 2008.
Angular to linear scales (in parsecs and in Schwarzschild radii $R_{\rm S}$) are indicated for distances in the sky plane and for distances along the axis of the jet assuming it is oriented at $17\arcdeg$ to the line-of-sight.
The beam with resolution $0.43 \times 0.21$~mas elongated in position angle $-16\arcdeg$ is shown at the lower  left.
The off-source noise level is 62 \ujb\  and the image peak is 0.83 Jy.}
\label{Jet_VLBA}
\end{figure*}

\begin{figure*}   % Try fullsize.  Can be single column.  Twocol: try without [h!] ============
% \plotone{3D_stack23_profl.eps}
\epsscale{0.8}
\plotone{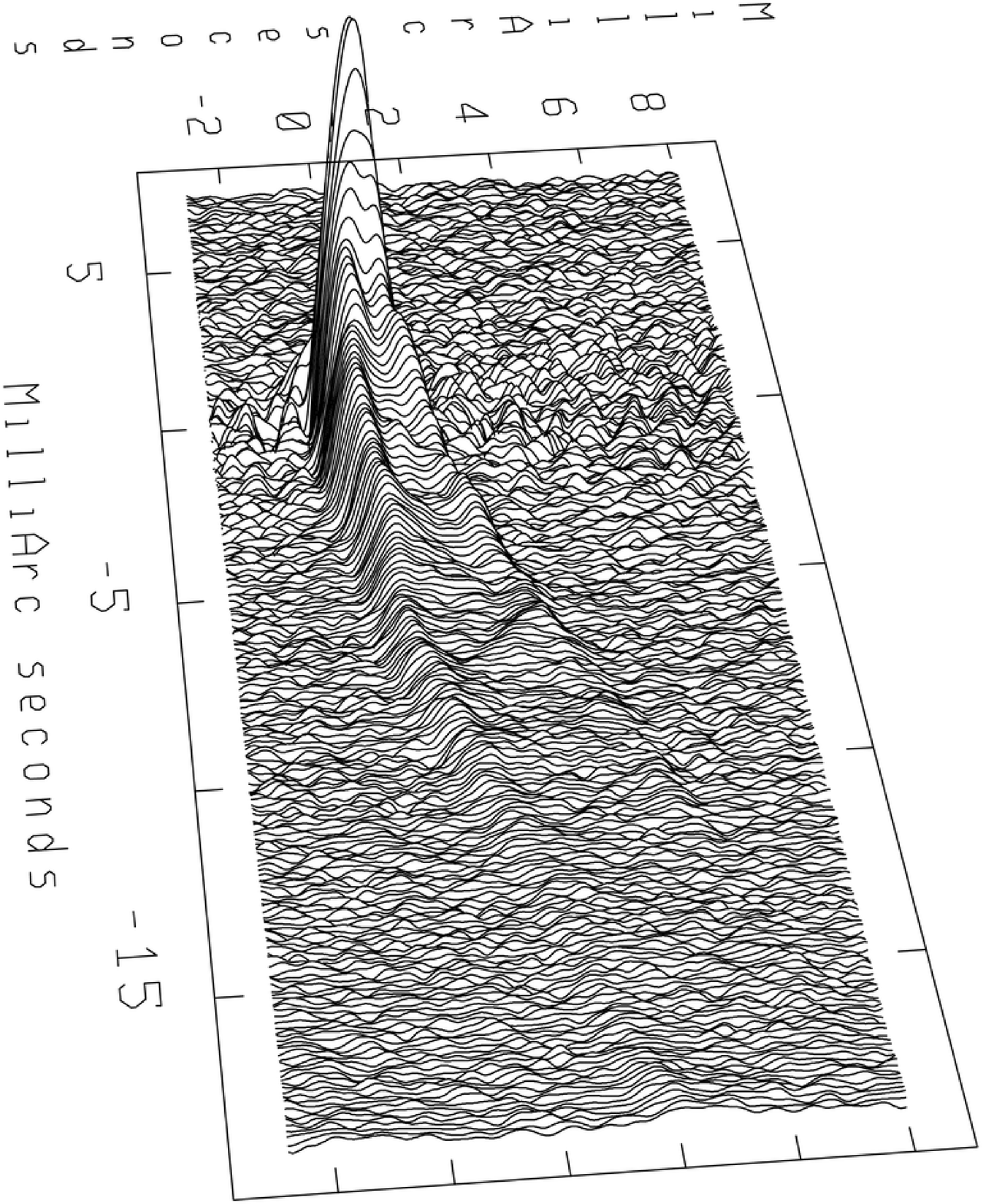}
\caption{A 3D slice profile rendering of the stacked 43 GHz VLBA image of M\,87 shown in Figure~\ref{Jet_VLBA}.
The amplitude scale is logarithmic to allow low level jet features to show without saturating the core region.
The long axis (near vertical) is the right ascension offset.
The short axis is the declination offset in mas.
The image is rotated by $75\arcdeg$ to provide a good perspective on the edge-brightened jet structure.
}
\label{slice_profile}
\end{figure*}

\begin{figure}   % Can be single column. Twocol try without [h!] ===============
\epsscale{1.17}
% \plotone{Slice_Location.eps}
\plotone{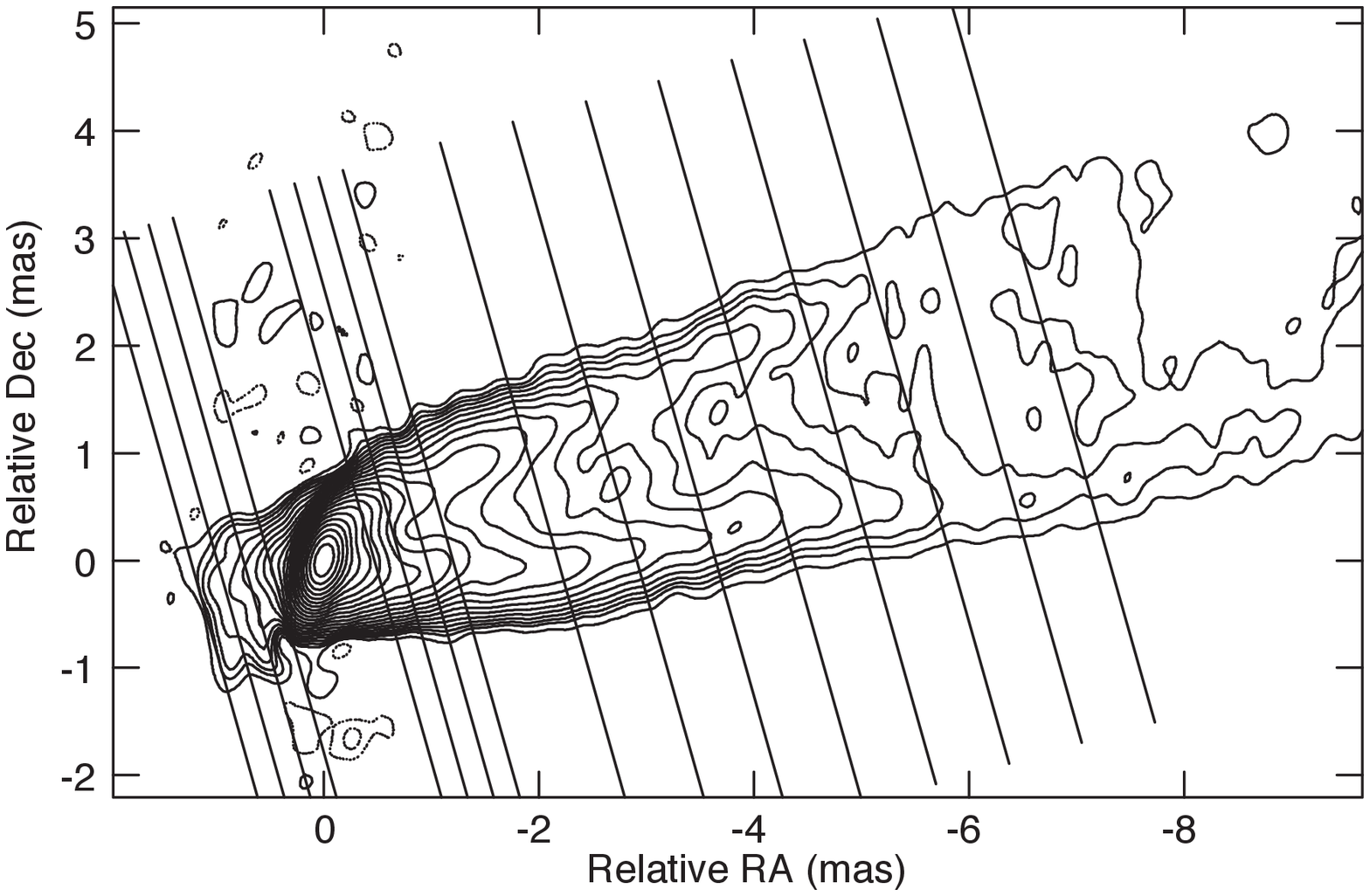}
\caption{The locations of the sample intensity slices across the jet and counter-jet shown in Figures~\ref{Slices_A} and \ref{Slices_B} are shown overlaid on a contour map 
version of the image in Figure~\ref{Jet_VLBA}.
The contour levels, in \mjb, are $-0.3$, 0.3, 0.6, 0.85, 1.2, 1.7, increasing from there by factors of $\sqrt{2}$.
The convolving beam for the clean image is $0.43 \times 0.21$~mas elongated in position angle $-16\arcdeg$.}
\label{Slice_Loc}
\end{figure}

A total of 23 images were used to make the movies showing the dynamics of the inner M\,87 jet that are presented in Section~\ref{SSec:RSE}.  
Those images have been averaged to improve the sensitivity and to show the persistent structure of the sub-parsec M\,87 jet as seen at 43~GHz over a period of 1.2 years in 2007 and 2008. 
The detailed structure that varies from epoch to epoch is smeared out by the averaging process. 
The average image is shown in Figure~\ref{Jet_VLBA}.
%
% \begin{figure*}[h!] % one column      %======================================
         %^^^^^^^^^^^^^^^^^^^^^^^^^^^^^^^^^^^^^^
At a viewing angle of $\theta = 17\arcdeg$, 1~mas in the sky plane corresponds to an intrinsic (ie, de-projected) distance along the jet of 0.27~pc or about 480~$R_{\rm S}$.
Structure on the counter-jet side, previously speculated to indicate a counter-jet \citep{Kov2007, Ly2007}, extends a little more than 1~mas to the other side of the radio core.
There are three positions located at $\unsim 1$~mas, $\unsim 3.5$~mas, and  $\unsim 13$~mas from the radio core where the jet appears to recollimate and then subsequently expand.
The image, along with those of Figures~\ref{slice_profile} and \ref{Slice_Loc}, suggests an oblique filament crossing the jet associated with a brighter feature near the jet center at $\unsim 2.5$~mas from the radio core and possibly some fainter filaments crossing the jet at larger distances from the radio core.
Overall the northern edge of the jet appears straighter than the southern edge.
An average of all the epochs to be presented in Section~\ref{SSec:LongTerm} is not used for the analysis here because of smearing of the otherwise sharp edges caused by the transverse motions reported in that section.
%
%  \begin{figure}[h!]   % Can be single column ======================
         %^^^^^^^^^^^^^^^^^^^^^^^^^^^^^^^^^^^^^^
\epsscale{1.0}
A 3D slice profile rendering of the 23-epoch total intensity image shown in Figure~\ref{slice_profile} shows more quantitatively the edge-brightening seen in Figure~\ref{Jet_VLBA}.
The southern edge is typically brighter than the northern edge by a few of tens of percent, but there is a wide scatter in the ratio with the north being brighter in a few locations.

\begin{figure*}       % Twocol ======================================
% \plotone{Slices_A.eps}
\plotone{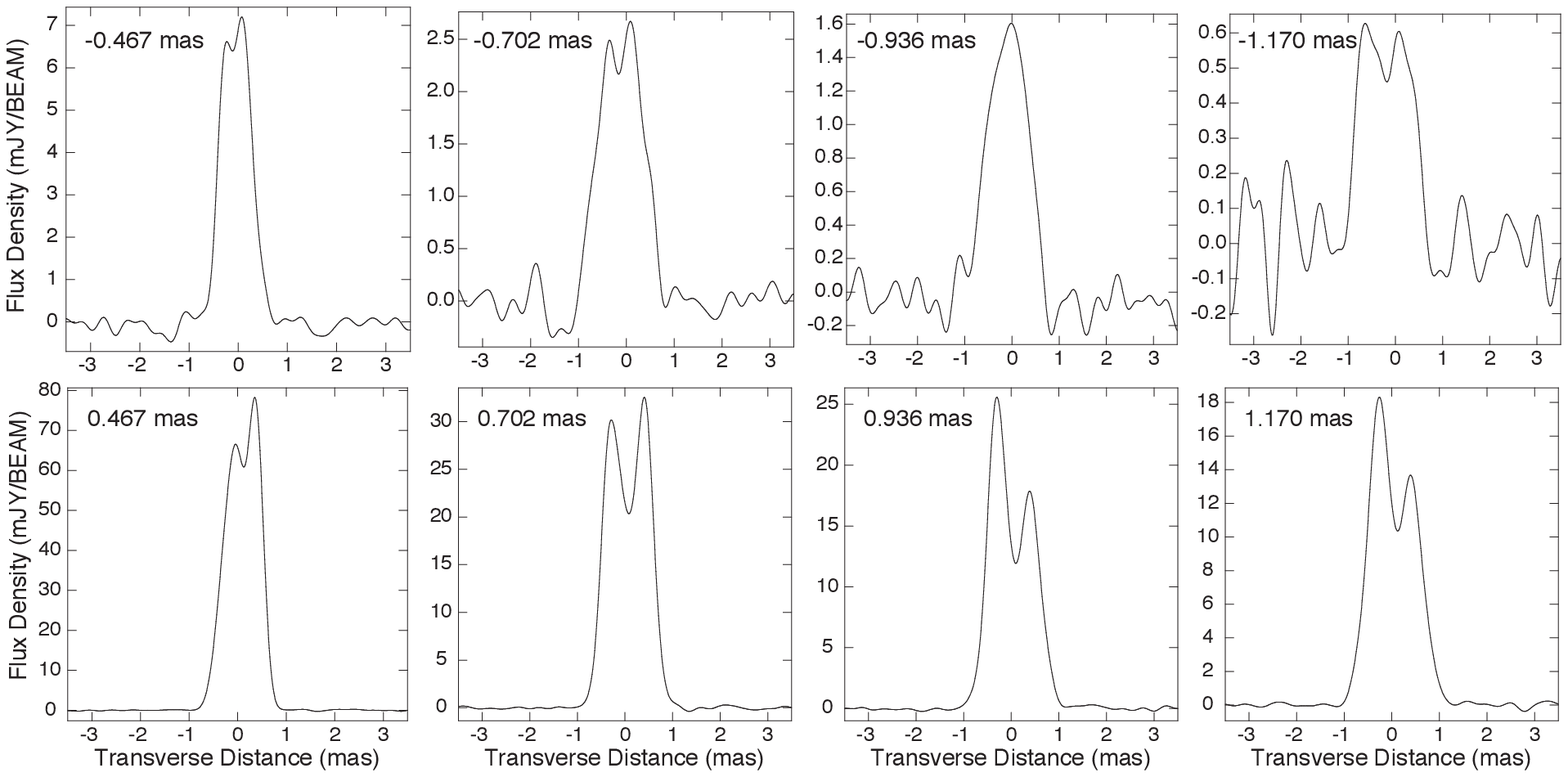}
\caption{Sample intensity slices at distances of 0.467, 0.702, 0.936 and 1.170~mas with respect to the radio core in the counter-jet (top row) and jet (bottom row) directions.
The slice locations are shown in Figure~\ref{Slice_Loc}.
The northern edge of the jet and counter-jet is at positive values on the transverse (mas) scale.}
\label{Slices_A}
% \end{figure}         %^^^^^^^^^^^^^^^^^^^^^^^^^^^^^^^^^^^^^^
\end{figure*}

\begin{figure*}       %======================================
% \plotone{Slices_B.eps}
\plotone{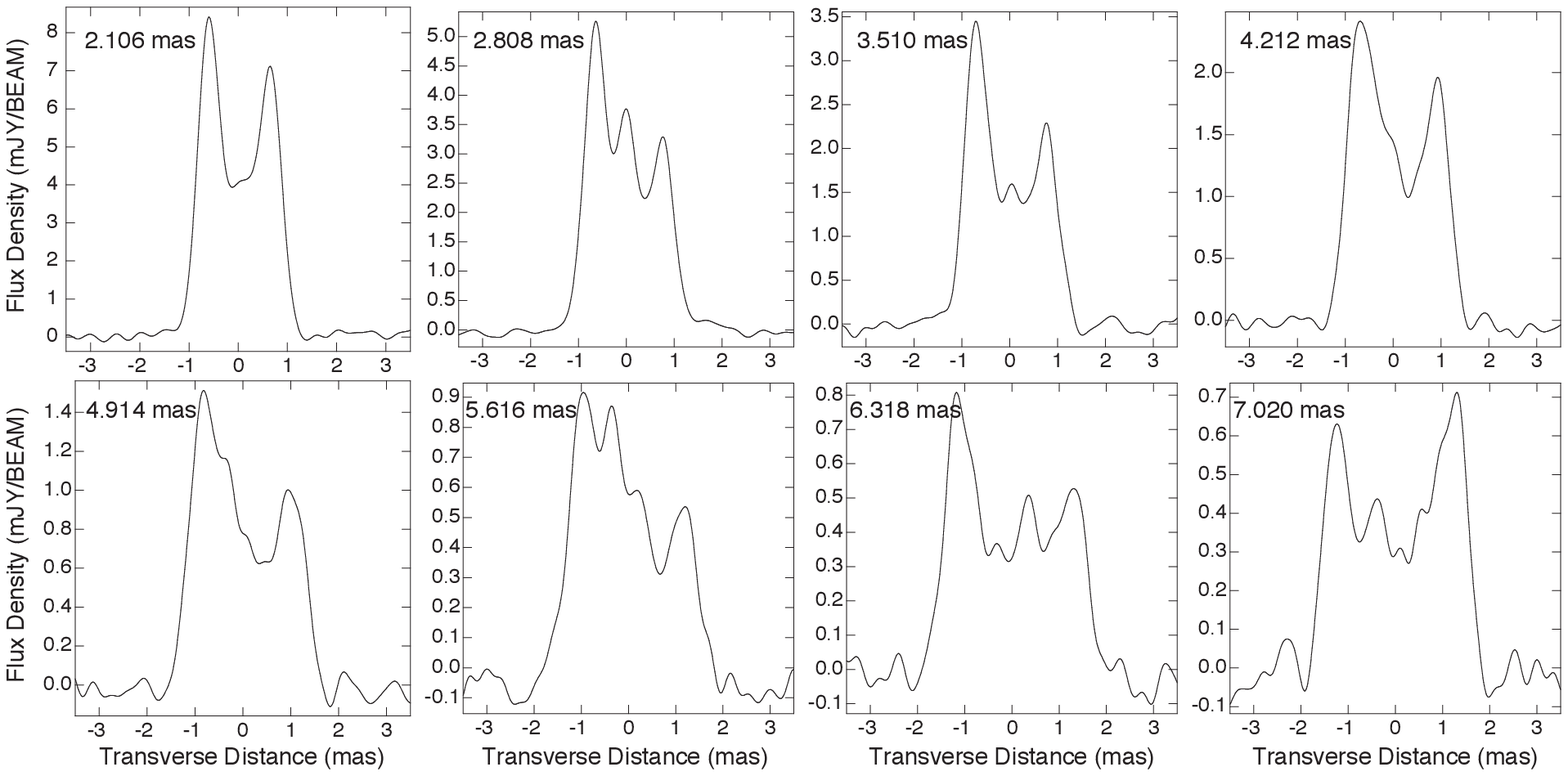}
\caption{Sample intensity slices across the jet (top row) at positions 2.106, 2.808, 3.510 \& 4.212~mas and (bottom row) at positions 4.914, 5.616, 6.318 \& 7.020~mas with respect to the core as shown in Figure~\ref{Slice_Loc}.} 
\label{Slices_B}
% \end{figure}         %^^^^^^^^^^^^^^^^^^^^^^^^^^^^^^^^^^^^^^
\end{figure*}

In order to examine in detail the transverse structure and compare counter-jet to jet sides of the core, slices of the intensity image separated by 0.117~mas were made from $-1.287$~mas to 7.956~mas along the counter-jet and jet axis, i.e., separated by about half a beamwidth.
%
% \begin{figure}[h!]   % Can be single column ==========================
         %^^^^^^^^^^^^^^^^^^^^^^^^^^^^^^^^^^^^^^
\epsscale{1.0}
Figure~\ref{Slice_Loc} shows an intensity contour plot of the inner 10 mas of the image shown in Figure~\ref{Jet_VLBA} with lines superimposed that mark the locations of the sample intensity slices shown in Figures~\ref{Slices_A} and \ref{Slices_B}. 
 These sample slices are a subset of all of those calculated and used in the analysis.
Here the slice axis is located at an angle of $15.76\arcdeg$ with respect to the R.A. axis.
Figure~\ref{Slices_A} shows sample transverse intensity slices at 0.234~mas intervals in the inner  $\pm 1.2$~mas  centered on the 43~GHz core.
%
% \begin{figure}       %======================================
       % Twocol ======================================
Similar transverse edge-brightened structure is evident on both jet and counter-jet sides of the core, albeit on the counter-jet side inside $-1.2$~mas there is one position where the maximum intensity is interior to the bright ridgelines (at $-0.819$~mas; not shown in Figure~\ref{Slices_A}) and one position where the intensity is centrally peaked (at $-0.936$~mas; shown in  Figure~\ref{Slices_A}).
On the jet side all slices inside 1.2~mas are edge-brightened.
 The northern edge is brighter nearer the core along both jet and counter-jet in the two innermost slices.
Along the jet the southern edge becomes brighter in the two slices farthest from the core, and there is weak indication of similar behavior along the counter-jet in the slice farthest from the core.
In the 23-epoch intensity image we cannot follow the counter-jet farther from the core.

Figure~\ref{Slices_B} shows sample transverse intensity slices along the jet from $\unsim 2.11$~mas to $\unsim 7.02$~mas in 0.702~mas (approximately three beamwidth) intervals.
The slices show the edge-brightening apparent in the 23-epoch image (Figure~\ref{Jet_VLBA}) and 3D slice profiles (Figure~\ref{slice_profile}).
Additionally, these slices show that the transverse intensity profile is more complicated than double edge peaked, in particular between $\unsim 2.3$ and $\unsim 3.5$~mas (2.808~mas; shown in Figure~\ref{Slices_B}) and also between $\unsim 4.5$ and $\unsim 7.0$~mas (bottom row in Figure~\ref{Slices_B}).
The structures responsible for these more complicated intensity profiles might be associated with filaments crossing the jet diagonally. Diagonal filaments are suggested by faintly visible structure in Figure~\ref{Jet_VLBA}, and are somewhat easier to identify in the 3D slice profiles of Figure~\ref{slice_profile} and in the
contour plot underlying Figure~\ref{Slice_Loc}.

%
% \begin{figure}       %======================================
         %^^^^^^^^^^^^^^^^^^^^^^^^^^^^^^^^^^^^^^

\begin{figure*}    %  Twocol try without [h!]  ==============================
\epsscale{0.8}
% \plotone{jtcomb17.eps}
\plotone{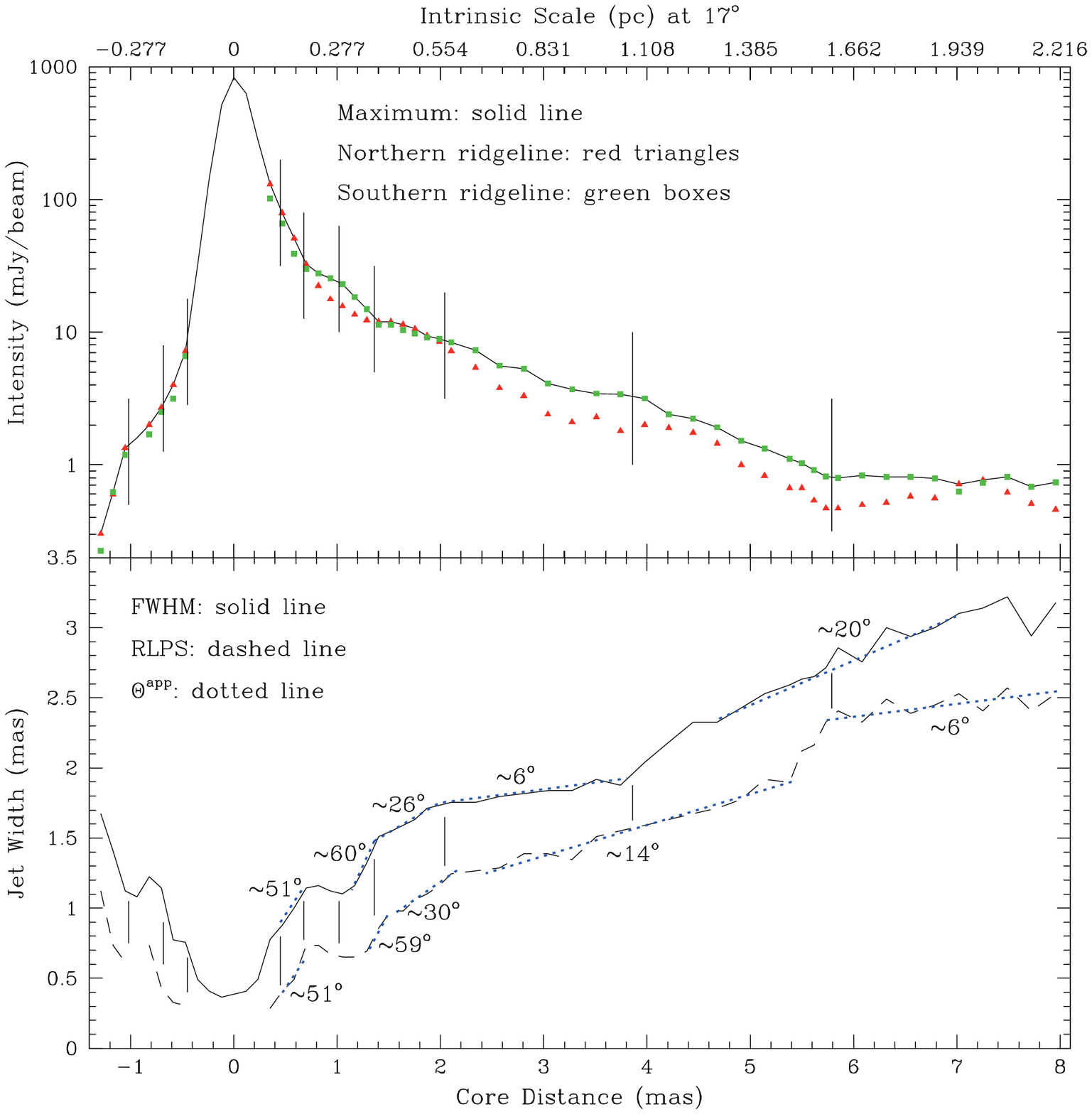}
\caption{Top panel: Maximum intensity (solid line) and maximum intensity associated with the northern (red triangles) and southern (green squares) ridgelines are plotted along with vertical lines indicating the first location not dominated by the unresolved core and subsequently the locations of change in the intensity decline along the jet.
Vertical lines along the counter-jet side are placed at the same core distance as the first three lines along the jet.
Bottom panel: Full width at half maximum (FWHM) and the ridgeline peak separation (RLPS) in mas along with vertical lines transferred from the top panel are shown.
Simple linear fits to the apparent jet full opening angle, $\Theta^{\rm app}$, for both FWHM and RLPS are  indicated by the dotted lines.
Angular scale is indicated along the bottom axis, and intrinsic scale for the jet at $17\arcdeg$ to the line-of-sight is indicated along the top axis.}
\label{Jet_Fits}
% \end{figure}         %^^^^^^^^^^^^^^^^^^^^^^^^^^^^^^^^^^^^^^
\end{figure*}

The intensities associated with the bright northern and southern jet edge ridgelines along with the maximum intensity at each transverse slice position are shown in the top panel in Figure~\ref{Jet_Fits}.
%
% \begin{figure}[h!]       %======================================
         %^^^^^^^^^^^^^^^^^^^^^^^^^^^^^^^^^^^^^^
\epsscale{1.0}
Along the jet, the maximum intensity is always associated with one of the bright ridgelines.
In the innermost $\pm 0.3$~mas the intensity profile is dominated by the unresolved core although even at $\pm 0.2$~mas the intensity on the jet side significantly exceeds the counter-jet side.
Vertical lines in the intensity plot are placed across the jet at the first location where the unresolved radio core does not significantly affect the intensity ($\unsim 0.4$~mas), and at larger distances where the intensity decline with distance changes slope significantly.
Three vertical lines across the counter-jet side are placed at the same radio core distance as the first three lines along the jet.
Along the jet and counter-jet sides of the core the second vertical line at $\unsim \pm 0.7$~mas is associated with a significant reduction in the intensity decline and the third vertical line at $\unsim 1$~mas is associated with a significant increase in the intensity decline.
  
The bottom panel in Figure~\ref{Jet_Fits} shows that change in the slope of the intensity decline can be identified with change in the expansion rate along the jet and the counter-jet side.
The bottom panel shows the full-width half maximum (FWHM) width, the ridgeline peak separation (RLPS) between the intensity peaks associated with the edge-brightened ridgelines, and simple linear fits to the apparent jet full opening angle, $\Theta^{\rm app}$, along the jet where the apparent full opening angle is defined as
$$
\Theta^{\rm app} \equiv \arctan \left( {\Delta W / \Delta z} \right)~,
$$
where $\Delta W$ and $\Delta z$ are the change in jet width and in core distance in mas.
The symmetry in the opening angle between the jet and counter-jet sides along with the similar edge-brightening confirms the identification of the faint structure as the counter-jet.
In both jet and counter-jet we see an initial rapid opening, $\Theta^{\rm app} \approx 51\arcdeg$, followed by a region in which the width decreases slightly,  $\Theta^{\rm app} \lesssim 0\arcdeg$.
Subsequently both jet and counter-jet increase in width rapidly with $\Theta^{\rm app} \approx 59\arcdeg - 60\arcdeg$, and the counter-jet becomes too faint to follow farther.
The obvious symmetry on either side of the radio core in the jet and counter-jet structure at our 0.117~mas slice separation means that the radio core is located no more than half the slice interval, i.e., one quarter the beamwidth, and $< 0.06$~mas from the central engine.
Thus the radio core is located at $< 8.5$~$R_{\rm S}$ from the central engine, consistent with the results of \citet{Hada2011}.

Following the second rapid increase in width the jet opening angle decreases first to $\Theta^{\rm app} \approx 26\arcdeg - 30\arcdeg$ and subsequently to $\Theta^{\rm app} \approx 6\arcdeg - 14\arcdeg$.
Up to about 3.9~mas from the radio core both FWHM and RLPS widths change their slopes at about the same radio core distance and to within the uncertainties $\Theta^{\rm app}$ is about the same for FWHM and RLPS determined widths.
This spatially synchronized behavior is broken beginning at $\unsim 3.9$~mas where the FWHM width increases rapidly out to $\unsim 4.4$~mas without a similar increase in the RLPS width.
The RLPS width shows a similar rapid increase beginning at  $\unsim 5.5$~mas extending to  $\unsim 5.8$~mas.  
It is unclear whether these differences between the two width measurements are significant or just reflect uncertainty in the measurement methods, especially given the weakness of the northern ridge relative to the center, to the large brightness difference between the sides, and changes occurring along the southern envelope in this region.
Beyond 5.8~mas the RLPS width indicates $\Theta^{\rm app} \approx 6\arcdeg$, similar to the parsec- and kpc-scale jet opening angle.
The implications of these structures will be discussed in Section~\ref{SSec:JER}.

\subsection{Rapid Structure Evolution}
\label{SSec:RSE}
In this section, we present individual epoch images and movies made from them that allow detailed studies of the structural evolution of the M\,87 jet near the core.
The structure is complex and rapidly evolving, and is not easily described as a series of moving features.
The most convincing way to see that there is significant outward motion is to view the movies which give a strong visual impression of such motion.
Despite the difficulties, we perform a traditional component motion analysis for the 2007 and 2008 data and show the results here.
For a full velocity analysis of our 2007 data using more sophisticated techniques that can identify multiple, overlapping, velocity fields, see \citetalias{MLWH2016}.

\subsubsection{Intensity Images}

% \onecolumngrid      %  Failed attempt to center sidewaysfigure with two columns.
% \begin{center}      %  Failed attempt to center sidewaysfigure with two columns.
% \begin{sidewaysfigure*}[p]       %======================================
\begin{figure*}   % twocolumn?     %======================================
\epsscale{1.17}
\centering
% \plotone{montage_cntr_comp_2007a6.ps}
%  Used the following at the command line to make smaller figures:
% convert -density 300 -background white -flatten f7.eps -resize 1024x1024 f7_1024.gif
% convert f7_1024.gif f7_1024.eps
\plotone{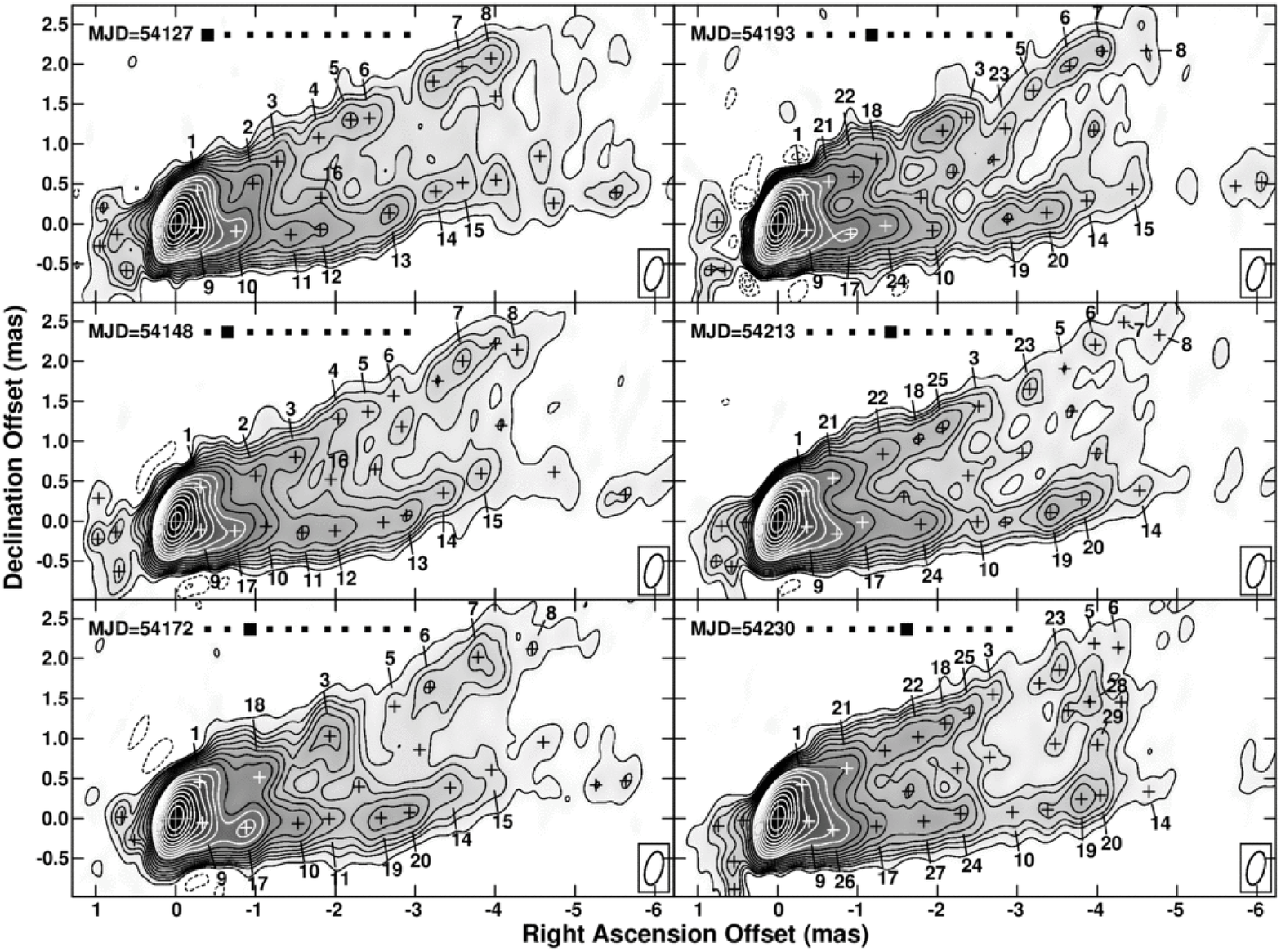}
\caption{Contour/gray scale plots of observations 1-6 of those that were made of M\,87 with the VLBA at 43 GHz during 2007 at about three week intervals.
 The convolving beam size used was $0.43 \times 0.21$ mas elongated along position angle $-16\arcdeg$.
The contour levels (\mjb) are $-1.4$, $-1.2$, $-1$, 1, 1.2, 1.4, 1.7, 2, 2.4, 2.8, 3.4, 4, 4.8, 5.7, 6.7, 8, 9.5, 11.3, 13.4, 16, 19, 23, 27, 32, 38, 45, 91, 181, 362, and 724.
The image off-source noise level is between about 0.17 and 0.25 \mjb.
Components, identified by eye, are marked with crosses.
Numbers are for components that appear to be related between epochs based on proximity and other cues in the structure.
}
\label{Epochs_A1}
\end{figure*}  % twocolumn?     %^^^^^^^^^^^^^^^^^^^^^^^^
% \end{sidewaysfigure*}         %^^^^^^^^^^^^^^^^^^^^^^^^^^^^^^^^^^^^^^
% \end{center}
% \twocolumngrid
\epsscale{1.0}

% \begin{sidewaysfigure*}       %======================================
\begin{figure*}   % twocolumn?     %======================================
\epsscale{1.17}
\centering
% \plotone{montage_cntr_comp_2007b6.ps}
\plotone{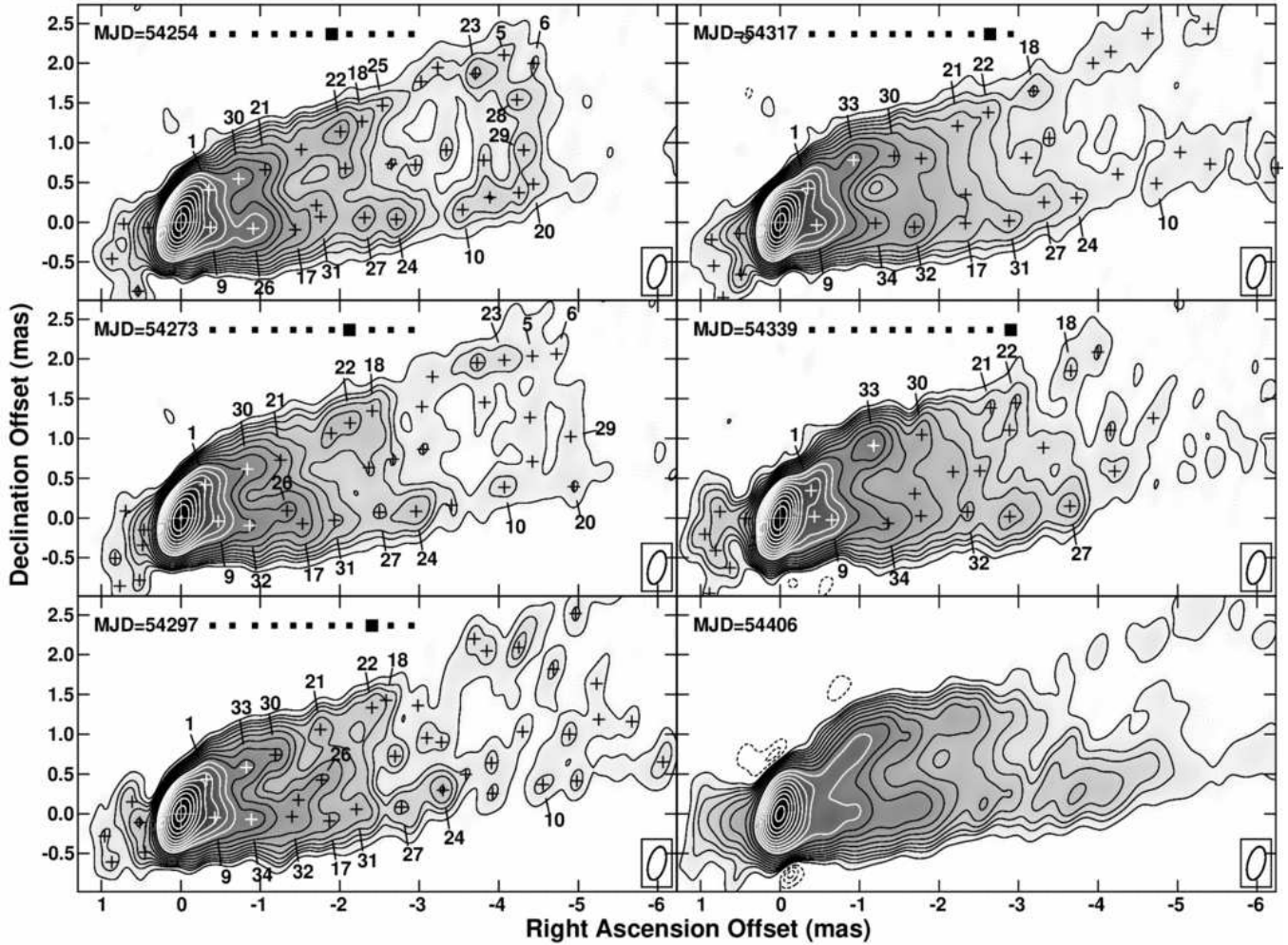}
\caption{Contour/gray scale plots of observations 7-11 of those that were made of M\,87 with the VLBA at 43 GHz during 2007 at about three week intervals.
The 11 images were used to make the movie shown in online animation Figure~\ref{Mov07}.
The final image (bottom right panel) is from late in 2007.
It is included for completeness but other data sets near it in time have problems and are not yet fully reduced so this image from late in 2007 was not used in the movie, the stack, and most of the analysis.
The convolving beam size, contour levels, and marked components for all of the images shown here are the same as in Figure~\ref{Epochs_A1}.
}
\label{Epochs_A2}
\end{figure*}  % twocolumn?     %^^^^^^^^^^^^^^^^^^^^^^^^
% \end{sidewaysfigure*}         %^^^^^^^^^^^^^^^^^^^^^^^^^^^^^^^^^^^^^^
\epsscale{1.0}

% \begin{figure}[h!]       %======================================
\begin{figure*}      %  Twocol try without [h!]  =====================
\epsscale{0.80}
% \plotone{BW088_RGB1_MOV_0.0_12.PS} later: BW088_RGB1_MOV_FRE18_0.0_12.PS
\plotone{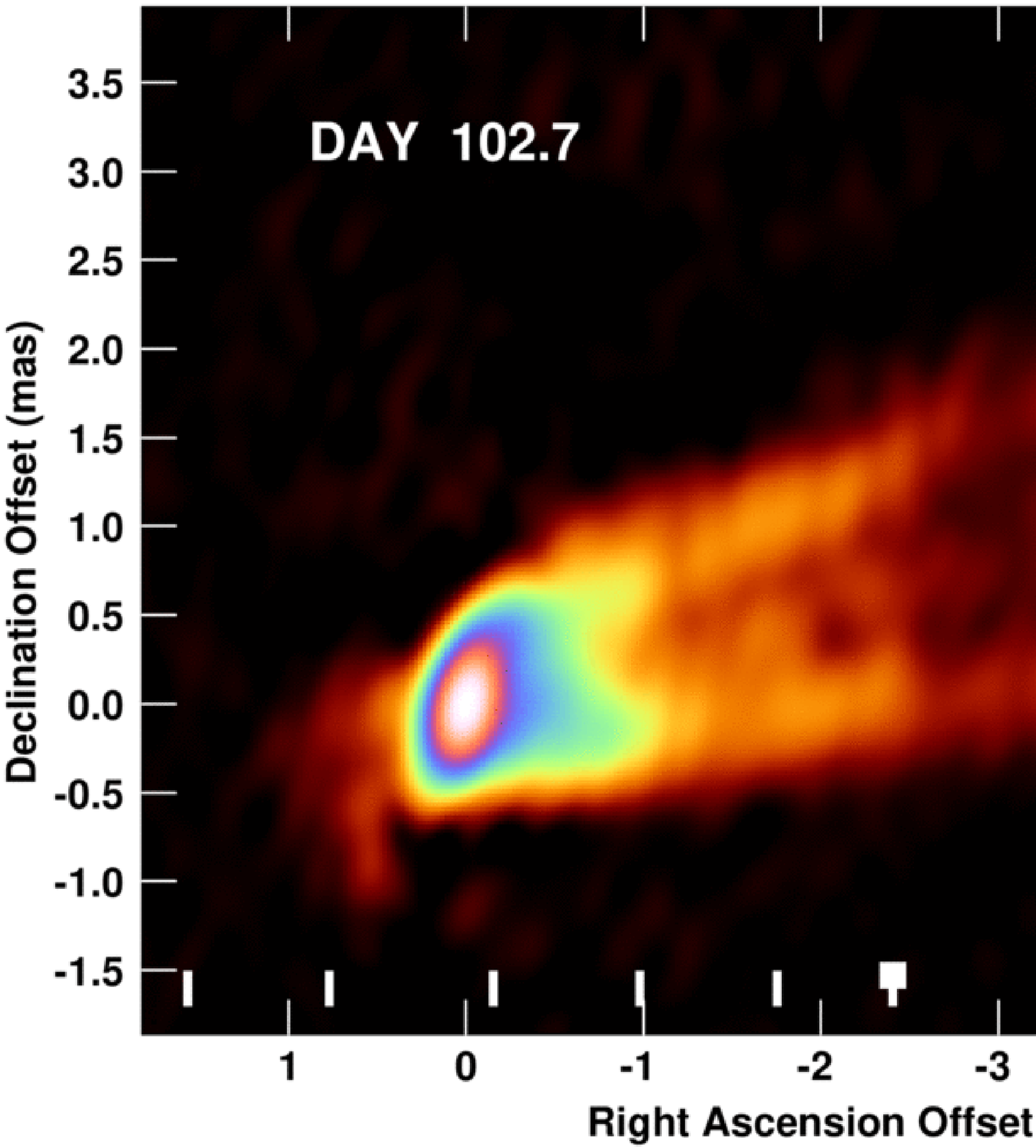}
%  The animation file is M87_VLBA_2007.mp4
\caption{For the offline versions of this paper, this figure is a colorized version of one frame from the 2007 movie.
Online at the journal, or in astro-ph ancillary file M87\_VLBA\_2007.mp4, the figure is the movie, showing the first 11 images in sequence, with pauses proportional in time to the interval between epochs (typically three weeks).
The underlying images are the same as those in Figures~\ref{Epochs_A1} and \ref{Epochs_A2}.
The series of ticks along the bottom indicates elapsed time and the moving box indicates which epoch is being displayed.
The movie shows the actual observed epochs because confusing artifacts appear when interpolating undersampled data.
The movie provides a clear visual impression of outward motion, but also of a rapidly evolving pattern that makes it hard to follow individual features over long periods.
}
\label{Mov07}
% \end{figure}            %^^^^^^^^^^^^^^^^^^^^^^^^^^^^^^^^^^^^^^
\end{figure*}            % Twocol ^^^^^^^^^^^^^^^^^^^^^^^^^^^^^^^^^^^^^^
\epsscale{1.0}

Figures~\ref{Epochs_A1} and \ref{Epochs_A2} show the total intensity contour plots from the 11 epochs at about three week intervals between 2007 January 27 (MJD 54127) and 2007 August 26 (MJD 54339).
The line of squares next to the dates gives a visual representation for the time of each epoch, especially if the images are viewed as a movie.
The filled square representing the particular epoch shown in each panel is enlarged.
Over this period the standard deviation of the 43~GHz flux from the core (inner 1.2~mas) was $\unsim 5\%$, not significantly above the noise and uncertainties.
There are significant changes to the intensity along the edges and down the centerline of the jet.
Significant change is also seen in the intensity of the counter-jet.
Some of the variation is the result of instrumental effects but other variation is the result of true intensity changes that may be related to motions along the jet.
True structural changes are large enough that it is difficult to connect features from epoch to epoch when sampled at three week intervals.

% \begin{sidewaysfigure*}       %======================================
\begin{figure*}   % twocolumn?     %======================================
\centering
\epsscale{1.17}
% \plotone{montage_cntr_comp_2008a6.ps}
\plotone{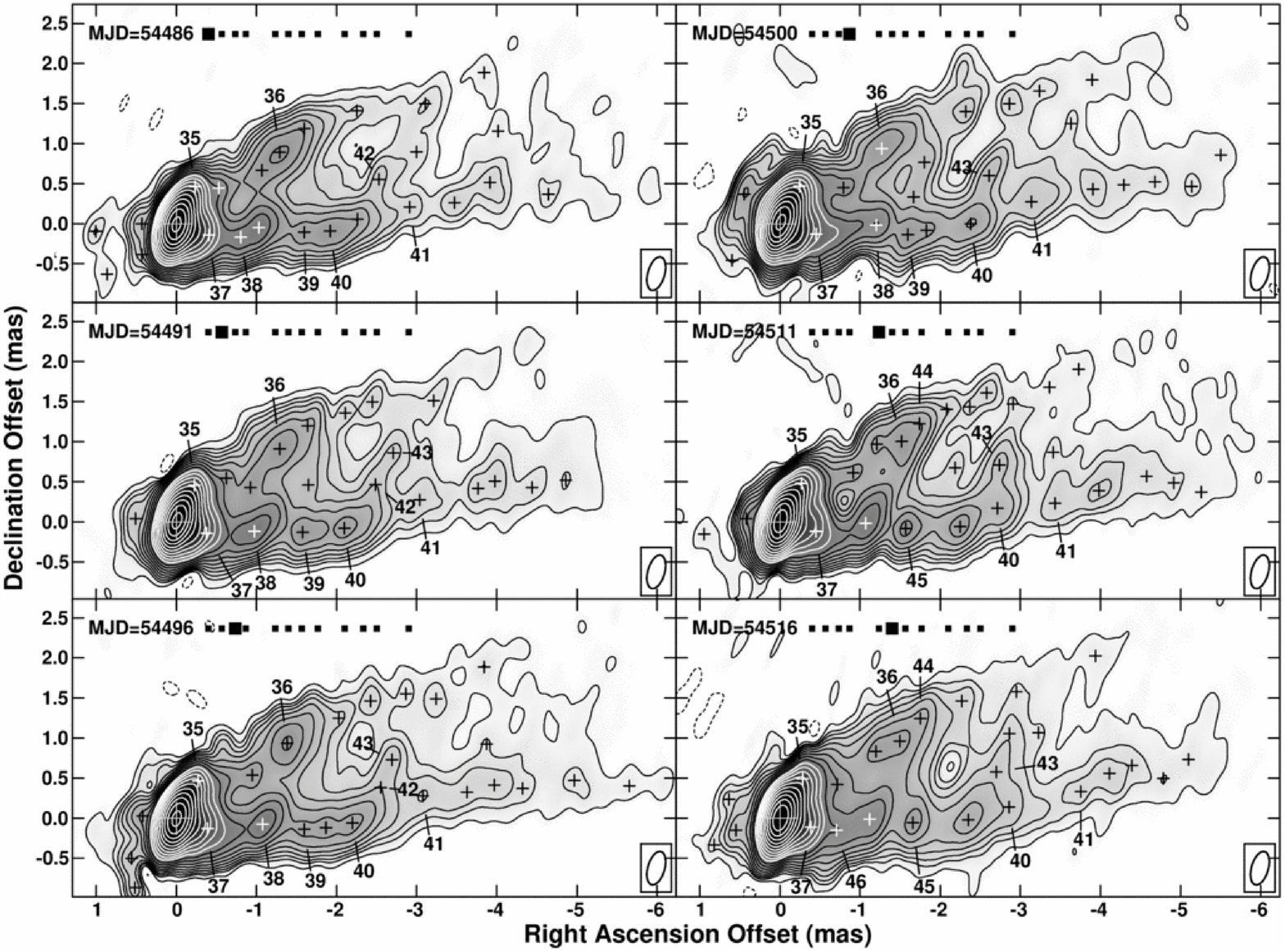}
\caption{Contour plots of the total intensity of M\,87 at 43 GHz for first 6 epochs at about 1 week intervals observed in 2008.
This set starts on 2008 January 21 and ends on 2008 February 20.
The beam dimensions, contour levels, and meaning of the marks are the same as for Figure~\ref{Epochs_A1}.
}
\label{Epochs_B1}
\end{figure*}  % twocolumn?     %^^^^^^^^^^^^^^^^^^^^^^^^
% \end{sidewaysfigure*}         %^^^^^^^^^^^^^^^^^^^^^^^^^^^^^^^^^^^^^^
\epsscale{1.00}

% \begin{sidewaysfigure*}       %======================================
\begin{figure*}   % twocolumn?     %======================================
\centering
\epsscale{1.17}
% \plotone{montage_cntr_comp_2008b6.ps}
\plotone{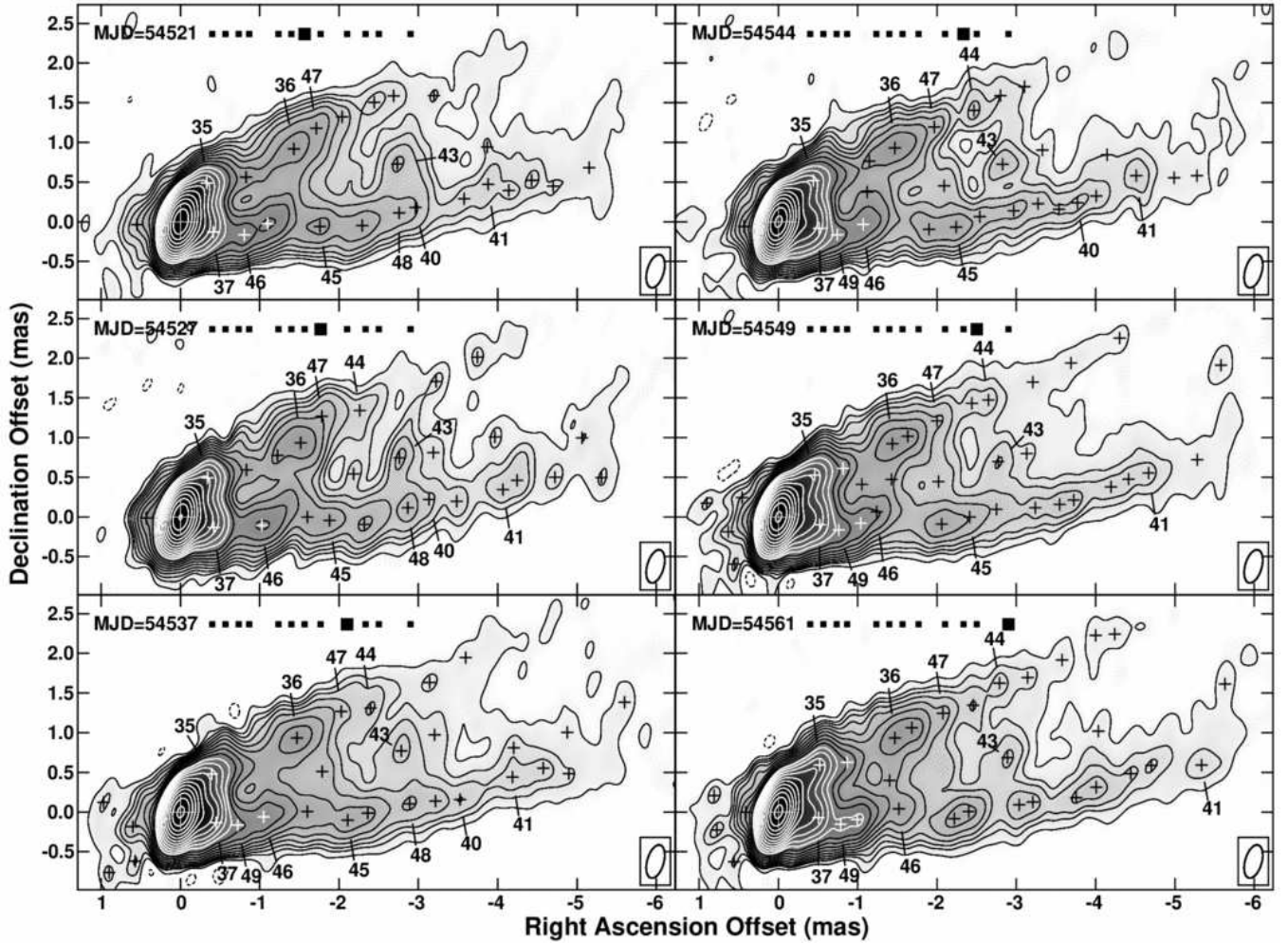}
\caption{Contour plots of the total intensity of M\,87 at 43 GHz for the last 6 epochs at about 1 week intervals observed in 2008.
This set starts on 2008 February 25 and ends on 2008 April 5.
The beam dimensions, contour levels, and meaning of the marks are the same as for Figure~\ref{Epochs_A1}.}
\label{Epochs_B2}
\end{figure*}  % twocolumn?     %^^^^^^^^^^^^^^^^^^^^^^^^
% \end{sidewaysfigure*}         %^^^^^^^^^^^^^^^^^^^^^^^^^^^^^^^
\epsscale{1.00}

% \begin{figure}       %======================================
\epsscale{0.80}
\begin{figure*}       % Twocol ======================================
% \plotone{BW090_RGB1_MOV_0.0_19.PS}  Later:  BW090_RGB1_MOV_FRE18_19.PS
\plotone{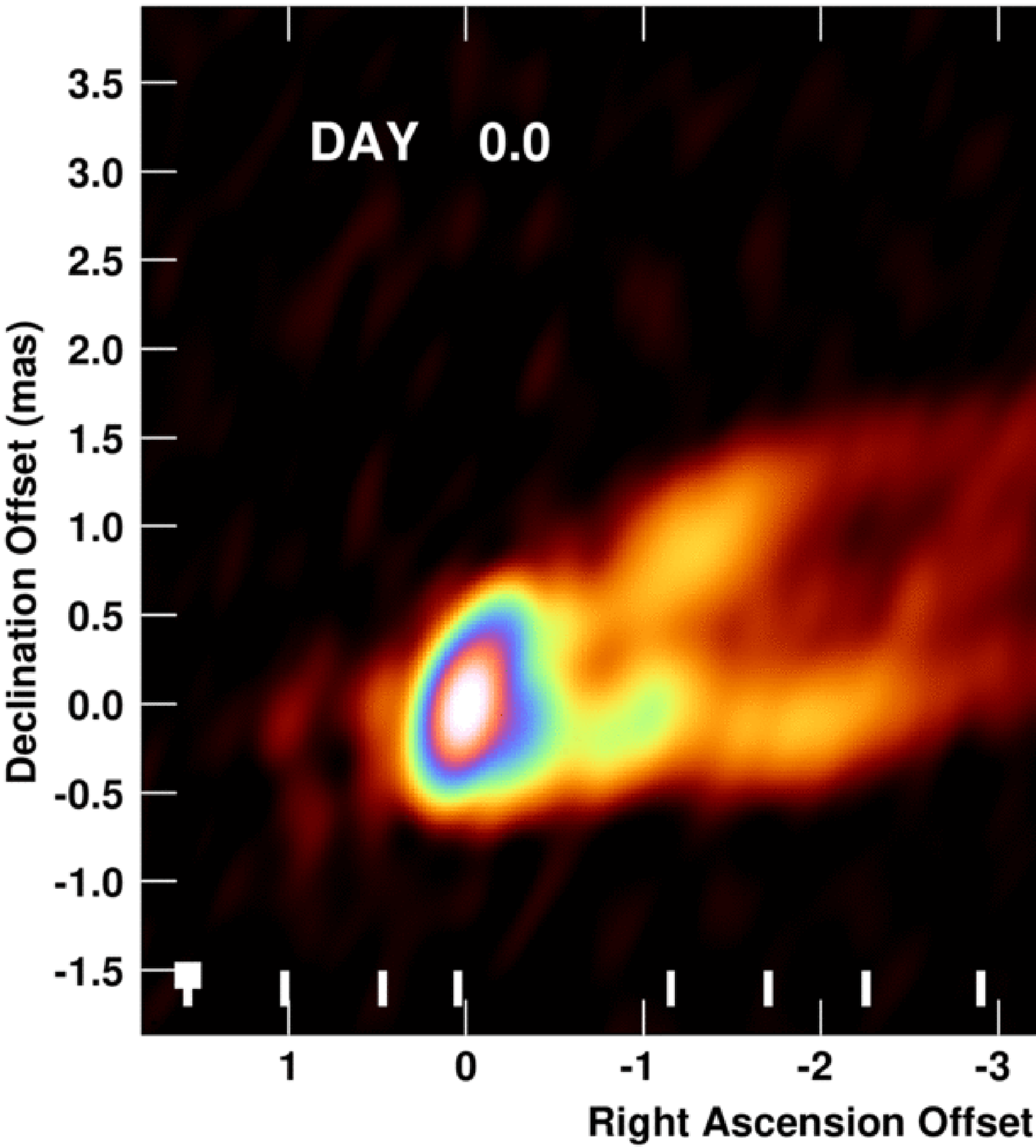}
%  The animation file is M87_VLBA_2008.mp4
\caption{For the offline versions of this paper, this figure is a colorized version of one frame from the movie made from the 2008 data.
Online at the journal, or in astro-ph ancillary file M87\_VLBA\_2008.mp4, this figure is a movie showing the 12 images in sequence, with pauses proportional in time to the interval between epochs (typically a week).
The underlying images are the same as those in Figures~\ref{Epochs_B1} and \ref{Epochs_B2}.
The series of ticks along the bottom indicate elapsed time and the moving box indicates which epoch is being displayed.
The data quality was not as good as in 2007 because tight observing date constraints precluded waiting for better weather.
Three epochs were so degraded that the images were not included, as evidenced by gaps in the time sequence.
Despite these degradations, the movie still gives a clear visual impression of outward motion.}
\label{Mov08}
% \end{figure}         %^^^^^^^^^^^^^^^^^^^^^^^^^^^^^^^^^^^^^^
\end{figure*}       % Twocol ======================================
\epsscale{1.0}         %  Must reset (outside fig env) or affects next figure.

A movie of the color image versions of the 2007 images is shown in online animation Figure~\ref{Mov07}.  
Offline, that figure is a sample frame from the movie.
Interpolation is not used for the movie because the typical motion in 2007 between epochs is about 2.5 times the beam width and interpolation leads to distracting artifacts in the structure or the noise depending on how the interpolating is done.
Viewing the movie is the best way to see the jet motions.

Observations continued between MJD 54363 and 54464 at three week intervals, but most have issues that complicate processing.
Those issues have been partially overcome for the observation on 2007 October 26 (MJD 54406), which is shown as the last panel in Figure~\ref{Epochs_A2}.
This epoch is sufficiently isolated in time from the other epochs that it is not used in the time series analysis or the 23-image stack.
It is included in the 49-image stack in Section~\ref{SSec:CJ} on the counter-jet.
Technically this image is of interest because the observations were made at four frequencies (39.4, 41.3, 43.1, and 44.8 GHz) in an attempt to improve the UV coverage.
The benefit of four frequencies was not fully realized because the two stations giving the longest baselines were degraded --- one was out for maintenance, the other had poor weather.

In Figures~\ref{Epochs_A1} and \ref{Epochs_A2}, crosses indicate the visually-determined positions of local maxima in the total intensity.
An attempt to relate these features from epoch to epoch was made by blinking rapidly back and forth between the epochs.
The attempt is hampered, as noted above, by the large motions between epochs.
But a variety of visual cues including the peak location, and also aspects such as major dips in the ridge line, made it possible to identify a significant number of components.
Each feature that appeared related in three or more successive epochs was given a number.
The motions of these components are analyzed in Section~\ref{SSSec:CompMotion}.

Figures~\ref{Epochs_B1} and \ref{Epochs_B2} show total intensity contour plots of M\,87 for 12 epochs at intervals of about 1 week from 2008 January 21 to 2008 April 05.
These plots are similar to the images shown in Figures~\ref{Epochs_A1} and \ref{Epochs_A2} for the 2007 data, although the observing interval is significantly shorter.
Feature identification was done in the same manner as for the 2007 data and the results are included in our motion analysis.
In principle, with the shorter interval, it should have been easier to relate features between epochs.
But, as noted in Section~\ref{Sec:Obs}, many of the images were of lower quality than those in 2007 so feature identification was as or more difficult.
A movie of the color image versions of the 2008 images is shown in online animation Figure~\ref{Mov08}.  
Offline, that figure is a sample frame from the movie.
Due to the lower quality of many of the images, the impression of motion is not as strong as with Figure~\ref{Mov07}, but the impression of motion is still present.

Serendipitously, TeV flaring occurred during the first two weeks of February 2008 and it is around this time that a significant long term increase in the core intensity at 43 GHz began.
By April 2008 the core intensity had increased by 60\% relative to the pre-TeV flaring average.
Over the same time period the jet edges first brighten near to the core with brightening extending outwards over time.
The increased brightening along the jet edges inside 1~mas can be seen by comparing the panel for January 21 in Figure~\ref{Epochs_B1}, from before the TeV flaring, to the panel for April 5 in Figure~\ref{Epochs_B2}.
It is even more clear in the difference images shown in Figure~3 of \citet{Acc09}.

\subsubsection{Component Motion Analysis}
\label{SSSec:CompMotion}

Figures~\ref{Epochs_A1}-\ref{Epochs_A2} and \ref{Epochs_B1}-\ref{Epochs_B2} show a jet structure that is complex and rapidly evolving.
%
% \begin{figure*}[h!]      %======================================
\begin{figure*}    % twocolumn     %======================================
\epsscale{0.8}
% \plotone{south.ps}
\plotone{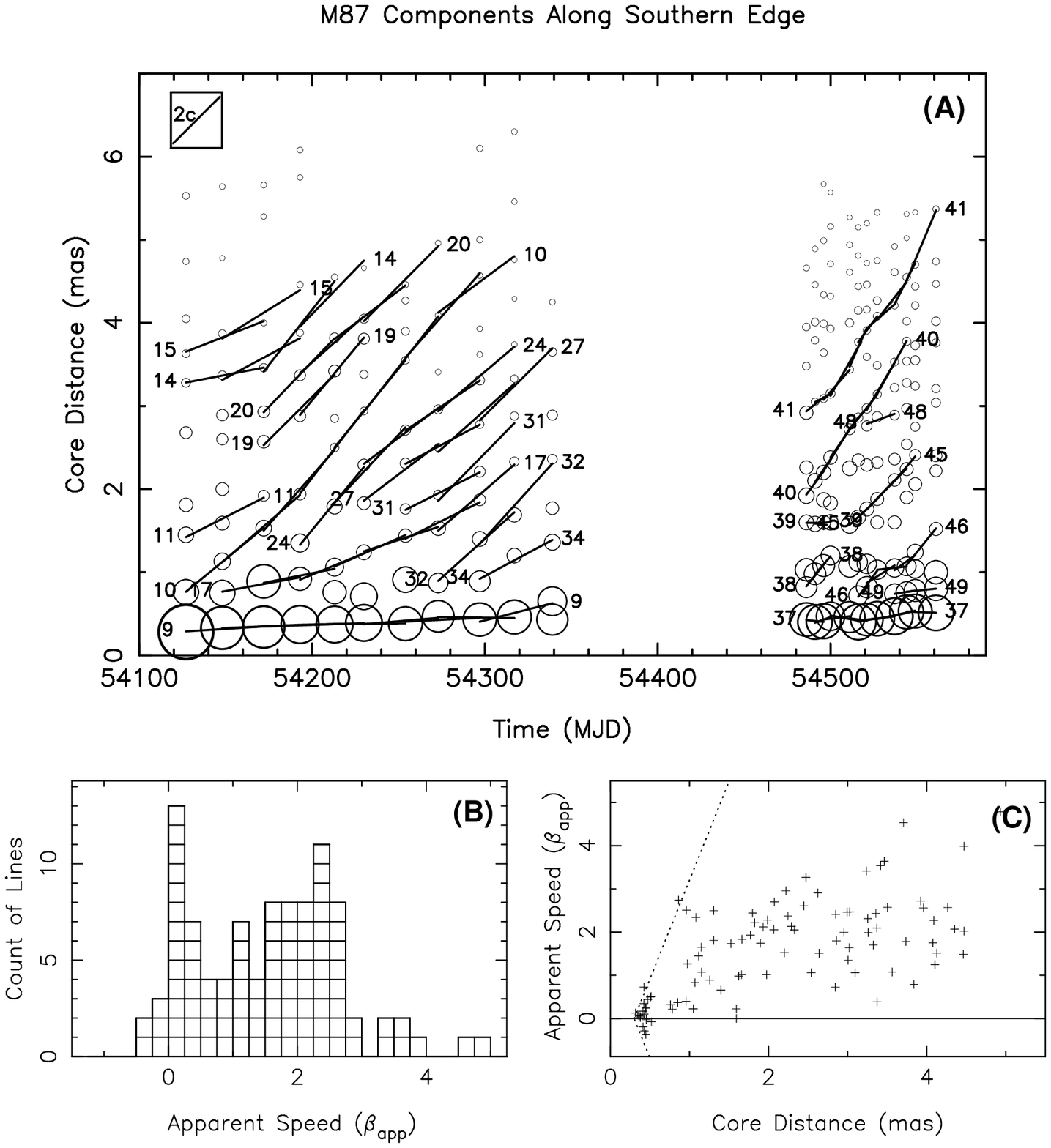}
\caption{Panel A: Positions of components that lie along the southern edge of the M\,87 jet as indicated in Figures~\ref{Epochs_A1}, \ref{Epochs_A2}, \ref{Epochs_B1}, and \ref{Epochs_B2} as a function of time (For reference, 2007 January 1 is MJD 54101) are marked by circles whose area is proportional to the component peak flux density.
Where three components have the same number in successive epochs, a fit was done for speed and position.
Each such fit is represented by a line segment in the figure.
Panel B:  A histogram of the number of line segments in Panel~A as a function of apparent speed $\beta^{\rm app} = v^{\rm app}/c$.
There are concentrations of speeds near zero, mostly from positions near the core, and concentrations at $\gtrsim 2$c, although intermediate and some faster speeds are also seen.  Panel C:  A plot of apparent speed with core distance (mid-point of line segment).
The dotted line is the right boundary of the exclusion region resulting from the requirement that the first of the three points in a line segment be resolved from the core.
The boundary shown is for the three-week cadence of the 2007 observations.
 As is apparent from looking at Panel~A, higher velocities are at larger core distances.
 For distances less than about 2 mas, an acceleration region, as identified by \citetalias{MLWH2016}, is seen.
}
\label{SouthMotions}
\end{figure*}         %^^^^^^^^^^^^^^^^^^^^^^^^^^^^^^^^^^^^^^
\epsscale{1.0}
The details of the structure, especially in the regions farther from the core, are somewhat uncertain because of the high dynamic range required  to image the jet in the presence of the bright core, the complexity of the overall structure being imaged with limited UV coverage, and the unfortunate loss of one or more antennas in some epochs, especially during 2008.
The fact that many of the marked features in the figures are not numbered is an indication that the structure is not easily described by long-lived discrete ``components''.
However, we can present a component motion analysis based on the marked and numbered features.
\begin{figure*}    %======================================
\epsscale{0.8}
% \plotone{north.ps}
\plotone{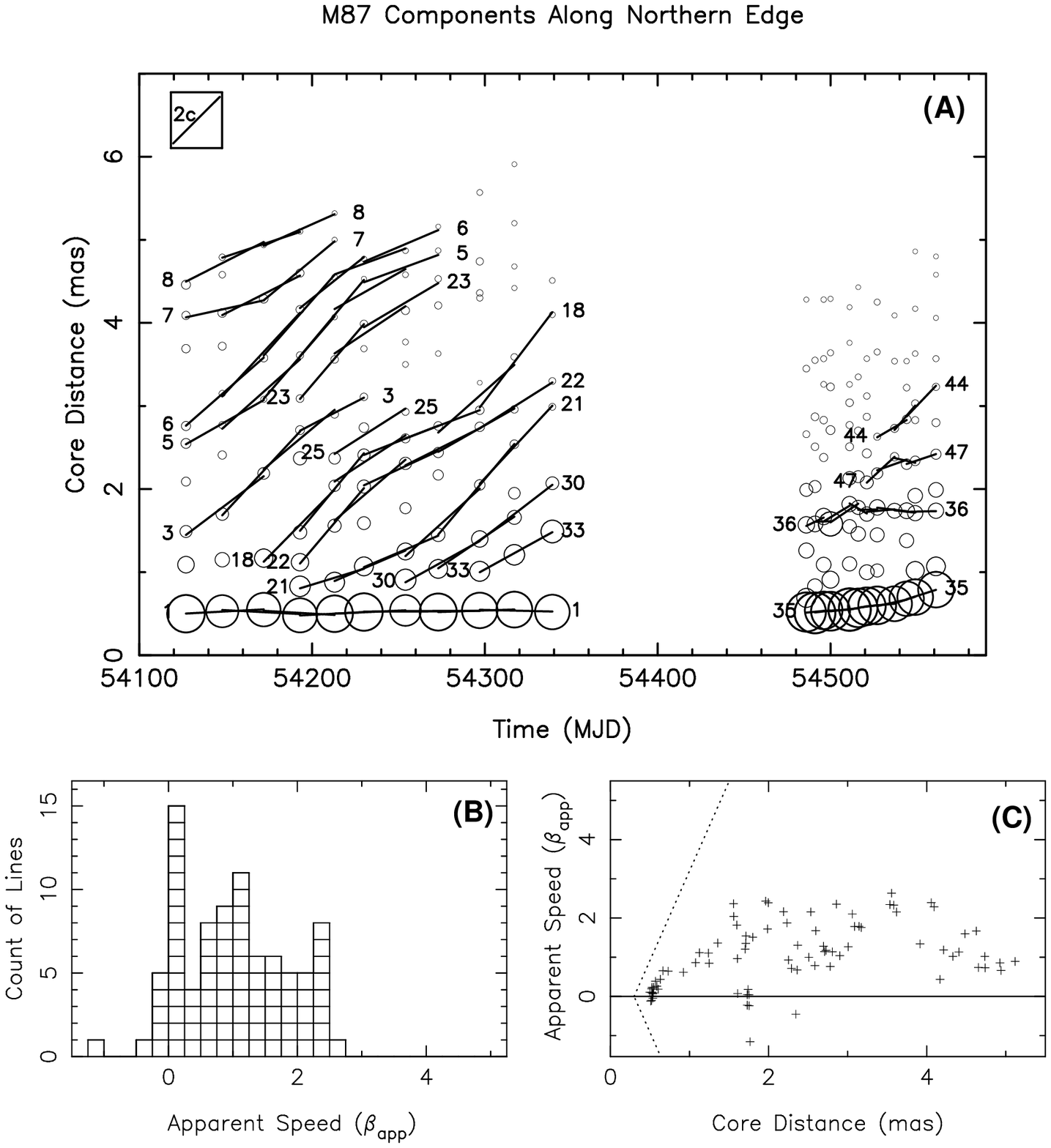}
\caption{A figure identical to Figure~\ref{SouthMotions}, but using data from components along the northern edge of the jet.
See the caption of Figure~\ref{SouthMotions} for details.
The speeds seen in both edges are discussed in the text along with comparison with results from \citetalias{MLWH2016}.
Compared to the southern edge, the northern edge shows more of a preponderance of intermediate speeds with a broad peak near $1c$.
As along the southern edge, an acceleration region is seen at distances less than about 2 mas.}
\label{NorthMotions}
\end{figure*}      %^^^^^^^^^^^^^^^^^^^^^^^^^^^^^^^^^^^^^^
\epsscale{1.0}

Figures~\ref{SouthMotions}~and~\ref{NorthMotions} show the results from such an analysis of the features seen in Figures~\ref{Epochs_A1} through \ref{Epochs_B2}.
The two figures show results for the southern and northern edges of the jet, respectively.
There were insufficient features for a similar analysis along the jet centerline or in the counter-jet.
In each figure, Panel~A shows a time series of the positions of all the marked features.
When three features have the same number in successive epochs the feature is assumed to correspond to the same moving component.
The feature is then connected by a line segment where the line segment is the result of a linear fit to the position and apparent speed of the component.
The other two panels are based on those line segment results.
Panel~B shows a histogram of fitted speeds for the line segments.
Panel C shows the speed plotted as a function of core distance where the position of the mid point of a line segment is used for the core distance.
There is an exclusion region for fast components near the core because the first feature closest to the core must be resolved from the core.
The dotted line indicates the boundary of the exclusion region assuming that the first point must be 0.4 mas from the core and that the observing cadence is once per three weeks, which is appropriate for 2007.
For 2008 with an observing cadence of once per week, the boundary of the exclusion region is steeper.

The data show a wide scatter in apparent speeds.
The speeds range from slightly negative to near $5c$.
Near the core, the speeds are slower with faster speeds at larger core distances.
The histograms show concentrations near zero and near $2.3c$ for both edges with an additional concentration near $1.0c$ along the northern edge.
The feature closest to the core is moving very slowly if at all, but that feature is generally just the inner part of the jet that is not well resolved from the core.
Toward the end of the data set, in 2008, the innermost northern feature (\#35) does move away more decisively.
That feature appears to be related to the radio flare in the core region that occurred around the time of the TeV flares seen between MJD 54497 and 54509, and is shown in difference images in \citet{Acc09}.
Here this feature is seen to have accelerated to $\beta^{\rm app} \approx 0.64$ at $\unsim 0.8$~mas from the core.
For features farther from the core, there is a general trend for speeds to increase out to $\unsim 2$~mas, reaching speeds $\beta^{\rm app} \approx 2$ albeit with large scatter.
Simply averaging all the points more than 1.8 mas from the core for both years gives results for the north and south ridges of $1.43c$ and $2.19c$ with standard deviations (SD) of $0.67c$ and $0.90c$, respectively.
These averages correspond to 5.44 and 8.30 \masr.
But averages may be too simplistic if, for example, additional acceleration is involved.

Our results can be compared to those presented for the 2007 data by \citetalias{MLWH2016}.
That study used the Wavelet-based Image Segmentation and Evaluation (WISE) velocity field analysis combined with a stacked cross correlation analysis \citep{ML2015} to analyze the velocity field in a manner that was able to pick out overlapping fast and slow motions.
Velocities from the WISE analysis, as shown in Figure~3 of \citetalias{MLWH2016}, show a wide scatter with predominantly slow speeds near the core and speeds over a wide range up to nearly $3c$ beyond about 2~mas.
Visual inspection doesn't show clear evidence for multiple components.
The velocity plots shown in our Figures~\ref{SouthMotions} and \ref{NorthMotions} look similar in character.
It was the stacked cross correlation analysis in \citetalias{MLWH2016} that helped delineate the velocity structure of the jet.
The analysis found that, at core distances between 0.5 and 1 mas, the motions are in the range of $\beta^{\rm app} = 0.2-0.5$ with the higher speeds in the north.
Between 1 and 4 mas, there are two velocity components.
The slower component has $\beta^{\rm app} = 0.17$ (north) and $0.49$ (south).
The faster component has $\beta^{\rm app} = 2.41$ (north) and $2.20$ (south).
Speeds were also determined along the center of the jet with values found between those of the northern and southern edges.

%  Try to force figure or next section to earlier page.
% \begin{figure*} [h!]      %======================================
\begin{figure*}[ht!]   % twocolumn   %======================================
\epsscale{1.14}         %  Must reset (outside fig env) or affects next figure.
% \plottwo{P88A_3C274_2016A_MAG_PCNTR_FIG.PS}{P88G_3C274_17_2016A_MAG_PCNTR.PS}
\plottwo{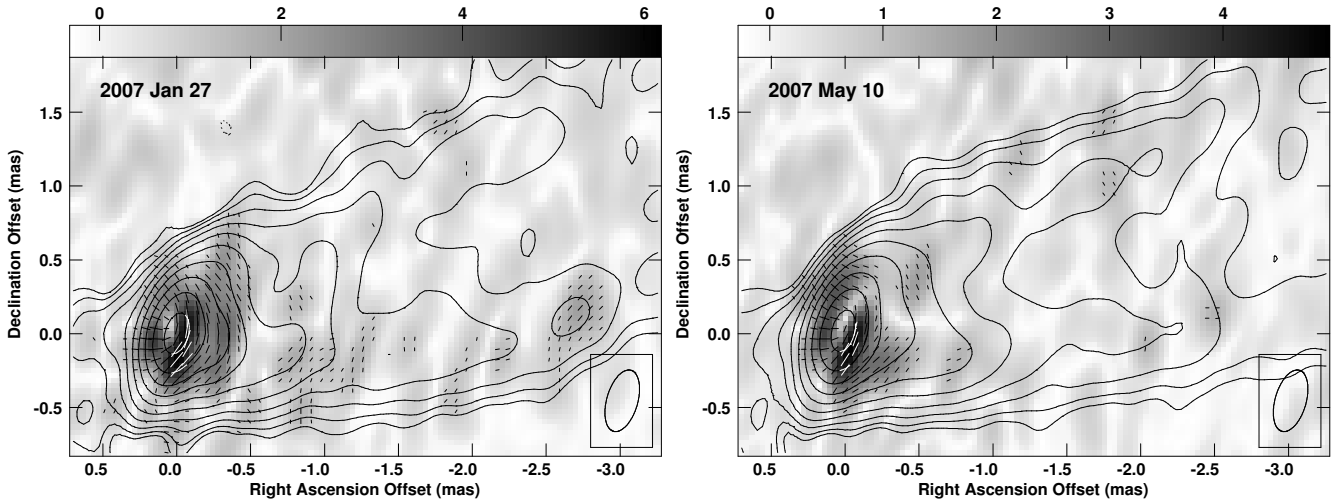}{f15b.eps}    % Single column mode
% \plotone{f15a.eps}     % Two column mode
% \plotone{f15b.eps}     % Two column mode
\caption{Magnetic field vectors overlaid on total intensity contours and a polarized intensity grey scale image for the 43 GHz VLBA observations made on 2007 January 27 (MJD 54127) and 2007 May 10 (MJD 54230).
The contour levels are 1, 2, 4, 8, 16, 32, 64, 128, 256, and 512 \mjb.
The convolving beam, with a resolution of $0.43 \times 0.21$~mas, elongated along position angle $-16\arcdeg$, is shown at the lower right.
The peak total flux densities are 666 and 660 \mjb, respectively, while the peak polarized flux densities, at slightly different positions from the peak total flux densities, are 6.15 and 4.91 \mjb, respectively.
The percent polarizations at the positions of the peak polarized intensity are 1.5\% and 1.1\% at the two epochs and rises to about 4\% at about 0.4 mas along the jet ridges.  
The cutoff levels for plotting of the polarization vectors are 0.5 \mjb\  for
the total intensity (2.5 and 3.1 times the RMS noise for the two epochs) and
1.0 \mjb\  for polarized intensity (4.3 and 5.0 times the RMS noise of the
Q and U images for the two epochs).
Both epochs show a consistent polarization structure near the core.  
On the jet side, the magnetic vectors wrap around the core.  
There is a minimum in the polarized intensity near the peak of the total intensity where the polarzation direction jumps by about $90\arcdeg$.
}
\label{Jet_Pol}
\end{figure*}         %^^^^^^^^^^^^^^^^^^^^^^^^^^^^^^^^^^^^^^
\epsscale{1.0}         %  Must reset (outside fig env) or affects next figure.

The stacked cross correlation analysis found a clear division at about $\beta^{\rm app} = 1.5$ between the fast and slow components.
For comparison, using only 2007 data (the only data used by \citetalias{MLWH2016}), our visually-determined components at core distances greater than 1.8 mas, have an average $\beta_{\rm N}^{\rm app} = 1.47$ with SD of $0.60$ for the northern edge and $\beta_{\rm S}^{\rm app} =1.79$ with SD of $0.56$ for the southern edge.
Similar averages for the WISE data in \citetalias{MLWH2016} Figure~3 for all points beyond 1.8 mas are $\beta_{\rm N}^{\rm app} = 0.91$ and $\beta_{\rm S}^{\rm app} =1.26$.
The WISE analysis clearly picks out more slow components than our visual analysis.
Restricting the average to components that have individual speeds above $\beta^{\rm app} = 1.5$ to try to focus on the fast component seen by \citetalias{MLWH2016}, the averages for our data are $\beta_{\rm N}^{\rm app} = 2.04$ and $\beta_{\rm S}^{\rm app} =2.06$ while they are $\beta_{\rm N}^{\rm app} = 2.05$ and $\beta_{\rm S}^{\rm app} =2.10$ for the \citetalias{MLWH2016} Figure~3 data.
Thus, the data sets give essentially the same result for the faster components.
In fact, they are remarkably close given the limitations of the data and of our visual analysis method.
For further discussion of the velocity results, the reader is referred to Section~\ref{SSec:JetFlow} and to \citetalias{MLWH2016}.

\subsection{Polarization Images}
\label{SSec:Pol}

Images of the polarized emission of a radio jet can provide much information about the magnetic fields in the jet.  
For that reason, all of our observations included the necessary cross-hand correlations and calibration observations to allow imaging of the polarization.  
Difficulties were encountered with the calibration and a full polarization analysis has been deferred.  
But two  epochs, 2007 January 27 and 2007 May 10, were successfully calibrated and imaged.
The polarization structure within 3~mas of the core is shown in Figure~\ref{Jet_Pol} for those two epochs.
Magnetic field vectors are shown under the assumption that they are rotated $90\arcdeg$ from the measured electric field vectors and that there is no significant relativistic aberration or rotation measure.
 These assumptions will need to be checked because rotation measures high enough to matter (greater than about 5000 radians m$^-2$) have been observed at other positions and frequencies in M\,87 \citep{ZT2002,Kuo2014}.
It may be possible to extract rotation measures from our data but that analysis has not yet been completed.
The magnetic field vectors are superimposed on total intensity contours and a grey scale image indicating the polarized intensity.
The polarization percentage varies from near zero close to the core where the field direction flips, to about 4\% about 0.4 mas from the core along the southern edge of the jet, and along the northern edge of the jet in the first epoch.
Farther along the jet, and closer to the core along the northern edge in the second epoch, the polarized emission is too weak for reliable detection in these two data sets.
The peak polarized flux densities (noted in the figure caption) are seen about 0.15 mas southwest of the core.
At those locations, the polarization percentages are 1.5\% and 1.1\% at the first and second epochs.
On the counter-jet (east) side of the core, within the wings of the beam, the polarization is 1\% to 2\%.

The magnetic field orientation shows a consistent pattern at the two epochs.
This might be expected as there were no large changes to the intensity structure during this period.
On the jet side of the core the magnetic field vectors are approximately perpendicular to the local jet axis.
Rotation of the vectors between north and south edges reflects the wide jet opening angle.
On the counter-jet side of the core, the magnetic field vectors are approximately parallel to the counter-jet axis, again rotation between north and south edges reflects the wide opening angle.
While the $\unsim 90\arcdeg$ polarization angle change near the core may reflect the magnetic field structure, it is also  possible to get such an effect in VLBI sources as a result of the transition from optically thick to optically thin emission.
A related opacity effect is expected to displace the radio core from the actual black hole location.
Such a shift for 43 GHz in M\,87 was measured by \citet{Hada2011} to be $41 \pm 12$ \uas.
Given that the polarized emission on the counter-jet side of the core is within the wings of the convolving beam, it is possible that it originates in the portion of the jet between the black hole and the radio core, i.e., not from the counter-jet.
As noted above, the polarized intensity drops rapidly away from the core but the percentage polarization rises over the short distance for which it is measured reliably.
There is some suggestion that the magnetic field vectors are more flow aligned along the northern and southern edges of the jet at $\unsim 2.5$~mas from the core, although this result needs to be confirmed with lower noise polarization data.

Magnetic field vectors that are wrapped around the core, reflecting the initial large jet opening angle, suggest a toroidal jet magnetic field geometry.
 However, modeling is needed to take into account the relatively small viewing angle to the line of sight, along with possible optical depth, Faraday rotation, and relativistic aberration effects.
The polarization data from our observations that have not yet been reduced should provide much higher sensitivity, allowing detection of polarization farther from the core. 
The data taken since 2009 will provide 24 GHz data to allow measurement of Faraday rotation.
These improvements should eventually provide more information on the magnetic field structure.

\subsection{Long-Term Evolution}
\label{SSec:LongTerm}

This project began as an effort to measure possible superluminal motions near the core of M\,87 by sampling significantly faster than had been done in previous projects.  
But slower, longer-term sampling has the potential to provide information on aspects of the jet behavior other than component speeds.  
For example, relatively slow changes in the jet envelope or direction could be observed, providing information about jet propagation and the external medium.  
As described in Section~\ref{Sec:Obs}, we have, somewhat fortuitously, roughly annual observations covering 17 years.  
Prior to our pilot project, a few archival observations done for other reasons are available.  
Between 2009 and 2016, annual observations were made as part of an effort to observe radio counterparts to TeV flares.  
In this section, we first present the images from these roughly annual observations and subsequently determine displacement motions from these images.  
A clear finding is that the jet moves transversely on time scales of several years.
%  peh suggested leaving this out: That motion is characterized.}

\subsubsection{Intensity Images}

\begin{figure*}     % Twocol trhy without [h!] ======================
\centering
% ----- Force the width to keep figures 16, 17, and 18 the same width  -----------
% \includegraphics[width=12.5cm]{montage_annual_0_g.eps}
\includegraphics[width=13.0cm]{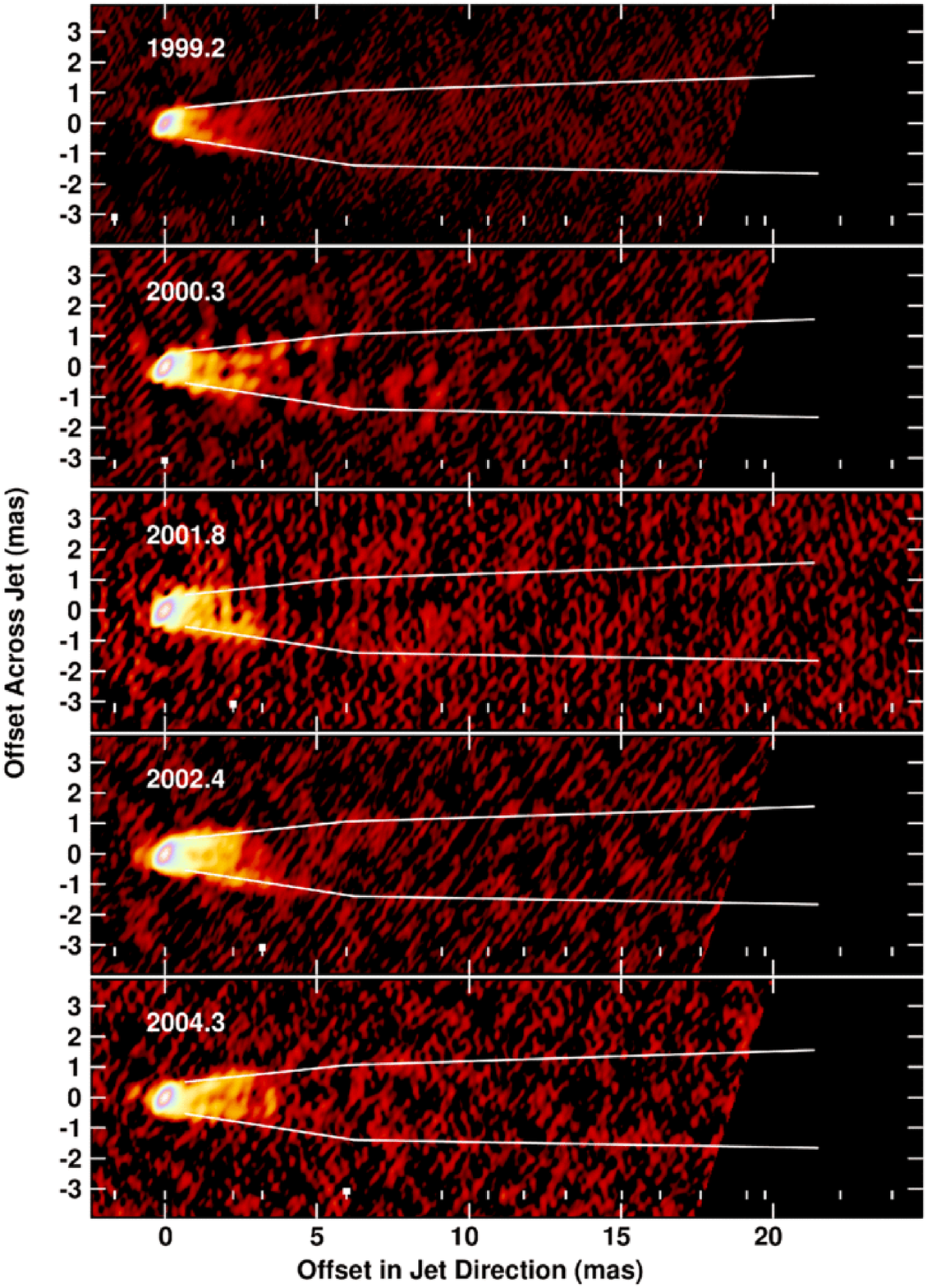}
\caption{First five previously published epochs between 1999 to 2004 \citep{Junor1999, Ly2007} of the roughly annual observations of M\,87 made between 1999 and 2016.
All images have been rotated by $-18\arcdeg$.
The white lines indicating the jet edges on the 2007 image are there to aid visual detection of changes at different epochs.
The images in Figures~\ref{LTE1}, \ref{LTE2}, and \ref{LTE3} are available as a movie in online animated Figure~\ref{LTE_movie}.  
}
\label{LTE1}
\end{figure*}

\begin{figure*}     % Twocol try without [h!] =================================
\centering
\includegraphics[width=13.0cm]{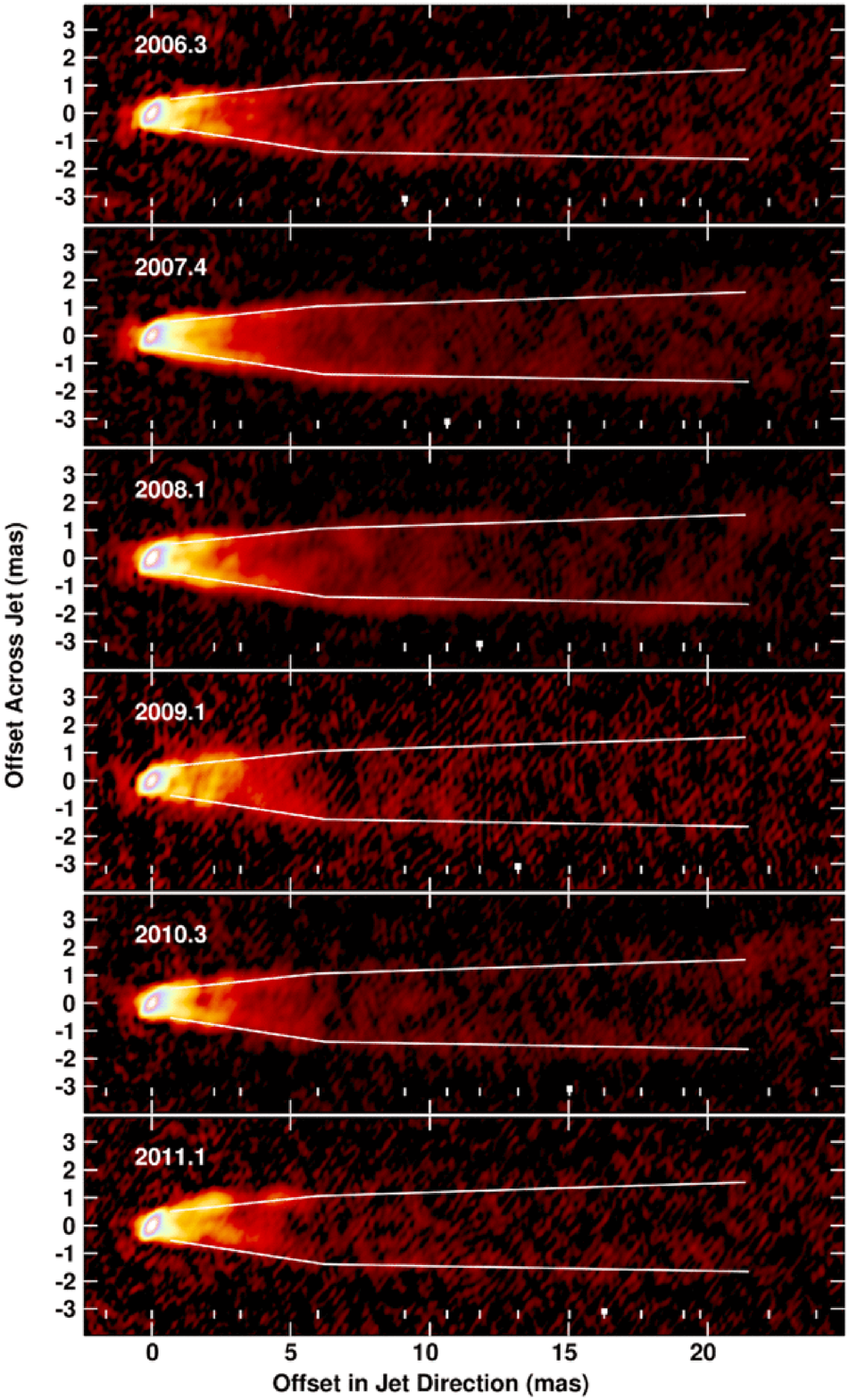}
\caption{Middle six epochs of the roughly annual observations of M\,87 made between 1999 and 2016.
The 2006.3, 2007.4, 2008.1, and 2010.3 images are actually stacks (noise-weighted means) of 6, 11, 12, and 6 images, respectively, taken within a few months of the indicated mean time.
As in Figure~\ref{LTE1}, the white lines indicate the jet edges on the 2007 intensity image.
 Recall that the stacks will smear out variable structure and emphasize persistent structures.
The images in Figures~\ref{LTE1}, \ref{LTE2}, and \ref{LTE3} are available as a movie in online animated Figure~\ref{LTE_movie}. 
}
\label{LTE2}
\end{figure*}

\begin{figure*}     % Twocol try without [h!] =======================
\centering
\includegraphics[width=13.0cm]{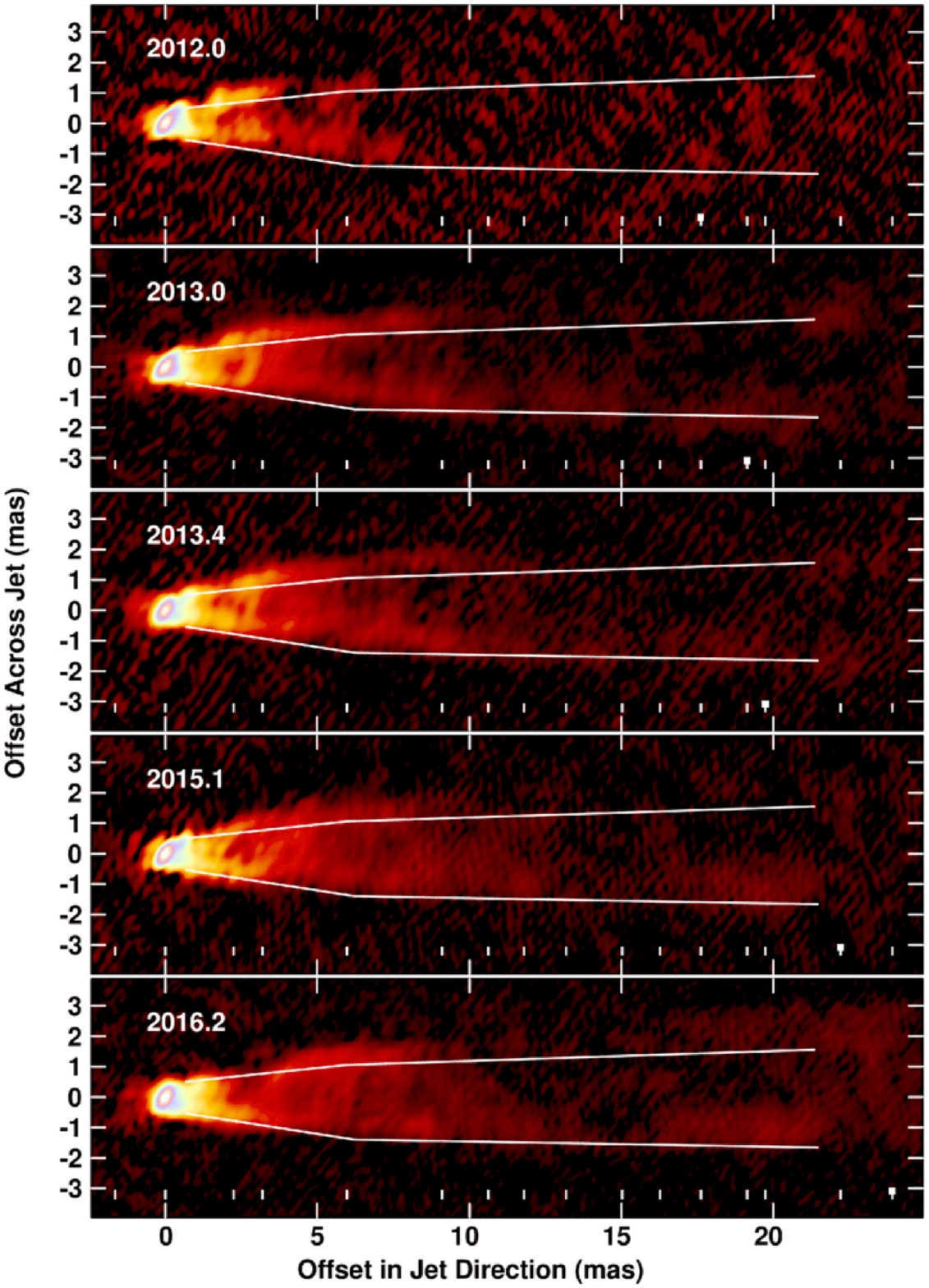}
\caption{Final five epochs of the roughly annual observations of M\,87 made between 1999 and 2016.
This group consists of single-epoch images (not stacks).
The final 4 are  based on observations made after the VLBA bandwidth upgrade.
With typically 8 times the bandwidth or nearly three times the sensitivity of the older observations, the individual epochs have noise levels comparable to the stacks in Figure~\ref{LTE2}.
As single-epoch images, transient details are not smeared.
Again, the white lines indicate the jet edges on the 2007 intensity image.
The images in Figures~\ref{LTE1}, \ref{LTE2}, and \ref{LTE3} are available as a movie in online animated Figure~\ref{LTE_movie}. 
}
\label{LTE3}
\end{figure*}

The total intensity evolution of the sub-parsec scale radio jet in M\,87, spanning a 17 year period from 1999 to 2016,  is shown in Figures~\ref{LTE1}, \ref{LTE2}, and \ref{LTE3}.
Approximately annual observations at 43 GHz involving the VLBA are shown.
%
% \begin{figure*}[h!]     % ======================================
         %^^^^^^^^^^^^^^^^^^^^^^^^^^^^^^^^^^^^^^
%
%
% \begin{figure*}[h!]     % ======================================
         %^^^^^^^^^^^^^^^^^^^^^^^^^^^^^^^^^^^^^^
Where multiple observations were made in less than a year, average images are shown.
Image quality improves significantly with time as the VLBA hardware improved and as more average images became available.
See Table~\ref{ObsTable} for details of the observations, including the bit rate and the  off-source image noise.
%
% \begin{figure*}[h!]     % ======================================
         %^^^^^^^^^^^^^^^^^^^^^^^^^^^^^^^^^^^^^^

The images in Figure~\ref{LTE1} are early observations and have limited sensitivity.
In some of the very early observations, the actual observing time was short because M\,87 was a phase calibrator, not the primary target of the archival observations shown here.
Average images are available from the pilot project in 2006, the main monitoring observations in 2007 and 2008, and from the trigger response in 2010.

Significant transverse displacement of the jet, especially between about 2 and 8 mas from the core, is seen in Figures~\ref{LTE1}, \ref{LTE2}, and \ref{LTE3}, particularly in the epochs after 2011.1 relative to earlier epochs.
Lines have been drawn on the figures that follow the bright jet edges in the 2007.4 average image.
These lines help make the northern shift of the jet clear, especially in the later epochs.
The transverse displacements are discussed more extensively below.

Figures~\ref{LTE1}, \ref{LTE2}, and \ref{LTE3} also show that the relative brightness of southern and northern edges experiences significant changes over the 17 year period in the innermost few mas.
There are epochs when the northern side is brighter, and epochs when the southern side is brighter.
Which is brighter can be different as a function of position along the jet.
It does appear that, when there is a transverse shift, the side toward which the shift happens brightens in the innermost $2 - 3$ mas.
The best case is for the large shift to the north which started in 2011.1.
For that epoch, the northern side is brighter by a factor of 2 to 3 near 2 mas.
The northern side remains brighter as the shift propagates along the jet, but, as the jet starts to return toward its original position, the southern side becomes brighter.

\subsubsection{Displacement Motion}
\label{DispMotion}

The results of an effort to measure the propagation speed of the jet offset are shown in Figures~\ref{LTVfit} and \ref{LTVcor}.
The transverse offset of the measured center of the jet relative to a nominal jet center line, that extends at position angle ${-72}\arcdeg$ from the core, is shown in Figure~\ref{LTVfit} for core distances at 2 to 8 mas at 1 mas increments (the offset is actually in the y direction in the rotated images).
The offsets show a general drift north superimposed on what appears as a quasi-sinusoidal variation.
The patterns for each distance from the core show a lag, indicating a propagation speed that is addressed in this section.
For core distances at and beyond 3 mas, the offsets were measured from images that had been blurred along the jet direction (convolved with a beam of $0.42 \times 1.0$ mas elongated along ${\rm PA} = {-72}\arcdeg$) to reduce effects of fine scale structure.
At 2 mas, the full resolution images were used.
Slices perpendicular to the jet were made at the core distances noted.

%
% \begin{figure*}[h!]     % ======================================
\begin{figure*}     % twocolumn  ======================================
\centering
% \plotone{E_ANNUAL_0.0_13.PS}  latest:  ANNUAL_0.0_FRE18_53.PS
\plotone{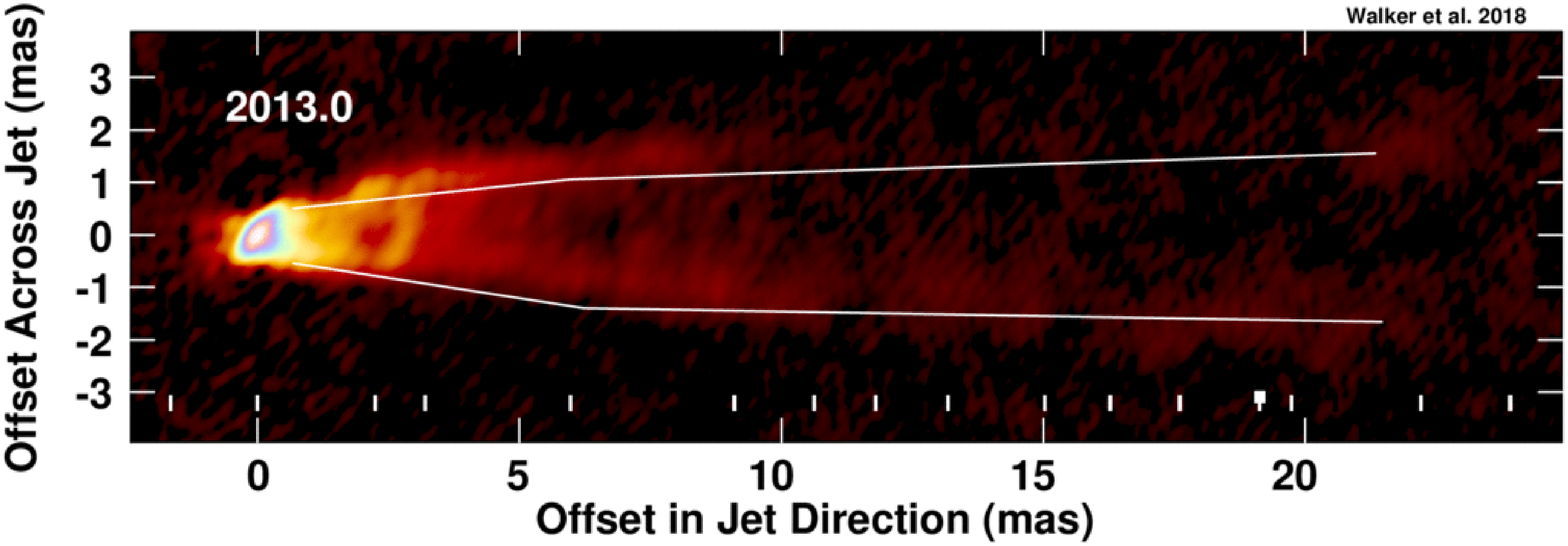}
%  The animation file is M87_VLBA_Long_Term.mp4
\caption{In the online version of this paper at the journal, or in astro-ph ancillary file M87\_VLBA\_Long\_Term.mp4, this figure is a movie made from the
roughly annual images shown in Figures~\ref{LTE1}, \ref{LTE2}, and \ref{LTE3}.  
It gives a visual sense of the changes that occurred in the source between 1999 and 2016, including the side-to-side displacements.  
Those shifts are described in detail in Section~\ref{DispMotion} in terms of a linear shift and a sine wave, both functions of time, both of which propagate down the jet at a pattern speed.
A fit to the offsets of the jet center from a nominal center line gives the parameters of these patterns and their pattern speed, which is fast, but not as fast as the faster component motions.  A pattern speed is also derived using a correlation analysis giving a similar result.
For the offline versions of this paper, the image from 2013.0 is shown.}
\label{LTE_movie}
\end{figure*}         %^^^^^^^^^^^^^^^^^^^^^^^^^^^^^^^^^^^^^^

The offsets of the jet center from the  PA$ = {-72}\arcdeg$ line were determined from the slices by the following algorithm.
First the peak intensity south of the PA$ = {-72}\arcdeg$ line was determined.
Then, the locations of the most southern interpolated points at one quarter and one half of the that peak intensity were found.
The average of those two points was treated as the position of the southern side and one half of the difference was treated as an ``error bar''.
The process was repeated for the northern side using the most northern points at a quarter and a half of the peak value north of the PA$ = {-72}\arcdeg$ line.
The separate determinations of the north and south peaks prevents spurious results when one ridge is significantly brighter than the other --- a frequent case.
The center was taken to be the average position of the northern and southern side points and the number used for weighting in the analysis was the quadratic sum of the ``error bars''.
In practice, these ``error bars'' did not vary much between points and, to avoid clutter, only those for the 2 mas and 8 mas points are shown in Figure~\ref{LTVfit}.

%
% \begin{figure}[h!]  % =========================================
\begin{figure}   %  Twocol try without [h!]  ===================================
\centering
\includegraphics[angle=-90, width=3.3in]{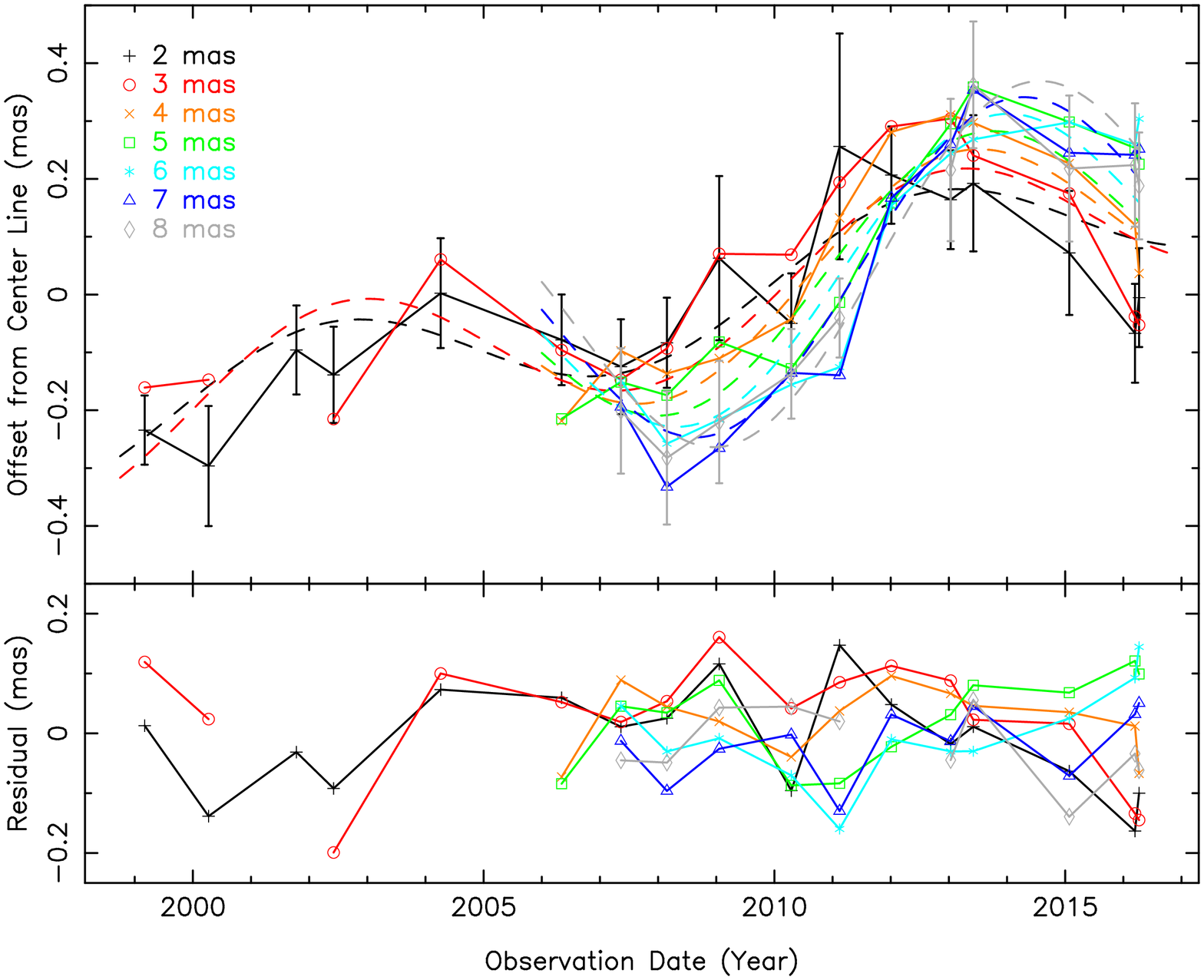}    % Two column mode.
\caption{Measured offsets of the M\,87 jet center from a line extending from the core along position angle of ${-72}\arcdeg$.
The data used are the images shown in Figures~\ref{LTE1}, \ref{LTE2}, and \ref{LTE3} with one extra observation (2016 April 10 in Table~\ref{ObsTable}) included at the end.
Offset measurements are based on finding the outermost half and quarter power points on each side, relative to the peak on that side.
``Error'' bars are based on the separation of the two points on each side and do not vary much so only those for the points at 2 and 8 mas are shown.
Dashed lines are the results, for different core distances as indicated by color, of a least squares fit for a model, described in the text, whose main components are a linear relation and a sine function, both with time, both propagating down the jet. 
The ``errors'' were used as the weights in the fit.
The apparent propagation speed, with its formal error from the fit, is $\beta^{\rm app}_{\rm p} = 0.89 \pm 0.18$.
The bottom panel shows the residual data which are the measured data points with the fitted model subtracted.
}
\label{LTVfit}
\end{figure}  % ^^^^^^^^^^^^^^^^^^^^^^^^^^^^^^^^^

Two methods were used to determine a speed of propagation of the transverse displacement pattern.
The first method is a least squares fit for the parameters of a simple equation that roughly describes the data in Figure~\ref{LTVfit}.
The purely empirical equation is not based on any particular physical model and is given by
\begin{align}
 t &= t_{\rm yr} - 2000.0  \nonumber \\
 t_{\rm d} &= t - z / v_{\rm p} \nonumber  \\
R_{\rm c} &= R_0 + a_1 z + a_2 t_{\rm d} + a_3 z^{a_4} \sin (2 \pi f t_{\rm d} + \phi_0 )
\label{fiteqn}
\end{align}
where $R_{\rm c}$ is the offset from the nominal centerline in mas, $t_{\rm yr}$ is the date in  decimal years, $t$ is the number of years since 2000.0 (to reduce the chances of numerical problems), and $t_{\rm d}$ is the time a feature seen at time $t$ at core distance $z$ was at the core assuming linear motion from the core at speed $v_{\rm p}$.
 The form of $t_{\rm d}$ used to parameterize the equation is not meant to imply that linear pattern motion occurs or exists very close to the core.
The fitted parameter results for $ v_{\rm p}$, $R_0$, $a_1$, $a_2$, $a_3$, $a_4$, $f$, and $\phi_0$ are given in Table~\ref{LSQresults}.
The offset, $R_{\rm c}$, for each core distance given by Equation~\ref{fiteqn} is shown as dashed lines in Figure~\ref{LTVfit} with the colors matching the data colors.
The bottom panel shows the residuals after subtracting the model from the data.

\begin{deluxetable*}{clccl}   %====================================== 
\tablecaption{Fitted Parameters of the Jet Displacement Model
   \label{LSQresults}
}
\tablehead{
  \colhead{Parameter} &
  \colhead{Description} &
  \colhead{Fit value} &
  \colhead{Std. Error} &
  \colhead{Units}
}
\startdata
   $R_0$     & Offset at $z_{\rm ob} = 0$, $t_{\rm d} = 0$       & ${-0.196}$              &  0.024 & mas                              \\
   $a_1$     & Slope with distance                               &  0.005                  &  0.004 & mas offset per mas core distance \\
   $a_2$     & Linear change with time                           &  0.022                  &  0.002 & \masr                            \\
   $v_p$     & Angular pattern speed                             &  3.36\tablenotemark{a}  &  0.68  & \masr                            \\ 
   $a_3$     & Sine wave amplitude at $z = 0$                    &  0.061                  &  0.017 & mas                              \\
   $a_4$     & Index of sine wave amp dependence on $z_{\rm ob}$ &  0.69                   &  0.16  & --                               \\
   $f$       & Sine wave frequency.                              &  0.097\tablenotemark{b} &  0.003 & Cycles mas$^{-1}$                \\
   $\phi_0$  & Phase at $t_{\rm d} = 0$                          &  31                     &  15    & Degrees                          \\
\enddata 
\tablenotetext{a}{For the assumed distance of M\,87, the angular pattern speed ($v_p$) converts to an apparent pattern speed of $\beta_{\rm p}^{\rm app} = 0.89 \pm 0.18$.}
\tablenotetext{b}{The period of the sine wave (inverse of the frequency) is $10.3 \pm 0.3$ years.}
\end{deluxetable*}          %^^^^^^^^^^^^^^^^^^^^^^^^^^^^^^^^^^^^^^

 A linear change in offset with core distance at a fixed time can be created in at least two ways.
First, an error in the choice of the position angle assumed for the jet center relative to the core would produce such an effect because the resulting linear offset would grow with core distance.
Second, if the offset changes with time at a constant rate, and that change propagates down the jet at finite speed, the same linear change in offset with distance  will be produced.
Both effects are included as terms in Equation~\ref{fiteqn}.
This would not normally work for the fit because the terms are completely correlated.
However, we assume that the pattern propagation speeds for the linear change in the jet center offset and for the sine wave are the same.  
This causes the pattern propagation speed to be controlled by the sine wave, leaving the error in the choice of jet position angle to absorb any additional linear change in offset with distance.  
This assumption decorrelates the terms.
The fitted result that the linear change with distance ($a_1$) is very small suggests that the assumed position angle was properly chosen for the reference epoch (2000.0).

This first method gives a period result of $10.3 \pm 0.3$ years within the errors.
However visual inspection of Figure~\ref{LTVfit}, concentrating on the 2 and 3 mas data points, suggests a shorter period.
 These two distances are the only ones measured from images before 2006 and have an extra  seven-year time baseline.
Indeed, a fit to just the 2 and 3 mas data gives a period result of $7.6 \pm 0.3$ years.
Nevertheless, adding the data from farther down the jet strongly discourages the shorter period.
The differences seem to result from the 2006 and 2016 data --- the extremes of the time range for the more distant data.
As a check of the 2016 March 14 data, and an indication of the reliability of the rest of the data, a 2016 April 10 epoch (the result of a high energy trigger and included at the end of Table~\ref{ObsTable}), not shown in Figure~\ref{LTE3}, was included for the fit.
The results using this 2016 April 10 epoch show good agreement with those ending with the 2016 March 14 epoch.
A few more years of good data, covering more than a full period, should provide some clarification.
In any event, this method gives an apparent pattern speed of $3.36 \pm 0.68$ \masr\ for $\beta_{\rm p}^{\rm app} = 0.89 \pm 0.18$.
At a viewing angle $\theta = 17\arcdeg$ the apparent pattern speed would imply an intrinsic pattern speed of $\beta_{\rm p} \approx 0.78 \pm 0.04$.

%
%  \begin{figure}[h!]  % ===================================
\begin{figure}   %  Twocol try without [h!]   =========================
\centering
\includegraphics[angle=-90, width=3.3in]{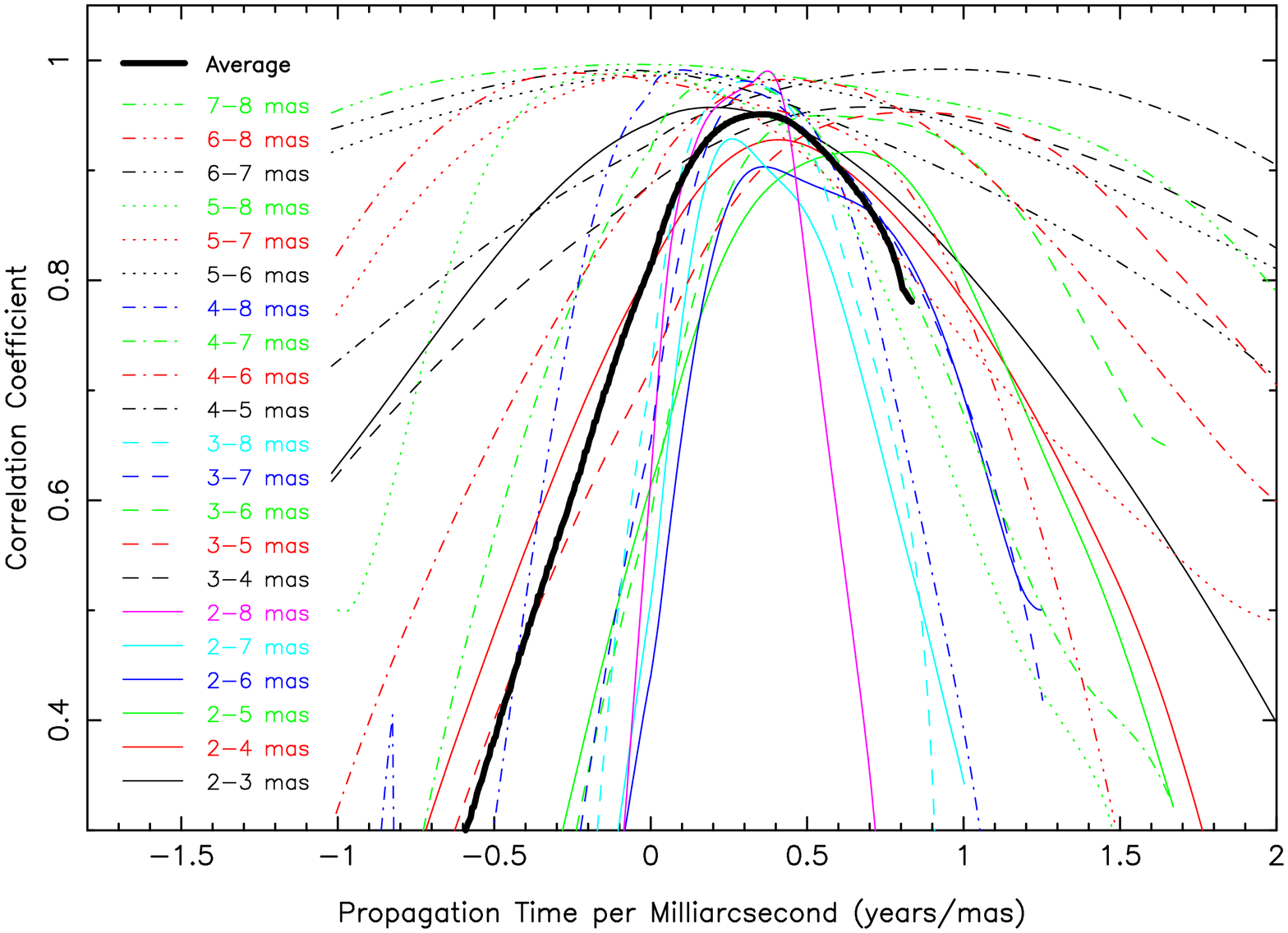}     %  Two column mode.
\caption{The correlation functions used to determine the propagation speed of the side-to-side variations in the M\,87 jet center position.
These provide an alternate determination to that from the least squares fit shown in Figure~\ref{LTVfit}.
The data used are the subset of those shown in Figure~\ref{LTVfit} which were taken between 2006 and 2016.
Data for each core distance, measured at 1 mas intervals in core distance, were correlated with that of each other distance as a function of time lag.
The time axis is scaled by dividing by the separation in core distance of the pair so that the axis becomes the propagation time per mas (years mas$^{-1}$).
The correlation function for each pair of core distances is identified by a unique combination of color and line type as shown in the legend.  
For most pairs, the line type indicates the initial core distance of each pair, while the color indicates the separation in core distance of the pair.
The $7-8$~mas pair does not conform to the pattern because  the plotting software used provided too few line types.
The average over all the correlation curves is shown by the bold black line.
The weighted average of the peaks of the individual curves is  $0.327 \pm 0.038$~year mas$^{-1}$ while the peak of the average curve is at $0.350$ year mas$^{-1}$.
}
\label{LTVcor}
\end{figure}   % ^^^^^^^^^^^^^^^^^^^^^^^^^^^^^^^^^^^^^^^^

In the second method used to measure the displacement pattern speed, for each core distance, the measured data points at the different epochs were linearly interpolated to a dense time series and each pair of time series (different distances) was cross correlated.  
This was done by calculating the ``Pearson's r'' statistic for time lags between $-5$ and 5 years.
The normalization is done separately for each lag to try to minimize the impact of end effects since the maximum lags are a noticeable fraction of the overall range.
The correlations should peak at a time lag related to the pattern propagation speed.
 This method of measuring the propagation speed does not assume any particular functional form for the offsets so it is insensitive to the time scale of any oscillations.
The correlation functions are plotted in Figure~\ref{LTVcor}.
The correlations are plotted against time lag divided by the core-distance separation of the two time series being correlated.
This turns the time lag axis into units of propagation time per mas, which is an inverse speed.
The actual speed is not a natural unit for plotting the correlation functions because there is an infinity and sign flip in speed at zero lag.
The correlations were then averaged as a function of the inverse speed with the result shown by the bold black line in Figure~\ref{LTVcor}.
The peak of the average curve is at 0.350 year mas$^{-1}$ which is 2.86 \masr\ or $\beta_{\rm p}^{\rm app} = 0.75$.
An average speed was also calculated by measuring the peaks of each correlation function, again scaled by the distance, and doing a weighted mean.
For the weights, the width to the 90\% amplitude points were used.
That gives reasonable relative weights and does not significantly affect the calculated standard error which takes into account the scatter of the points.
The result, determined in this manner, is $0.327 \pm 0.038$~year mas$^{-1}$ which is $3.1 \pm 0.4$ \masr\ or $\beta_{\rm p}^{\rm app} = 0.81 \pm 0.10$.
At a viewing angle $\theta = 17\arcdeg$ the apparent pattern speed (weighted average) would imply an intrinsic pattern speed of $\beta_{\rm p} = 0.76 \pm 0.025$.

The speed measured from the correlation analysis ($\beta_{\rm p}^{\rm app} = 0.81 \pm 0.10$) is slower than, but within the errors of, the speed measured with the least squares fitting ($\beta_{\rm p}^{\rm app} = 0.89 \pm 0.18$).  
This match between the speeds from the two methods lends credence to the results.
In either case, the speed is significantly less than the component speeds measured in the component motion analysis ($\beta^{\rm app} > 2$ for the fast components).
Thus the transverse shift of the jet is not propagating down the jet ballistically.
Another indication that the offset is not ballistic is that it does not grow linearly with distance.
The amplitude deviation of the fitted sine wave from the nominal axis grows with distance by only a power-law index of $\unsim \,^2\!/_3$.

\subsection{The Counter-Jet}
\label{SSec:CJ}

Essentially all of the images presented in this paper show evidence for a counter-jet.
A counter-jet can provide important clues to the nature of the jet.
Firstly, its mere existence shows that the M\,87 jet is two-sided as it is launched, which is commonly assumed, but rarely proven in superluminal sources.
This provides evidence against flip-flop models \citep[e.g.,][]{Rudnick1984} where the jet is one-sided at any given time.
Secondly, the ratio of brightness of the jet and counter-jet, if it is assumed to be the result of Doppler boosting, gives a measure of the jet speed.
Thirdly, similarities or differences in the structure of the jet and counter-jet help distinguish features related to fundamental dynamical structures in the jet and surrounding medium from more ephemeral, time-dependent structures based on details of the stability of the jet generation mechanism.

While all of our images show the counter-jet, those based on data that predates the recent upgrades to the VLBA did not reliably delineate the structure of the counter-jet.
The counter-jet feature is weak and located in the region near the bright core where artifacts from residual calibration errors are most pronounced.
Stacked images showed that it has a structure roughly symmetric with the main jet and some of the recent, high-sensitivity, observations show the structure more clearly.
Here we first push the sensitivity at the normal resolution as far as possible with our data by stacking all of  the images, with weighting based on the off-source noise.
Subsequently, we explore the structure close to the core by adjusting imaging parameters to emphasize resolution in the creation of images from two of the best high-sensitivity epochs.

\begin{figure}  % [hp!]Could be 1 column   Try without [h!]  =====================
\epsscale{1.17}
\centering
% \plotone{M87_SUM_ALL1_CMEAN_LN.PS}
\plotone{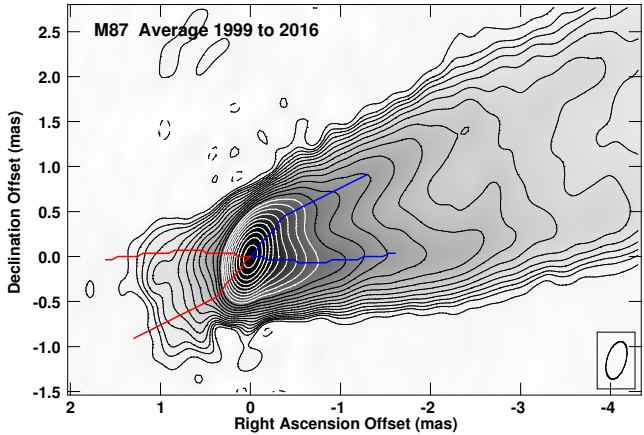}
\caption{A noise-weighted average image made from the first 49 epochs of the 50 epochs listed in Table~\ref{ObsTable}.
The image shows the average counter-jet structure at 43 GHz as far from the core as is allowed by our data.
The convolving beam was the usual $0.43 \times 0.21$~mas, elongated along position angle $-16\arcdeg$.
The contour levels, in \mjb, are $-0.2$, 0.2, 0.4, 5.6, 8, 11.3, 16, increasing from there by factors of $\sqrt{2}$.
The peak is 726 mJy beam$^{-1}$ and the off-source RMS noise level is 47 \ujb.
Blue lines (jet side) and red lines (counterjet side) show the locations where intensity measurements were made to give the sidedness data shown in Figure~\ref{Sided}.  The lines on the jet side follow the ridges while those on the counter-jet side are the jet-side lines reflected through the core.
}
\label{Stack49}
\end{figure}

An image based on stacking 49 images from the epochs listed in Table~\ref{ObsTable} in the appendix (excluding the last epoch), is shown in Figure~\ref{Stack49}.
%
% \begin{figure}[h!]  % [hp!]Could be 1 column ===========================
         %^^^^^^^^^^^^^^^^^^^^^^^^^^^^^^^^^^^^^^
\epsscale{1.00}         %  Must reset (outside fig env) or affects next figure.
Only the inner few mas are shown to emphasize the counter-jet and because the side-to-side variations discussed in Section~\ref{SSec:LongTerm} blur the otherwise sharp-edged structure on larger scales.
On these small scales, the side-to-side variation-induced blurring is relatively small compared to the resolution.
This image allows measurement of the sidedness intensity ratio to $\unsim 1.7$~mas from the core, provides cross jet resolution and shows the edge-brightening similar to that seen on the jet side.
The blue (jet side) and red (counterjet side) lines crossing the core in Figure~\ref{Stack49} show the locations of the pixels whose intensities are used to measure the sidedness that will be discussed in Section~\ref{SSSec:Iratio}.  % A figure reference here would put figures out of order.
These pixels are chosen to follow the north and south ridges on the jet side.
On the counter-jet side, each pixel chosen is simply at the symmetrically opposite position from the corresponding pixel on the jet side.
This builds in an assumption that the jet/counter-jet structure is symmetric which the image shows is close to, but not precisely, true.  Recall that, as with any stack spanning times longer than the time scales for changes in the source, the detailed changing structure is smoothed out, leaving the persistent structure.

Our recent higher-sensitivity, wide-bandwidth observations have provided an opportunity to delineate the structure of the counter-jet at individual epochs in more detail than was previously possible.
%
% \begin{figure}[h!]  % [hp!]Could be 1 column ===========================
\begin{figure*}  % [hp!]Could be 1 column ===========================
\epsscale{0.9}
\centering
% \plotone{SuperresolveFig.ps}
\plotone{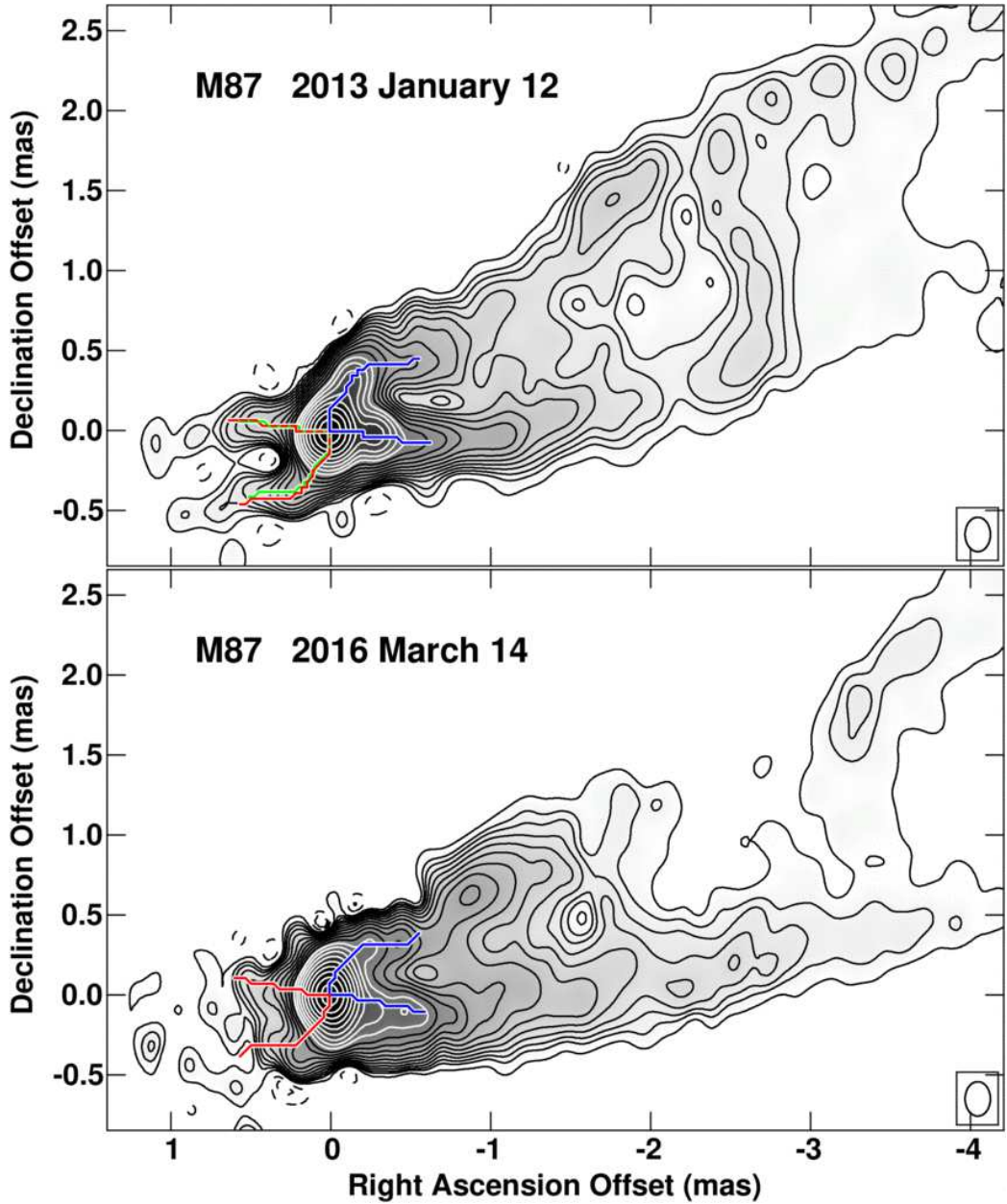}
\caption{Mild super-resolution 2013 and 2016 images using the wide bandwidth system on the VLBA.
The convolving beam of $0.21 \times 0.16$ mas, elongated north-south, is shown  in the lower right corner.
It is $\unsim 70$\% the size of the fitted beam in the north-south direction and $\unsim 100$\% in the east-west direction.
The contour levels, in \mjb, are $-2.8$, $-2$, $-1$, 1, 2, 2.8 and increase from there by factors of $\sqrt{2}$.
Jagged blue (jet side) and red (counterjet side) lines connect pixel center points where intensity measurements were made for the sidedness data used in Figure~\ref{Sided}.
Green lines on the counter-jet side in the 2013 image connect counter-jet points plotted at 90\% of the core distance of corresponding jet side points.
Those green lines are dotted where they overlap the red lines.
Note the symmetry between the jet and counter-jet, especially in 2013.
At very low levels there are residual calibration artifacts near the core, particularly in the 2016 image.}
\label{MRES}
\end{figure*}  % twocolumn       %^^^^^^^^^^^^^^^^^^^^^^^^^^^^^^^^^^^^^^
% \end{figure}         %^^^^^^^^^^^^^^^^^^^^^^^^^^^^^^^^^^^^^^
\epsscale{1.0}
To examine the counter-jet structure close to the core with two of the best data sets, 2013 January 12 and 2016 March 16, we have used imaging parameters that lead to higher resolution than our other images at the cost of degraded image quality on larger scales.
This was accomplished by using a ``robustness" \citep{Briggs} approximating uniform weighting.
Uniform weighting emphasizes the long baselines even though they are a relatively small fraction of the data.
Then the point source CLEAN components were convolved with a beam smaller than that fitted to the dirty beam.
This ``super-resolution'' increases the noise level and can be risky if overdone.
Here the fitted beams were $0.31 \times 0.16$ mas elongated in PA $= -5.3\arcdeg$ in 2013 and $0.30 \times 0.17$ mas elongated in PA $= -6.2\arcdeg$ in 2016.
The final convolving beam used was $0.21 \times 0.16$ mas in PA $= 0\arcdeg$.
Thus the super-resolution reduced the beam area by about 50\%, mostly in the north-south direction.
The peak flux densities are 0.62 and 0.48 \jb ~in 2013 and 2016,  respectively.

The resulting fairly mild super-resolution images of the inner jet from the two data sets are shown in Figure~\ref{MRES}.
The images show a structure in which the counter-jet is a reflection of the main jet, but at lower intensity.
The bright edges of the main jet that exit the core at an apparent opening angle of about $70\arcdeg$ have corresponding bright edges on the counter-jet side.
As with the stacked image, the blue lines (jet side) and red lines (counderjet side) show the locations where measurements of intensity along the jet were made for the sidedness data discussed in Section~\ref{SSSec:Iratio}.  % A Figure reference here would put figures out of order.
The blue and red lines follow the jet ridge line on the jet side and are reflected across the core for the counter-jet side.
They are jagged because measurements were made at pixel centers.
% rcw new paragraph break

 The northwest jet edge and the southeast counter-jet edge, especially on 2013 January 12, show a distinct, sharp bend in both jet and counter-jet at about 0.44~mas from the core.
Such a feature is not common in our other images, so its appearance in both jet and counter-jet suggests a common origin.
If the bend were moving outwards at a significant fraction of the speed of light, one would expect to see the bend closer to the core on the counter-jet side because of differential light travel time.
That allows limits to be placed on the propagation speed of the bend.
The green lines in the top panel in Figure~\ref{MRES} (2013 January 12 epoch) show a $\unsim 10$\% core-distance reduction compared to the red lines (which are symmetric with the blue lines on the jet side).
This small reduction does appear to be a somewhat better match to the counter-jet ridge line.
A 10\% difference requires that the speed associated with bend motion be no more than about $0.05c$.
 This inner jet and counter-jet region is where  the slowest moving components are seen (see Figures~\ref{SouthMotions}~and~\ref{NorthMotions}).
On the other hand, there is a significant sidedness intensity ratio in this region (see Section~\ref{SSSec:Iratio}) which suggests higher material speeds.
Thus, we conclude that material is moving through the bend.

The results of the sidedness measurements and their implications are discussed in Section~\ref{SSSec:Iratio}.

\section{FLOW DYNAMICS AND EMISSION}
\label{Sec:Implications}

In this section, we explore what can be learned about the flow within the jet and about the structure of the jet based on the observational results of Section~\ref{Sec:Results}.  In particular:

% SSec:JER  4.1  Expansion & Recolimation
Section~\ref{SSec:JER} examines the regions of expansion and recollimation of the jet envelope in the inner 7 mas of the jet, shown previously in Section~\ref{SSec:Avg} based on the 23-epoch average.
Here we provide intrinsic opening angle values based those data, on the 49-epoch average image data of Section~\ref{SSec:Avg} and on the images presented to study the counter-jet in Section~\ref{SSec:CJ}. 
This information can be used to provide an indication of the interaction between the jet and the external medium.  
It is distinct from Section~\ref{SSec:IMSS} which is concerned with the internal flow structure of the jet.

% JetFlow   4.2.  Jet Flow
Section~\ref{SSec:JetFlow} addresses the implications of the proper motion measurements and of the ratio of the intensities of the jet and counter-jet for jet flow speed and acceleration. 
Additionally, the different proper motions and intensities along north and south sides of the jet and counter-jet are used to explore poloidal versus helical flow models.

%  See 2017sep23 or earlier version for more filled out outline from here to the next subsection.

%  SSec:IMSS  4.3  Jet Structure
Section~\ref{SSec:IMSS} addresses implications for the internal jet spine/sheath structure from the observed slice profiles, and discusses potential differential Doppler boosting effects resulting from helical flow. 
These issues are distinct from Section~\ref{SSec:JER} which addresses the jet envelope.

%  PatMotion  4.4  Pattern Motion
Section~\ref{SSec:PatMotion} explores the implications of propagation down the jet of the long-term changes in jet transverse position.  
Here we explore the constraints placed on the jet to external medium interface that would lead to the observed propagation assuming that non-ballistic propagation can be described by the Kelvin-Helmholtz helical mode.

\subsection{Expansion \& Recollimation}
\label{SSec:JER}

\begin{figure*}    % Twocol try without [h!]  ============================
\epsscale{0.90}
% \plotone{Open_Angllle_17.eps}
\plotone{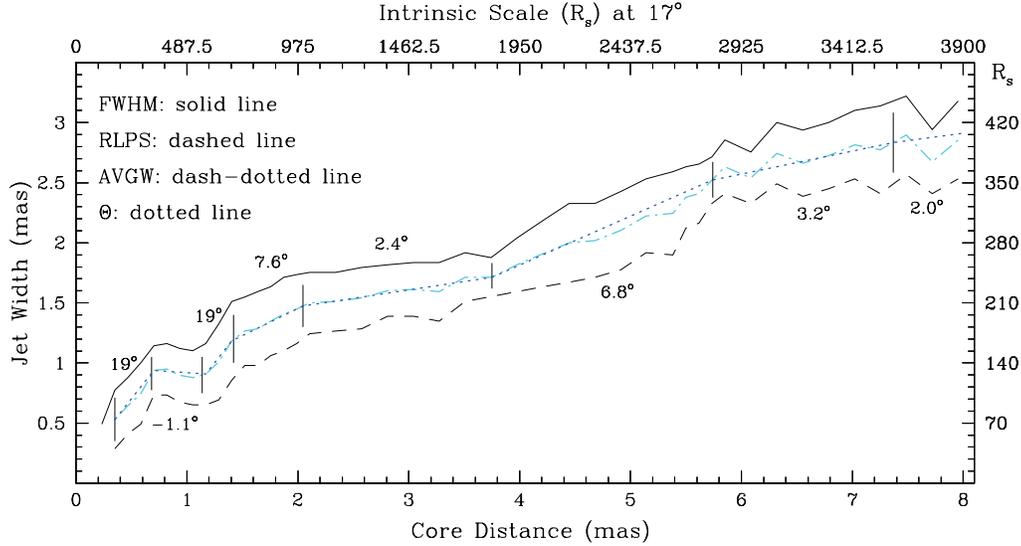}
\caption{Average (AVGW) of Full width at half maximum (FWHM) and the ridgeline peak separation (RLPS) determined widths (cyan dash--dotted line) along with linear fits to the intrinsic opening angle, $\Theta$ (blue dotted line), between the vertical lines from the 23-epoch image.
First and third vertical lines are shifted by $-0.05$~mas and $+0.1$~mas relative to the locations indicated in Figure~\ref{Jet_Fits}  in order to better accomodate the linear fitting.
Numerical labels show the intrinsic full opening angle associated with the linear fits between the vertical lines.
Beyond 7.4~mas the dotted line and label indicate the intrinsic opening angle of the kpc-scale jet (see Section~\ref{Sec:Intro}).
The width along the right axis and the intrinsic core distance along the upper axis for the jet oriented at $17\arcdeg$ to the line of sight are given in units of $R_{\rm S}$. 
}
\label{Jet_width}
\end{figure*}

Here the shape of the jet, especially regions of expansion and recollimation, is explored.  
That shape can provide constraints on the interaction of the jet and the external medium.
In Figure~\ref{Jet_width} we redisplay the FWHM and RLPS widths from Figure~\ref{Jet_Fits}, which are based on the 23-image stack of 2007 and 2008 data, and average the two to give a more reliable indicator of the behavior of the jet width as a function of the intrinsic distance in units of $R_{\rm S}$.
As implied in the earlier discussion of Figure~\ref{Jet_Fits}, there are sufficient uncertainties in the separate width measurements, especially in the region of greatest
differences between about 5 and 6 mas, that we will not attempt to interpret those differences in terms of jet physics.
Linear fits to the average width, indicated by the dotted line, are made between the vertical lines which are shifted up to $\unsim 0.1$~mas relative to the locations indicated in Figure~\ref{Jet_Fits} in order to better accommodate the linear fitting.
The resulting intrinsic full opening angle associated with the linear fits is defined as 
$
\Theta \equiv \arctan \left[  \sin \theta \left( {\Delta W  / \Delta z} \right) \right]
$
where the viewing angle $\theta = 17\arcdeg$ and $\Delta W$ and $\Delta z$ are the change in apparent width and apparent distance.
Beyond 7.4~mas  the dotted line in Figure~\ref{Jet_width} indicates the $\unsim 2.0\arcdeg$ intrinsic opening angle of the jet beyond HST-1 from \citet{OHC89}.
Our results suggest three collimation regimes:  (1)  an initial rapid expansion out to $\unsim 340 R_{\rm S}$ ($\unsim 0.7 ~{\rm mas}$) followed by jet contraction over a distance $\unsim 195 R_{\rm S}$ ($\unsim 0.4 ~{\rm mas}$), (2) a second equally rapid expansion beginning at $\unsim 535 R_{\rm S}$ ($\unsim 1.1 ~{\rm mas}$) with recollimation occurring over a distance $\unsim 1265 R_{\rm S}$ ($\unsim 2.6 ~{\rm mas}$), and (3) a third slower expansion beginning at $\unsim 1805 R_{\rm S}$ ($\unsim 3.7 ~{\rm mas}$) with recollimation still occurring at $\unsim 3605 R_{\rm S}$ ($\unsim 7.4 ~{\rm mas}$) where intrinsic distances in $R_{\rm S}$ assume a viewing angle of $\theta = 17\arcdeg$.   
%
%  \begin{figure}[h!]    %======================================
         %^^^^^^^^^^^^^^^^^^^^^^^^^^^^^^^^^^^^^^
\epsscale{1.00}         %  Must reset (outside fig env) or affects next figure.
On average our expansion is consistent with the parabolic structure out to about HST-1 found by \citet{AN2012}.
Our results confirm the initial rapid parabolic expansion found by \citet{Hada2013} inside $\unsim 300 R_{\rm S}$.
Our results also show that the structure consists of a subsequent contraction followed by additional expansions and recollimations beyond $\unsim 535 R_{\rm S}$ that is more complicated than the slower single parabolic expansion beyond a few hundred $R_{\rm S}$ found in \citet{Hada2013}.  

The 49-epoch image (Figure~\ref{Stack49}), that shows the jet and counter-jet and is based on 17 years of data, also reveals different opening angle regimes for the jet.
Motions blur the structure in the 49-epoch image but the image still reveals a 
high initial opening angle near the core, followed by a narrower opening angle indicating
slower expansion with distance.
If we use the blue lines to determine the opening angle, we find $\Theta^{\rm app} \approx 55\arcdeg$ inside 0.620~mas and $\Theta^{\rm app} \approx 20\arcdeg$ beyond 0.620~mas and out to $\unsim 1.5$~mas.
The corresponding intrinsic opening angles are $\Theta \approx 23\arcdeg$ and $\Theta \approx 6\arcdeg$, respectively.

The mild super-resolution single epoch 2013 and 2016 images (Figure~\ref{MRES}) that show the inner jet and counter-jet structures provide a more detailed view of the inner structure not blurred by motions.
Now it is possible to see a rapid initial expansion, followed by a slower expansion in both jet and counter-jet.
In the 2013 image the blue lines indicate $\Theta^{\rm app} \approx 58\arcdeg$ inside 0.5~mas and  $\Theta^{\rm app} \approx 10\arcdeg$ between $\unsim 0.5$ and $\unsim 0.6$~mas.
In fact the initial opening angle inside 0.5~mas must be even larger:
we find an initial opening angle of $\Theta^{\rm app} \approx 65\arcdeg$ if we connect the location of the sharp bend in the lines along the south counter-jet and north jet to the core position.
The corresponding intrinsic opening angles are $\Theta \ge 25\arcdeg$ inside 0.5~mas and $\Theta \approx 3\arcdeg$ between $\unsim 0.5$ and $\unsim 0.6$~mas.
In the 2016 image the blue lines indicate $\Theta^{\rm app} \approx 56\arcdeg$ inside 0.4~mas and $\Theta^{\rm app} \approx 12\arcdeg$ between $\unsim 0.4$ and $\unsim 0.6$~mas.
Again if we connect the core position to the sharp bend in the south counter-jet and north jet we find an initial opening angle of $\Theta^{\rm app} \approx 68\arcdeg$.
The corresponding intrinsic opening angles are $\Theta \ge 23\arcdeg$ inside 0.4~mas and $\Theta \approx 3.6\arcdeg$ between $\unsim 0.4$ and $\unsim 0.6$~mas. 

At even smaller scales than we resolve, \citet{Hada2016}, using data at 86 GHz, find a region of jet expansion with apparent opening angle of $\Theta^{\rm app} \approx 100\arcdeg$ inside 0.2~mas. 
They also find a subsequent contraction at a core distance of about $\unsim 0.25$ mas.  
At a viewing angle of $\theta=17\arcdeg$, the corresponding intrinsic opening angle is $\Theta \approx 38\arcdeg$ inside 0.2~mas.

The multiple expansion and collimation regions indicate a jet launching region in which equilibrium parabolic expansion is achieved only after 5 to 7~mas, i.e., several thousand $R_{\rm S}$. 
The non-equilibrium behavior found here and at smaller scales by \citet{Hada2016} should provide important information for theoretical/numerical modeling of the jet launching region.  
For example, we may be seeing the interaction between a black hole launched jet spine, velocity shearing sheath region and disk wind cocooning medium \citep[see also][]{Hada2016} as well as the extent of the jet launching and acceleration region.  
We note that the initial intrinsic opening angle  near the jet base is such that a portion of the jet flow comes directly into the line-of-sight at a viewing angle of $\theta = 17\arcdeg$ to the jet axis. 
That some jet flow can be directly into the line-of-sight near to the jet base has consequences for models of the location and mechanism responsible for the TeV and associated radio flaring observed from M\,87.

\subsection{Jet Flow}
\label{SSec:JetFlow}

We now turn to the flow dynamics of the jet as revealed by the component motion measurements and by the observed sidedness ratios described in Section~\ref{SSec:RSE}.
The component motions represent the material flow speed if their origin is some feature of the underlying flow, such as a density enhancement, that moves with the jet material.  They can also be slower than the underlying material if they are a pattern speed, such as might occur with instabilities, shocks, or interactions with external influences of some sort.  
The components are unlikely to be faster than the material speed.  
The sidedness ratio is most likely the result of Doppler boosting and is a direct consequence of the material flow speed.
A sidedness ratio higher than implied by the true material speed would require some asymmetry in the physical parameters of the jet/counter-jet system.  
A sidedness ratio lower than expected for the faster material motions also requires some complexity to the jet, such as the presence of some slower material that is contributing some of the emission.
In the presence of Doppler boosting, which is assumed when there are superluminal apparent motions or high sidedness ratios, there can also be beaming-induced brightness differences between similar regions with different flow angles.
In this section, the speeds implied by the component motions and sidedness are explored.  
Differences in speed and brightness between the sides of the jet are interpreted in terms of the possible flow patterns such as a helical flow.
This discussion complements the extensive proper motion study along the jet using the WISE analysis technique that can be found in \citetalias{MLWH2016}.

\subsubsection{Proper Motion Implications}
\label{SSec:ProperMotionImplications}

Apparent jet motions obtained from the wavelet analysis given in \citetalias{MLWH2016} Tables 2 and 3 indicate averages of $\langle\beta^{\rm app}_{\rm edge}\rangle = 0.48 \pm 0.06$ (north) and $0.21  \pm 0.04$ (south) between 0.5 and 1~mas.
Assuming pure poloidal flow along the jet edges, this would indicate intrinsic speeds of $\beta_{\rm edge} = 0.64 \pm 0.03$ (north) and $0.43 \pm 0.05$ (south) at a viewing angle of $\theta = 17\arcdeg$.
Between 1 and 4~mas  there are two velocity components.
The slower component has  $\langle\beta^{\rm app}_{\rm edge}\rangle = 0.17 \pm 0.16$ (north) and $0.49 \pm 0.24$ (south).
The faster component has $\langle\beta^{\rm app}_{\rm edge}\rangle = 2.41 \pm 0.05$ (north) and $2.20 \pm 0.15$ (south).
Following the same viewing angle and flow assumptions, the slower velocity component would indicate  $\beta_{\rm edg} = 0.37^{+0.17}_{- 0.34}$ (north) and $0.64^{+0.10}_{-0.17}$  (south), and the faster velocity component would indicate $\beta_{\rm edge} = 0.928 \pm 0.002$ $[\gamma \approx 2.68]$ (north) and $0.918 \pm 0.008$ $[\gamma \approx 2.52]$ (south).

It is hard to understand the velocity differences between opposite jet edges found assuming a purely poloidal flow.
On the other hand, different values along opposite jet edges can naturally result from flow rotation about the jet axis.
With rotation, the flow on each edge will be at different angles to the line-of-sight.  
Faster apparent motions will be seen along the edge where the flow is at an angle to the line-of-sight that comes closer to the angle at which apparent motions are maximized.

The critical angle at which apparent motions are maximized, $\theta_{\rm crit} \equiv \cos^{-1} \beta$, for $\beta \le 0.928$ is $\theta_{\rm crit} \ge 22\arcdeg$.
This angle is greater than the poloidal flow angle to the line-of-sight along the jet edges, which is slightly larger than the jet axis angle to the line-of-sight (see Equation~\ref{edgang}).
Thus, the higher apparent speeds along the northern jet edge between 0.5 and 1~mas and the higher apparent speeds along the northern jet edge of the faster component between 1 and 4~mas would indicate clockwise flow rotation about the jet axis.
Here the ``clockwise'' terminology for the sense of rotation is for an observer looking towards M\,87 along the jet (not counter-jet) axis and thus remains the same for a jet and counter-jet that share a common source of rotation.
Such rotation places the northern edge flow closer to the critical angle.  
The higher apparent speeds along the southern edge of the jet for the slower component between 1 and 4~mas would indicate counter-clockwise flow rotation for that component.

If the faster component motion along the northern edge is interpreted as resulting from clockwise rotation, then the speed of the implied clockwise helical flow can be given approximately by
\begin{equation}
\label{hspeed}
\begin{split}
\beta_{\rm h} &\approx {\beta_{\rm N}^{\rm app} \over \vert \sin (\theta_{\rm edge} + \Delta \phi_{\rm h}) \vert + \beta_{\rm N}^{\rm app} \cos (\theta_{\rm edge} + \Delta \phi_{\rm h})} \\
&\approx
{\beta_{\rm S}^{\rm app} \over \vert \sin (\theta_{\rm edge} - \Delta \phi_{\rm h}) \vert + \beta_{\rm S}^{\rm app} \cos (\theta_{\rm edge} - \Delta \phi_{\rm h})}~,
\end{split}
\end{equation}
where $\beta_{\rm N}^{\rm app}$ and $\beta_{\rm S}^{\rm app}$ are the apparent north (N) and south (S) edge motions, and $\Delta \phi_{\rm h}$ provides an estimate of the difference in flow angle relative to pure poloidal flow along the jet edges at line-of-sight angle $\theta_{\rm edge}$. 
Here  $\Delta \phi_{\rm h} > 0$ indicates clockwise flow rotation, and we have made the simplifying assumption that the flow angle to the line-of-sight can be expressed as $\theta_{\rm edge} \pm \Delta \phi_{\rm h}$.
A little manipulation allows us to solve for $\Delta \phi_{\rm h}$ which is found from 
\begin{equation}
\label{hangle}
\begin{split}
\tan & \Delta \phi_{\rm h} = \\
&{(\beta_{\rm N}^{\rm app} - \beta_{\rm S}^{\rm app}) \sin \theta_{\rm edge} \over (\beta_{\rm S}^{\rm app} + \beta_{\rm N}^{\rm app})
\cos \theta_{\rm edge} - 2 \beta_{\rm S}^{\rm app} \beta_{\rm N}^{\rm app} \sin \theta_{\rm edge}}~.
\end{split}
\end{equation}
In Equations (\ref{hspeed}) \& (\ref{hangle}), $\theta_{\rm edge}$ is the viewing angle to the jet edges and is related to the viewing angle to the jet axis by
\begin{equation}
\label{edgang}
\cos \theta_{\rm edge} = {\cos \theta \over (\tan^2 \psi + 1)^{1/2}} \approx \left(1 - {\psi^2 \over 2}\right) \cos \theta~,
\end{equation}
where here $\theta = 17\arcdeg$ is the viewing angle to the jet axis and $\psi \equiv \Theta/2$ is the intrinsic half opening angle.  

Between 0.5 and 1~mas the intrinsic half opening angle $\psi$ declines from about $9.5\arcdeg$ to about $-0.5\arcdeg$ (see Figure~\ref{Jet_width}).
Here we use an average of  $\psi\approx 4.5\arcdeg$ so the viewing angle to the jet edges is 
$\theta_{\rm edge} \approx 17.6\arcdeg$.
Now using the apparent edge motions reported above from \citetalias{MLWH2016} so that $\beta_{\rm N}^{\rm app} = 0.48 \pm 0.06$ and $\beta_{\rm S}^{\rm app} = 0.21 \pm 0.04$, we find that 
$$
\Delta \phi_{\rm h} = 7.8\arcdeg \pm 1.8\arcdeg ~{\rm and}~ \beta_{\rm h} = 0.56 \pm 0.03
$$
provides an indication of intrinsic clockwise helical flow less than 1~mas from the core.  
The likely range of values was deterimined by offsetting $\beta_{\rm N}^{\rm app}$ and $\beta_{\rm S}^{\rm app}$, separately, by plus and minus one sigma, then taking the geometric sum of the positive changes and of the negative changes.  
The results were sufficiently symmetric to use an average absolute offset as the quoted error.

Between 1 and 4~mas, $\psi$ declines from $\unsim 9.5\arcdeg$ to $\unsim 1.2\arcdeg$ (see Figure~\ref{Jet_width}) but is at the smaller angle over two thirds of this range.
Here we use a distance-weighted average of $\psi \approx 3\arcdeg$ so that the viewing angle to the jet edges is $\theta_{\rm edge} \approx 17.3\arcdeg$.
Now using the faster apparent component motions from \citetalias{MLWH2016} so that $\beta^{\rm app}_{\rm N} = 2.41 \pm 0.05$ and $\beta^{\rm app}_{\rm S} = 2.20 \pm 0.15$, we find that
$$
\Delta \phi_{\rm h} = 2.9\arcdeg \pm 2.0\arcdeg ~{\rm and}~ \beta_{\rm h} = 0.924 \pm 0.003
$$
provides an indication of intrinsic clockwise helical flow a few mas from the core.
Thus, in a flow rotation scenario we find evidence for a clockwise helical flow accelerating from $\beta_{\rm h} \approx 0.56$ inside 1~mas to $\beta_{\rm h} \approx 0.92$ at larger distances with the flow becoming more poloidal at larger distances.

For the slower-velocity components from \citetalias{MLWH2016}, if the edge motions, $\beta^{\rm app}_{\rm N} = 0.17 \pm 0.16$ and $\beta^{\rm app}_{\rm S} = 0.49 \pm 0.24$, are interpreted as resulting from counter-clockwise rotation, then the flow angle and speed of the implied counter-clockwise helical flow beyond 1~mas  is $\Delta \phi_{\rm h} = -9\arcdeg \pm 8\arcdeg$ and $\beta_{\rm h} = 0.55 {\pm 0.12}$.  
Here we note that angles closer to the poloidal case ($\Delta \phi_{\rm h} = 0$) result in a smaller value for the speed $\beta_{\rm h}$.
Our results for the fast and slow component flow angles allow for poloidal flow of both components within less than 1.5 times the errors.  
Nevertheless, our results suggest the possibility of a complex velocity shearing region with opposite sense of rotation for the fast and slow components found by \citetalias{MLWH2016}.

\subsubsection{Jet/Counter-jet Intensity Ratio Implications}
\label{SSSec:Iratio}

\begin{figure*} % twocolumn  %======================================
% \plotone{jcjratio49crop.eps}
\plotone{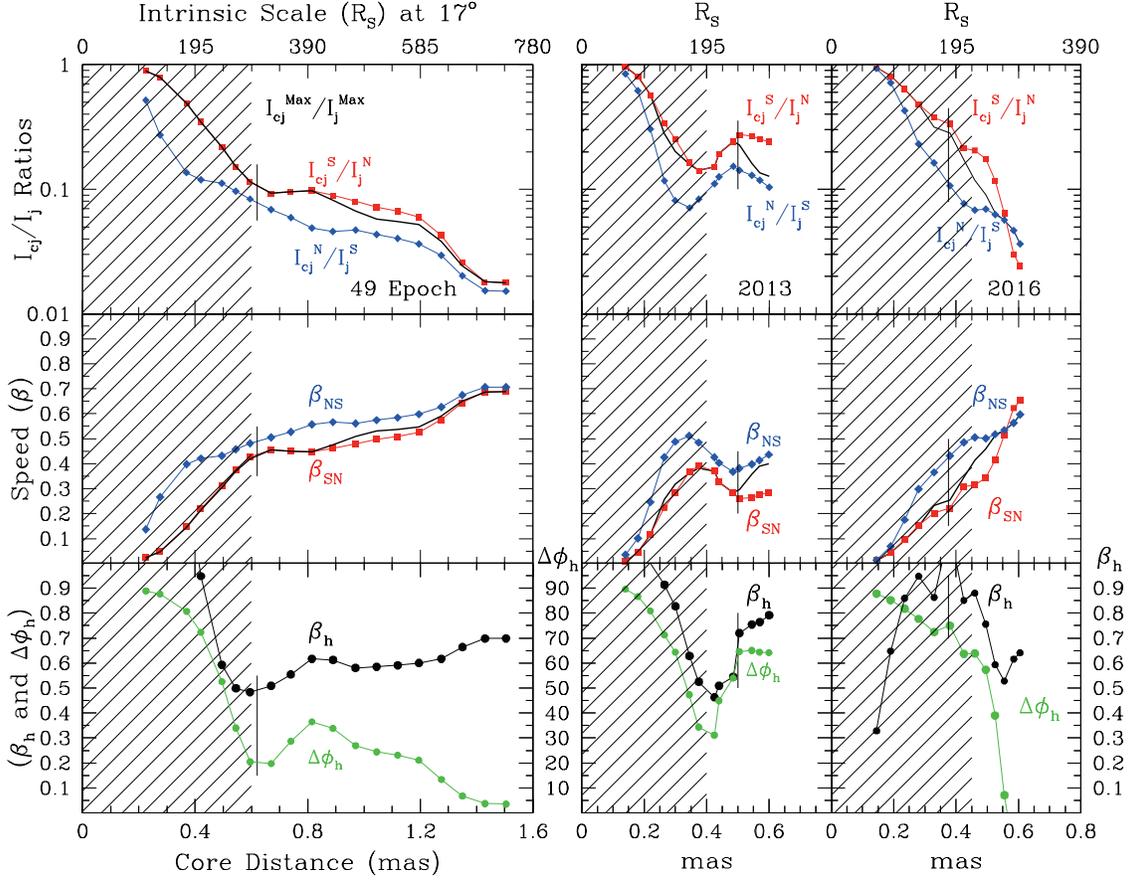}
\caption{Upper panels: Counter-jet/jet intensity ratios at image pixel positions along the blue and red lines in the M\,87 images of Figures~\ref{Stack49}~and~\ref{MRES}.
Each set is the ratio between intensities at positions on the counter-jet (cj) side and the main jet (j) side that are symmetrically opposite the core.
For comparison a black line  indicates the intensity ratio choosing the maximum counter-jet and jet edge intensities at each position.
The positions were chosen to be along the northern (N) and southern (S) ridges on the jet side.
The shaded region indicates where south counter-jet and north jet ratios are dominated by the image resolution and core.
The intrinsic scale  in units of ($R_{\rm S}$) along the jet is shown at the top.
Middle panels:  Intrinsic jet speed implied by the ratios shown in the upper panel at the viewing angle appropriate to the jet edges, $\theta_{\rm edge}$.
An angle to the line-of-sight of the jet axis $\theta = 17\arcdeg$ and a spectral index of $\alpha = 0.55$ are assumed.
The vertical lines indicate the location where the jet opening angle changes significantly, seen mainly as a kink in the lines along the northern jet edge.
Bottom panels: Intrinsic flow speed, $\beta_{\rm h}$, and flow angle, $\Delta \phi_{\rm h}$, relative to pure poloidal motion along the jet edges.
}
\label{Sided}
\end{figure*}

The ratio of the intensities of the jet and counter-jet provides information about the jet flow,
under an assumption of an inherently symmetric system, that is independent of the component motions.
The average image of M\,87 shown in Figure~\ref{Stack49} and in the super-resolution images of M\,87 shown in Figure~\ref{MRES} have been used to obtain intensity ratios between counter-jet and jet edges that are symmetrically opposite the core.
The resulting intensity ratios are shown in the upper panels in Figure~\ref{Sided} where the data points are at image pixel positions along the blue and red lines shown in the images.
The ratios, $I^{\rm S}_{\rm cj}/I^{\rm N}_{\rm j}$ and $I^{\rm N}_{\rm cj}/I^{\rm S}_{\rm j}$ are between intensities at positions on the counter-jet (cj) side and at positions on the main jet (j) side that are symmetrically opposite the core.
The positions were chosen to be along the northern (N) and southern (S) ridges on the jet side.
The structures on the counter-jet side did not determine the positions, although the images show that the symmetry assumption is good but not exact.
In Figure~\ref{Sided} the ratio $I^{\rm Max}_{\rm cj}/I^{\rm Max}_{\rm j}$, which is the intensity ratio choosing the maximum counter-jet and jet edge intensities at each core distance independent of symmetry considerations, is included for comparison purposes.
 Recall that the intensity ratios are dominated by the beam until the jet and counter-jet features are resolved from the core. 
% 
% \begin{figure*} [h!]   %======================================
         %^^^^^^^^^^^^^^^^^^^^^^^^^^^^^^^^^^^^^^
Since the beam is elongated more along the inner south ridge on the counter-jet side and north ridge on the jet side, these intensity ratios will remain high farther from the core, as is evident in all the intensity ratio panels.
For  example, the 49-epoch $I^{\rm N}_{\rm cj}/I^S_{\rm j}$ ratio stops its steep core-dominated decline at $\unsim 0.4$~mas whereas the $I^{\rm S}_{\rm cj}/I^{\rm N}_{\rm j}$ ratio stops its steep core dominated decline at $\unsim 0.6$~mas, and that the 2013 $I^{\rm N}_{\rm cj}/I^{\rm S}_{\rm j}$ ratio stops its steep core dominated decline at $\unsim 0.3$~mas whereas the $I^{\rm S}_{\rm cj}/I^{\rm N}_{\rm j}$ ratio stops its steep core dominated decline at $\unsim 0.4$~mas.
 Also the 2016 $I^{\rm N}_{\rm cj}/I^{\rm S}_{\rm j}$ and $I^{\rm S}_{\rm cj}/I^{\rm N}_{\rm j}$ ratios do not appear beyond core dominated influence until $\unsim 0.45$~mas.  
The region of core influence is shaded in the figure panels.
Beyond the region of core influence, the fact that the regions sampled are symmetric about the core, as is the beam, should ensure that the ratios are not significantly sensitive to the beam shape.

The persistence of larger ratios between the south counter-jet and north jet edges, $I^{\rm S}_{\rm cj}/I^{\rm N}_{\rm j}$, than between the north counter-jet and south jet edges, $I^{\rm N}_{\rm cj}/I^{\rm S}_{\rm j}$, can be interpreted in terms of Doppler boosting and suggests that the northern side of the counter-jet and jet is deboosted relative to the south side of the  counter-jet and jet.
In Figure~\ref{Sided} the middle panels show the computed values of a purely poloidal speed along the north counter-jet edge and south jet edge, $\beta_{\rm NS}$, and along the south counter-jet edge and north jet edge, $\beta_{\rm SN}$, required by the intensity ratios.
Speeds are computed from
\begin{equation}    %======================================
\beta \cos \theta_{\rm edge}  = \left( {1 - (I_{\rm cj} /I_{\rm j})^{1/(2 + \alpha)} } \over {1 + (I_{\rm cj}/ I_{\rm j})^{1/(2 + \alpha)} } \right)~.
\end{equation}         %^^^^^^^^^^^^^^^^^^^^^^^^^^^^^^^^^^^^^^
where, as previously stated,
$
\cos \theta_{\rm edge}  \simeq (1 - {\psi^2 /2}) \cos \theta .
$
 Values are computed using the viewing angle to the jet axis $\theta = 17\arcdeg$, the intrinsic half opening angle $\psi \equiv \Theta/2$, and the spectral index $\alpha = 0.55$ ($I \propto \nu^{-\alpha}$; $\alpha$ appropriate to the kpc-scale jet in the radio band; Frazer Owen, private communication).
In the panels the vertical line indicates the location where the opening angle changes.
For all computations, the intrinsic opening angles for the average 49-epoch image are assumed to be
$\Theta = 23\arcdeg$ inside 0.620~mas and  $\Theta = 6\arcdeg$ beyond 0.620~mas.
The intrinsic opening angles for the 2013 and 2016 images are assumed to be: (2013)  $\Theta = 25\arcdeg$ inside 0.5~mas and $\Theta = 3\arcdeg$ beyond 0.5~mas and (2016) $\Theta = 23\arcdeg$ inside 0.4~mas and $\Theta = 3.6\arcdeg$ beyond 0.4~mas. 

For an assumed purely poloidal flow, the intensity ratios associated with the 49-epoch intensity image require a higher speed along the north counter-jet and south jet edges and lower speed along the south counter-jet and north jet edges.
These relative speeds are not consistent with the faster motions along the north side of the jet and slower motions along the south side of the jet found from component motions by \citetalias{MLWH2016}.  
A resolution to this inconsistency for poloidal flow  will be suggested in the discussion of helical flows below.
The trend of intensity ratios with core distance provides evidence for speed changes, and, in particular, acceleration in resolved regions near the core.
On average the flow implied by the intensity ratios increases from $\beta \approx 0.3$ to $0.5$ between $\unsim 0.4$ and $\unsim 0.7$~mas ($\unsim 195$ to $\unsim 341$~$R_{\rm S}$), then slowly increases to $\beta \approx 0.55$ at 1.2~mas ($\unsim 585$~$R_{\rm S}$) and finally accelerates to $\beta \approx 0.7$ at $\unsim 1.4$~mas ($\unsim 682$~$R_{\rm S}$).
At larger distance the counter-jet intensity approaches the level of the noise and imaging artifacts so the intensity ratios cannot be determined from our data.
The counter-jet/jet intensity ratio implies that the flow speed increases most rapidly in the regions of about 0.4 to 0.7~mas and about 1.2 to 1.4~mas where the two rapid expansions of counter-jet and jet occur (see Figure~\ref{Jet_width}).
Recall that the counter-jet width behaves similarly to the jet width within the $\unsim 1.2$ mas of the core where it can be seen.

The average purely poloidal flow behavior inferred from the intensity ratios for the 2013 and 2016 super-resolution images are very different from each other.
In part the differences may result from small separation from the core along with noise and imaging artifacts particularly in the 2016 image.
Nevertheless, the 2016 image intensity ratios imply an intrinsic flow that on average rapidly accelerates from $\unsim 0.25c$ at 0.3~mas to $\gtrsim 0.6c$ at $\unsim 0.6$~mas that exceeds the speed $\unsim 0.5c$ found for the 49-epoch average image at $\unsim 0.6$~mas.
The better quality 2013 image intensity ratios imply a much more complex behavior with an intrinsic  flow that on average first increases from $\unsim 0.4c$ at 0.3~mas to $\gtrsim 0.45c$ at $\unsim 0.35$~mas, then decreases to $\unsim 0.3c$ at 0.5~mas where the opening angle changes from rapid to a slower expansion.
At larger distance the flow accelerates back to $\unsim 0.4c$ at 0.6~mas.   
Notice that typical speeds from the 2013, 2014 and 49-epoch images at 0.5~mas are in the range 0.4 to $0.6c$. 

We can compare typical poloidal flow speed values inferred from the counter-jet/jet intensity ratios to the proper motions associated with bright features within the jet shown in Figures~\ref{SouthMotions} and \ref{NorthMotions} and the apparent motions from the wavelet analysis reported by \citetalias{MLWH2016}.
The proper motions shown in Figures~\ref{SouthMotions} and \ref{NorthMotions} indicate $\beta^{\rm app} \lesssim 0.5$ at core distances $< 0.5$~mas with motions up to $\beta^{\rm app} \lesssim 2.2$ at core distances $> 1$~mas (somewhat less on the north side).
Figure 16 in \citetalias{MLWH2016} also shows clearly that the velocities of the fastest 10\% of features increase from less than $0.5c$ at $\unsim 0.5$~mas to over $2c$ beyond $\unsim 2$~mas.
Thus, tracking individual components and the wavelet analysis indicate acceleration in the same region indicated by the counter-jet/jet intensity ratios.
In particular, at a viewing angle of $\theta = 17\arcdeg$ the wavelet analysis indicates acceleration along the jet edges from $\langle\beta_{\rm edge}\rangle \approx 0.5$ at $< 0.5$~mas to $\langle\beta_{\rm edge}\rangle \approx 0.92$ at $> 1$~mas from the core.
Thus, for an assumed purely poloidal flow the intensity ratios provide qualitative but not quantitative agreement with the speeds and acceleration found by the component tracking and by the wavelet analysis.  

In Section~\ref{SSec:ProperMotionImplications} we found that faster northern edge motions along the jet can be associated with helical flow.  We examine the intensity ratio implications for helical flow in the lower panels in Figure~\ref{Sided}. 
Here we compute the difference in flow angle, $\Delta \phi_{\rm h}$, relative to pure poloidal motion along the jet edges, and the flow speed, $\beta_{\rm h}$, required to produce the observed counter-jet/jet intensity ratios, where again we have made the simplifying assumption that the flow angle to the line-of-sight can be expressed as $\theta_{\rm edge} \pm \Delta \phi_{\rm h}$. 
In these lower panels the flow speed is computed from
\begin{equation}
\label{betah}
\begin{split}
\beta_{\rm h} &\approx
{(1 - R_{\rm NS}) \over \cos (\theta_{\rm edge} - \Delta \phi_{\rm h}) + R_{\rm NS} \cos (\theta_{\rm edge} + \Delta \phi_{\rm h})} \\
&\approx {\{(1 - R_{\rm SN}) \over \cos (\theta_{\rm edge} + \Delta \phi_{\rm h}) + R_{\rm SN} \cos (\theta_{\rm edge} - \Delta \phi_{\rm h})}~,
\end{split}
\end{equation}
where
$$
R_{\rm NS} \equiv \left(I^{\rm N}_{\rm cj}/I^{\rm S}_{\rm j}\right)^{1/(2 + \alpha)}~{\rm and}~R_{\rm SN} \equiv \left(I^{\rm S}_{\rm cj}/I^{\rm N}_{\rm j}\right)^{1/(2 + \alpha)}~.
$$
A little manipulation allows us to solve for $\Delta \phi_{\rm h}$ which is found from
\begin{equation}
\tan \Delta \phi_{\rm h} = {(R_{\rm SN} - R_{\rm NS}) \over (1 - R_{\rm SN})(1 - R_{\rm NS})} \cot \theta_{\rm edge}~,
\end{equation}
and as previously, $\Delta \phi_{\rm h} > 0$, indicates clockwise helical flow.
Our helical flow analysis here depends on both north and south edge counter-jet and jet intensity ratios and thus can only be applied if both edge ratios are beyond the region of core influence as shown by the shaded region in Figure~\ref{Sided}.  These equations look very similar to Equations~\ref{hspeed} and \ref{hangle} but are in terms of intensity ratios rather than component speeds.

The assumption of helical flow solves the edge flow issues associated with the intensity ratios raised by the assumption of purely poloidal flow.
Now flow will appear faster on the northern jet edge and slower on the southern jet edge as observed.  
The helical flow speeds and acceleration associated with the 49-epoch image intensity ratios are comparable to the values indicated by the component tracking and by the wavelet analysis inside 1~mas, i.e., acceleration from $\beta_{\rm h} \approx 0.475$ to $\beta_{\rm h} \gtrsim 0.625$ between 0.6 and 0.8~mas, albeit with values of $\Delta \phi_{\rm h} \approx 20 - 35\arcdeg$, considerably larger than suggested by our estimate of $\Delta \phi_{\rm h} \approx 7.8\arcdeg$ based on average edge motions inside 1~mas.
 Inside 1~mas the 49-epoch image intensity ratios probe the counter-jet and jet region where the first rapid expansion ceases and the jet recollimates. 

Speeds and acceleration from component tracking and the wavelet analysis  exceed the values suggested by the 49-epoch image counter-jet/jet intensity ratios at core distances beyond 1~mas, although by 1.4~mas  $\Delta \phi_{\rm h}$ has decreased to a value of about $2.9\arcdeg$ that is comparable to our estimate based on average edge motion between 1 and 4~mas.
A possible explanation for slower motion based on intensity ratios beyond 1~mas is the slow components detected in the wavelet analysis.
Those components would not be subject to strong beaming so adding the emission from them and from the fast components would mimic an intermediate speed, which is what is seen.
 Recall that the intensity ratio is set by the material flow speed, in contrast to component speeds which can be set by patterns moving at different speeds from the material.
While it is possible for a pattern to move faster than the material speed, e.g., flow plus sound or Alfv\'en speed, it is unlikely that pattern effects can explain why the counter-jet is brighter than would be expected for the observed component speeds.  

At very small angular scales the super-resolution image intensity ratios  probe the first rapid counter-jet and jet expansion region.
The better quality 2013 intensity ratios would indicate a helical flow that accelerates from $\beta_{\rm h} \approx 0.5$ to $\beta_{\rm h} \approx 0.8$ between 0.4 and 0.6~mas in the initial rapid counter-jet and jet expansion region that extends to $\unsim 0.7$~mas (see Figure~\ref{Jet_width}).
This provides our only evidence for rapid acceleration associated with rapid counter-jet and jet expansion and towards flow values consistent with the faster component motions found by the WISE analysis between 1 and 4~mas.
The poorer quality 2016 intensity ratios could be taken to indicate the creation of a slower moving counter-clockwise helical flow, i.e. $\Delta \phi_{\rm h} < 0$, beyond 0.55~mas, consistent with the slower moving component motions found by the WISE analysis between 1 and 4~mas.
In any event, it is clear that at these small angular scales the situation is complex and any quantitative analysis is subject to noise and imaging artifacts in addition to any real temporal changes in the flow structure.

\subsection{Jet Structure}
\label{SSec:IMSS}

We now turn from jet shape and flow speeds to examine what can be learned about the internal structure of the jet from its intensity distribution.
The pronounced edge brightness is a clear sign that the jet is not uniform and that the main emission comes from the edge.
This raises questions such as how thick is the emitting region and what might be in the center.
Also Doppler boosting implied by the relativistic motions and helical jet flow have implications for the intensity distribution of the jet that are compared with the imaging results.
Interpreting the brightness distribution has to begin with an understanding of what part of 
the jet is actually being observed along a given line of sight in a jet oriented close to the line-of-sight.

\subsubsection{Lines of Sight}
\label{SSSec:LofS}
 
The line-of-sight through the jet axis passes through the jet nearside, center and farside at significantly different core distances.  
This complicates understanding the implications of the transverse intensity profiles for sheath and spine emission structure.  
It also complicates the use of mid-jet as opposed to jet edge proper motions for understanding the jet flow and Doppler beaming effects.
For example, at a viewing angle of 17\arcdeg, a sightline that passes through the jet axis at 1~mas from the core, passes through the farside and nearside of the jet at  $\unsim 0.7 ~{\rm and}~1.5$~mas from the core, respectively.
A typical line-of-sight through the jet axis crosses the nearside of the jet from 2 times (close to the core) to 1.5 times  farther from the core than it crosses the farside of the jet.
As a consequence the average farside edge intensity is 3 (close to the core) to 2 times the nearside edge intensity where the sightline leaves the jet (see Figure~\ref{Jet_Fits} to compare northern or southern ridgeline intensities at different distances from the core).
If emissivity is relatively uniform, the intensity where a sightline crosses the jet axis might be the average of the nearside and farside ``edge" intensities along the sightline.
 This intensity difference along a sightline through the jet axis is problematic for determination of the nearside, center and farside location of observed centerline structures.  
We will address this issue in what follows.

\subsubsection{Sheath \& Spine}
\label{SSSec:ST}

Transverse slice profiles (see Figures~\ref{Slices_A} and \ref{Slices_B}) indicate that the intensity at the jet centerline can be less than half that of the bright edges.
Production of such edge-brightened profiles requires a high sheath and low spine effective emissivity.
If we assume a cylindrical jet and that emission comes only from a sheath of thickness $\Delta R = R_{\rm j} - R_{\rm sp}$, where $R_{\rm j}$ and $R_{\rm sp}$ are the radius of the jet and spine respectively, then the sheath path length through the jet axis for viewing angle $\theta$ can be written as $\ell_{\rm cen} = 2 \Delta R/ \sin \theta$.
The path length through the sheath along a line-of-sight through the jet edge at $(R_{\rm j} - \Delta R/2)$ (i.e., the midpoint of the sheath), can be written as
%  $$
\begin{equation}
\begin{split}
\ell_{\rm edge}^2 &= 4 {R_{\rm j}^2\over \sin^2 \theta} \left[1 - \left({R_{\rm j} - \Delta R/2\over R_{\rm j}}\right)^2\right] \\
&= 4{\Delta R [R_{\rm j} - \Delta R/4]\over \sin^2 \theta},~{\rm and}\\
\ell_{\rm edge} & \approx 2 {(R_{\rm j} \Delta R)^{1/2} \over \sin \theta}~{\rm for}~\Delta R \ll 4R_{\rm j}.  \nonumber
\end{split}
\end{equation}
%  $$
Thus, a cylindrical jet with uniform sheath emissivity and a jet edge intensity twice the center intensity, as suggested by the transverse slice profiles, means
%  $$
\begin{equation}
\begin{split}
I_{\rm cen}/I_{\rm edge} &\approx \ell_{\rm cen}/\ell_{\rm edge} \approx  \left({\Delta R \over R_{\rm j}}\right) ^{1/2} = 0.5~~{\rm and}\\
\Delta R &\approx R_{\rm j}/4.   \nonumber
\end{split}
\end{equation}
%  $$
The actual jet is expanding, so the intrinsic emissivity and sheath thickness will change with the jet's radius and core distance. 
Also sightlines cross the jet edges and jet axis at very different core distances.
However, given that we found the average edge intensity at the distance where a line-of-sight crosses the jet centerline to be on the order of the nearside and farside average, we can still conclude that the jet consists of a low apparent emissivity jet spine and high apparent emissivity jet sheath with observed emission largely coming from a sheath with thickness on the order of $R_{\rm j}/4$.  Note that these arguments apply once the jet ridges are resolved from each other and the jet expansion is moderate --- say at core distances beyond about $0.5$~mas.

For intensity profiles to be as edge-brightened as they appear despite the path length along sightlines through the jet spine being on the order of three times the path length through the nearside and farside jet sheath, the ``effective" spine emissivity must be at least an order of magnitude less than the ``effective" sheath emissivity.
 Intrinsic spine emission can be Doppler deboosted relative to intrinsic sheath emission by a factor 
$$
\epsilon \equiv \left({\delta_{\rm sp} \over \delta_{\rm sh}}\right)^{2 + \alpha} = \left[ {\gamma_{\rm sh} (1 - \beta_{\rm sh} \cos \theta ) \over \gamma_{\rm sp} (1 - \beta_{\rm sp} \cos \theta ) }\right]^{2.55}~.
$$
To achieve an order of magnitude deboosting, $\epsilon \le 0.1$, for $\theta = 17\arcdeg$ with the intrinsic sheath speed and Lorentz factor of  $\beta_{\rm sh} \approx 0.924$ and $\gamma_{\rm sh} \approx 2.62$ appropriate between 1 and 4~mas, would require a spine Lorentz factor and speed of $\gamma_{\rm sp} \ge 16.5$ and $\beta_{\rm sp} \ge 0.998$.
This required minimum gives an apparent motion at $\theta = 17\arcdeg$ of $\beta^{\rm app} \ge 6.4$.
This lower limit is consistent with the $\beta^{\rm app} = (6.1 \pm 0.6)$ seen at HST-1 in the optical by \citet{BSM99} and the $\beta^{\rm app} = (6.4 \pm 0.8)$ of the fastest component seen in the radio by \citet{Giroletti2012}.
Thus, it is possible that a high speed Doppler deboosted spine lies within the slower moving sheath. 

\subsubsection{Rotation, Expansion \& Doppler Boosting}
\label{SSSec:JRot}

The presence of clockwise helical flow in the sheath would lead to differential Doppler boosting across the jet.
If we consider the values found above for the region between 0.5 and 1~mas (1 and 4~mas region in parentheses) of $\Delta \phi_{\rm h} \approx 7.8\arcdeg$ ($\approx$2.9\arcdeg) and $\beta_{\rm h} \approx 0.56$ ($\approx$0.924) with $\theta_{\rm edge} \approx 17.6\arcdeg$ ($\approx $17.3\arcdeg), then we would predict an enhanced Doppler boost of the southern jet edge relative to the northern jet edge in the range
$$
\left({\delta_{\rm S} \over \delta_{\rm N}}\right)^{2 + \alpha} = \left[{1.0 - \beta_{\rm h} \cos (\theta_{\rm edge}+ \Delta \phi_{\rm h}) \over 1.0 - \beta_{\rm h} \cos (\theta_{\rm edge} - \Delta \phi_{\rm h})}\right]^{2.55} \approx 1.3 -1.8~,
$$
where the numerical value range corresponds to the results for 0.5 to 1~mas and 1 to 4~mas in that order.

This predicted differential Doppler boosting is lower closer to the core.
A comparison of ridgeline intensities from Figure~\ref{Jet_Fits} shows a southern ridgeline intensity equal to or greater than the northern ridgeline intensity at core distances $> 0.8$~mas.
However, the northern ridgeline intensity is higher at core distances $ < 0.8$~mas in the 23-epoch image.
While the ratio is significantly less close to the core,  it would still predict a brighter southern edge.
It seems more likely that the maximum edge intensity is a function of change in jet direction associated with long-term epoch evolution (see Figures~\ref{LTE1} to \ref{LTE3}), and jet edge intensity differences are not likely the result of differential Doppler boosting resulting from rotation.

There will also be differential Doppler boosting between the nearside (ns) and farside (fs) of the jet along lines-of-sight through the jet axis because, thanks to the jet opening angle, the sides are at different angles to the line-of-sight.
If we consider $\beta_{\rm h} \approx 0.56$ with opening angle $\psi \approx 4.5\arcdeg$ between 0.5 and 1~mas, and  $\beta_{\rm h} \approx 0.924$ between 1 and 4~mas with $\psi \approx 3\arcdeg$ then we would predict an enhanced Doppler boost of the nearside relative to the farside in the range
$$
\left({\delta_{\rm ns} \over \delta_{\rm fs}}\right)^{2 + \alpha} = \left[{1.0 - \beta_{\rm h} \cos \theta_{\rm fs} \over 1.0 - \beta_{\rm h} \cos \theta_{\rm ns}}\right]^{2.55} \approx 1.15 - 1.85~,
$$
where the smaller value is closer to the core, and $\theta_{\rm fs} \approx \theta + \psi$ and  $\theta_{\rm ns} \approx \theta - \psi$.
This amount of differential Doppler boost between nearside and farside can partially compensate for higher farside intrinsic intensities along lines-of-sight through the jet axis. 

Clockwise helical flow around the jet will lead to apparent transverse motion across the jet centerline with northwards apparent transverse motions across the nearside of the jet and southwards apparent transverse motions across the farside of the jet.  \citetalias{MLWH2016} find a tendency towards northwards apparent transverse motions and this indicates a bias towards seeing motions on the nearside of the jet (see Figures~2 and 3 in \citetalias{MLWH2016}).

\subsection{Long-Term Pattern Motion}        
\label{SSec:PatMotion}

In Section~\ref{SSec:LongTerm}, we showed that the transverse position of the jet shifts on a quasi-periodic about 8 to 10~yr timescale.  
This period provides a characterization of the variation timescale but does not have an adequate time span to claim true periodic motion such as would be expected from precession.
In this section, we discuss the implications of these shifts for the properties of the interface between the jet and external medium if their propagation is described by the Kelvin-Helmholtz helical mode.
 
The structural variations observed are consistent with a jet base that changes direction on the timescale noted above, leading to a variation in flow direction that appears to propagate down the jet at apparent speeds  $\beta_{\rm p}^{\rm app} \approx 0.81 - 0.89$ where the range is set by the two different methods to determine the speed used in Section~\ref{DispMotion}.
At a viewing angle of $\theta = 17\arcdeg$, the intrinsic pattern motion range is $\beta_{\rm p} \approx  0.76 - 0.78$.
 The intrinsic pattern motion is significantly greater than the slower intrinsic component motions from \citetalias{MLWH2016} of $ \beta_{\rm h} \approx 0.553$ but is significantly less than the  faster intrinsic component motions from \citetalias{MLWH2016} of $\beta_{\rm h} \approx 0.924$ at distances beyond 1~mas.
\citetalias{MLWH2016} suggested that the fast and slow component motions associated with the bright jet edges indicated velocity shear associated with the jet-external medium interface.
Thus, we suggest that the pattern can be thought of as moving through an external (e) medium which itself can be moving with flow speed $0 < u_{\rm e} < 0.553c$ but that the pattern moves more slowly than the internal jet (j) flow speed which is $u_{\rm j} > 0.924c$. 

In what follows let us assume that the jet with flow speed $\beta_{\rm j} > 0.924$ ($\gamma_{\rm j} > 2.62$), determined by the \citetalias{MLWH2016} faster moving jet components beyond 1~mas, is separated by a sharp velocity discontinuity from an external medium with flow speed $\beta_{\rm e} < 0.553$ ($\gamma_{\rm e} < 1.20$), determined by the \citetalias{MLWH2016} slower moving components beyond 1~mas.
We assume a direction change at the jet base on a $\unsim 9$~yr timescale and that this direction change propagates with an intrinsic  pattern speed of $\beta_{\rm p} \approx 0.770$.

A non-ballistically moving bulk displacement of the jet can be described by the properties of the Kelvin-Helmholtz helical mode \citep[see][]{Hardee2007} provided the displacement frequency is less than the resonant frequency (see Equation~\ref{hmrfreq}).
See Appendix~\ref{HMtheory} for a brief review of the important equations.
Numerical study shows that the helical mode above resonance does not lead to bulk displacement of the jet as a whole \citep{HHRG2001}.
Below resonance the helical mode propagation properties are relatively insensitive to the presence of a velocity shear layer \citep[see][]{Birkinshaw1991}.
The important propagation properties are revealed by analytical approximations for helical wave propagation along a cylindrical jet with purely poloidal flow and magnetic field inside and outside a sharp velocity shear edge \citep[see][]{Hardee2007}).
 A simple linear displacement at the jet base can be described by two counter-rotating equal amplitude helical displacements.
Provided flow and magnetic helicity are relatively small beyond 1~mas, approximations to pattern motions assuming no helicity should prove adequate for estimating the jet to external medium interface properties required to produce the observed pattern motion.

The $\unsim 9$~yr timescale for the change in direction at the jet base implies an angular frequency of $\omega \approx 7 \times 10^{-9} \pi$~rad s$^{-1}$.
An estimate for the conditions required for the lower limit to the resonant frequency to exceed this angular frequency can be obtained by assuming that, in the ``sonic" limit, the sound speeds in the jet and external medium are the same or that, in the ``Alfv\'enic" limit, the Alfv\`en speeds in the jet and external medium are the same. 
Here the ``sonic" limit means that the sound speed greatly exceeds the Alfv\'en speed (weakly magnetized flow scenario) and the ``Alfv\'enic" limit means that the Alfv\'en speed greatly exceeds the sound speed (strongly magnetized flow scenario).
In these scenarios the resonant frequency (see Equation~\ref{resfreq})
$$
\omega^{\ast}  \gtrsim \frac{3 \pi}{4} \frac{v_{\rm w}}{R_{\rm j}} > \omega \approx 7 \times 10^{-9} \pi~{\rm rad s}^{-1}~,
$$ 
at say 3~mas where $R_{\rm j} \approx$ 0.9~mas $\approx 2.25 \times 10^{17}$~cm  provided $v_{\rm w} \ge 0.07c$. We have used the FWHM value for the jet width at 3~mas from Figure~\ref{Jet_width} to estimate $R_{\rm j}$, and $v_{\rm w}$ is the sound speed in the sonic limit or the Alfv\'en speed in the Alfv\'enic limit. 
This required lower limit to the sound speed or Alfv\'en speed in the jet and external medium at about 7\% of lightspeed lies well below the maximum possible sound speed or Alfv\'en speed of $c/\sqrt 3$ or $\lesssim c$, respectively. 
Thus, it is clearly possible for a quasi-periodic $\unsim 9$~yr timescale jet displacement to be described by the low to resonant frequency helical mode propagation properties.

Pattern motion, $v_{\rm p}$, is a function of  the jet speed $u_{\rm j}$, the external medium speed $u_{\rm e}$, and the enthalpy ratio, $W_{\rm j}/W_{\rm e}$ between the jet and external medium. Here the enthalpy $W\equiv \rho +\left[ \Gamma /\left( \Gamma -1\right) \right] P/c^{2} $ where $\rho$, $P$ and $\Gamma$ are the rest mass density, pressure and adiabatic index, respectively.

If pattern motion is described by helical mode propagation below the resonant frequency, the required enthalpy ratio is given by Equation~\ref{Wlowfreq}.
If magnetic fields are small, then in the sonic limit $W_{\rm j} \approx [(v_{\rm p} - u_{\rm e})/(u_{\rm j} - v_{\rm p})] (\gamma_{\rm e}/\gamma_{\rm j})^2 W_{\rm e}$ where $1.5 \le [(v_{\rm p} - u_{\rm e})/(u_{\rm j} - v_{\rm p})] \le 5$ and $ 0.21 \ge (\gamma_{\rm e}/\gamma_{\rm j})^2 \ge 0.14$ for $\beta_{\rm j} = 0.924$ ($\gamma_{\rm j} = 2.62$), and $\beta_{\rm e} < 0.553$ ($\gamma_{\rm e} < 1.20$).
If $\rho \gg P/c^2$ the jet is similar in density to the surrounding medium unless the faster moving jet components are not truly representative of the underlying jet flow.
A much faster jet flow of $\gamma_{\rm j} = \gamma_{\rm sp} \gtrsim 16.5$ ($\beta_{\rm sp} \gtrsim 0.998$) such as required to Doppler deboost the spine and produce the apparent $\unsim 6c$ superluminal motions in the optical at HST-1 implies a jet over an order of magnitude less dense than the surrounding external medium.
If magnetic fields are large, then in the Alfv\'enic limit $W_{\rm j} \approx [v_{\rm p}/(u_{\rm j} - v_{\rm p})] (\gamma_{\rm A}/\gamma_{\rm j})^2 W_{\rm e}$ where $[v_{\rm p}/(u_{\rm j} - v_{\rm p})] \approx  5$ and $\gamma_{\rm A} \approx [\gamma_{\rm e}^2(1 - u_{\rm e}/v_{\rm p}) + (B_{\rm e}^2 + B_{\rm j}^2)/ 4 \pi W_{\rm e} c^2]^{1/2}$.
The Alfv\'enic limit suggests somewhat higher jet density relative to the unmagnetized case as in general $(\gamma_{\rm A}/\gamma_{\rm j})^2 > (\gamma_{\rm e}/\gamma_{\rm j})^2$.

If pattern motion is more accurately described by helical mode propagation near the resonant frequency, the required enthalphy ratio is given by Equations~\ref{Wresfreq}.
The principal difference with the low frequency form lies in the square in the velocity term so that $W_{\rm j} \approx [(v_{\rm p} - u_{\rm e})/(u_{\rm j} - v_{\rm p})]^2 (\gamma_{\rm e}/\gamma_{\rm j})^2 W_{\rm e}$ where $2.25 \le [(v_{\rm p} - u_{\rm e})/(u_{\rm j} - v_{\rm p})]^2 \le 25$ and $ 0.21 \ge (\gamma_{\rm e}/\gamma_{\rm j})^2 \ge 0.14$ for $\beta_{\rm j} = 0.924$ ($\gamma_{\rm j} = 2.62$), and $\beta_{\rm e} < 0.553$ ($\gamma_{\rm e} < 1.20$).
Thus, if the pattern motion is described by a helical mode propagation near resonance and $\rho  \gg P/c^2$, a jet density comparable to the external medium density would provide the properties required by the observed pattern motion. 
Once again a much faster jet of $\gamma_{\rm j} = \gamma_{\rm sp} \gtrsim 16.5$ ($\beta_{\rm sp} \gtrsim 0.998$) implies a jet over an order of magnitude less dense than the surrounding external medium.

Overall this analysis would lead us to believe that a jet rest mass density somewhat less, to about an order of magnitude less, than the external medium density would provide the properties required by the observed pattern motion. 
 The external medium referred to here is not the interstellar medium but rather the immediate cocooning medium surrounding the jet and likely an outflowing magnetized wind from the accretion disk.
 
 There is no evidence for bulk displacement of the jet at arcsecond scales until knot A.
This implies that the bulk displacement that we observe at milli-arcsecond  scales is damped at larger scales.
Theory and numerical simulation show that damping of a displacement will occur in the presence of strong magnetic fields \citep[see Equation~\ref{damping} and][]{Mizuno2007}. 
Even if this condition is not met, a bulk displacement of the jet at the jet base that can be described by a wave frequency $\omega$ that is initially less than the local resonant frequency $\omega^{\ast}$ can propagate down the expanding jet to where $\omega$ is above the local resonant frequency because $\omega^{\ast} \propto R^{-1}$ declines along an expanding jet.
Thus, an initial bulk displacement at the jet base can become damped at larger distances.
 
% Force section heading onto next column.
\vspace{0.5cm}

\section{SUMMARY}
\label{Sec:SD}

Because of proximity and  large black hole mass, M\,87 is the best available source in which to study the region close, in terms of gravitational scales, to where a jet is launched.
This paper presents the main observational results from a program of intensive monitoring of the sub-pc to pc scale structure of M\,87 in 2007 and 2008 and from roughly annual observations from 1999 to 2016 using the VLBA at 43 GHz.
A total of 50 observations are included.

The major results of the study are:

\begin{itemize}

\item  The images clearly show a wide-opening-angle jet base and edge-brightened structure. 

\item  The jet has an overall parabolic shape as reported previously, but with regions of rapid 
expansion and recollimation.

\item  Mildly superluminal motions with apparent speeds of $\beta^{\rm app} \approx 2$ are seen beyond about 2 mas (0.16 pc in projection) from the core.

\item  A region of acceleration is seen in the inner 0.5 to 2~mas, confirming results presented in \citetalias{MLWH2016} which were based on a new analysis method applied to some of the observations reported in this paper.  This region is at an intrinsic (deprojected) core distance of about 0.14 to 0.55 pc or 240 to 960 $R_{S}$ along the jet for our assumed angle to the line-of-sight ($17\arcdeg$), distance (16.7 Mpc), and black hole mass ($6.1 \times 10^9$ M$_{\odot}$).

\item  The jet structure is consistent with an unseen spine with the radio emission coming from a sheath with a thickness of about a quarter of the jet radius.

\item  The counter-jet is clearly seen and has a structure similar to the main jet, but with brightness that decays more rapidly with distance, presumably because of Doppler boosting.

\item  The jet/counter-jet brightness ratios indicate an acceleration zone leading to high speeds.
However, the ratios do not get as high as expected for the component superluminal motions observed, perhaps because of less beamed emission from material providing the slower component motions seen by \citetalias{MLWH2016}.

\item  The north/south brightness ratios, the north/south edge speed ratios, and the jet/counter-jet brightness ratios for the north and south edges all point to a clockwise helical flow (as viewed by an observer on the jet side looking toward the core).

\item  The roughly annual observations over 17 years  reveal transverse motions of the inner parsec of the jet that are consistent with a slow change in jet position angle superimposed on an oscillation with a period of about 8 to 10 years.  These motions have not been  reported before to our knowledge.

\item  The oscillation pattern seen in the long-term observations propagates non-ballistically along the jet at an apparent speed somewhat less than half that of the fast components motions seen by \citetalias{MLWH2016}.

\item  The non-ballistic propagation speed of the long-term patterns is consistent with Kelvin-Helmholtz instability wave propagation speed along a jet with density comparable to or less dense than the cocooning environment.

\item  The polarization data shows a consistent pattern for two epochs and suggests a toroidal magnetic field geometry.

\end{itemize}

We will not attempt detailed comparisons of our data with existing jet launch models and simulations.  
But our observations are consistent with general classes of models where the radio emission region is in the outer portions --- the sheath --- of a symmetric jet launched by magnetic fields threading an accretion disk \citep[cf.][]{BP1982, BZ1977, McK2006, McK2009}.
The sheath material is accelerated over the inner 500--1000 $R_{\rm S}$ to mildly relativistic speeds ($\gamma_j > 2$).  
The dark center of the jet allows for a jet spine that might be launched from the black hole and could be faster than the sheath that we see.

Below we discuss the various results in more detail.

\subsection{Jet Structure:  Edge brightening and counter-jet}
All of the images show a strongly edge-brightened structure with a wide opening angle base as originally reported by \citet{Junor1999}.
They also show the existence of a counter-jet, also edge-brightened, that fades much more rapidly than the main jet.
Much of the analysis of jet structure presented here is based on an average image made from the 23 epochs of the intensive monitoring observations in 2007 and 2008.
That image smooths out the variable structure, giving a high sensitivity picture of the persistent aspects of the structure.  
Transverse intensity slices of the average image show that jet and counter-jet have similar width structure out to 1.2~mas beyond which the counter-jet becomes too faint to follow.
The symmetric structure between the jet and counter-jet places the radio core to within $0.06$~mas (half the spacing between slices and one quarter the beamwidth) of the central engine, consistent with the core location found by \citet{Hada2011} from the core shift with frequency.

%  Note suggested point for hyphenating "collimation" below.  Otherwise line too long.

\subsubsection{Expansion and Collimation Regions}
Overall the jet shows three intrinsic expansion/colli\-mation regimes: (1) a rapid $\unsim 19\arcdeg$ expansion followed by a contraction, (2) a second rapid $\unsim 19\arcdeg$ expansion beginning at about $530 R_{\rm S}$ followed by a  recollimation, and (3) a third slower expansion of $\unsim 6.5\arcdeg$ beginning at about $1800 R_{\rm S}$ followed by a slow recollimation.
Changes in the slope of the jet intensity decline are associated with changes in jet expansion.
The average expansion is consistent with  the parabolic expansion found by \citet{AN2012} to apply out to HST-1, and the initial parabolic expansion found by \citet{Hada2013} inside $300 R_{\rm S}$.
However, our results show more complex variable expansion than a single parabolic outside $300 R_{\rm S}$.
In retrospect, some evidence for this complex behavior can be found in \citet{Hada2013}.
Using  high-resolution observations at 86 GHz, \citet{Hada2016} find an additional region of expansion and contraction closer to the core than those seen at 43 GHz.  
The multiple expansion/collimation regimes indicate a non-equilibrium jet launching region in which equilibrium parabolic expansion is achieved only after several thousand $R_{\rm S}$.

\subsubsection{Spine/Sheath Structure}
The very edge-brightened jet and counter-jet suggest that most of the emission comes from a sheath surrounding a low ``effective'' emissivity spine.
On the jet side, the edge-brightened structure that we see from the unresolved radio core out to at least $20$~mas is comparable to edge-brightened structure along the pc-scale jet from about 40 to 200~mas 
\citep[see Figure 2 in][]{Reid1989} and is even more pronounced on the kpc-scale jet \citep[see][]{OHC89,HE2011}.
Transverse intensity profiles and line-of-sight effects imply a sheath with thickness of about a quarter of the jet radius.
Reduction of an intrinsic spine emissivity comparable to the intrinsic sheath emissivity by Doppler deboosting relative to the sheath would require a spine Lorentz factor $\gamma_{\rm sp} \ge 16.5$ relative to an observed typical sheath Lorentz factor of $\gamma_{\rm sh} \approx 2.6$.
This required spine Lorentz factor would result in an apparent motion at a 17\arcdeg\ viewing angle of $\beta^{\rm app} \ge 6.4$ that is consistent with observed optical proper motions of $\beta^{\rm app} = 6.1 \pm 0.6$ at HST-1.

% Force subsection heading onto next column.
\vspace{1.0cm}
\subsubsection{Filaments}
The average intensity image and its transverse profile structure indicate faint jet crossing filaments.
Faint cross jet filamentary structure inside 20~mas begins to show up more prominently on the parsec scale jet from about 40 to 200~mas \citep[see Figure 2 in][]{Reid1989} and is even more pronounced on the kpc-scale jet \citep[see][]{OHC89,HE2011}.

\subsubsection{Edge Brightness Variations}
There are systematic differences in the edge intensities between the sides of the jet, although there are short term variations in those differences.
In the innermost 0.7~mas  the northern edge of both jet and counter-jet is brighter than the southern edge, but farther out the jet's southern edge is usually equal to or brighter than the northern edge.
The most likely reason for intrinsic spatial edge intensity differences is the approximately 8 to 10~yr quasi-periodic intrinsic change in the flow direction at the jet base discussed below.
However, other more rapid events can also produce significant changes.
Within the 1.2 year time frame covered by the 23-epoch average image, significant intensity change within the inner few mas was observed.
In particular, the weekly observations in early 2008 tracked a core brightening, associated with a VHE flare event in early February 2008, that led to jet edge-brightening inside 1~mas over the following 2 months.
More recently, \citet{Hada2014} found a similar jet base brightening associated with an elevated VHE state in 2012.
The deviation of the brightness centroid to below the mean jet position angle inside 20~mas found by \citet{Reid1989} appears related to our finding that the southern edge is brighter than the northern edge.
The \citet{Reid1989} results suggest a quasi-periodic oscillation of the brightness centroid from about 5 to 80~mas that is likely related to a quasi-periodic variation in the northern and southern edge brightness.
Deviation of the brightness centroid between HST-1 and knot D at considerably larger angular scales, see \citep[see][]{OHC89,HE2011}, may or may not be related to that seen on the parsec scale jet.

\subsection{Jet Motions}
The individual images from 2007 and 2008 are used to study the dynamics of the inner jet, and reveal a jet with complicated motions that clearly accelerates to mildly superluminal speeds over the inner 2 mas.
Our component motion analysis gives results similar to an analysis by \citetalias{MLWH2016} of the motions in our 2007 data using a wavelet-based analysis scheme.  
The discrepancy between reports of fast motions based on our data \citep[e.g.,][\citetalias{MLWH2016}]{Walker2008} and recent KaVA results \citep{Hada2017} and slower motions seen at lower frequencies \citep[e.g.,]{Reid1989, Kov2007} may be the result of the presence of both faster and slower motions as indicated explicitly by the wavelet-based analysis of \citetalias{MLWH2016}.
 
The component motion analysis of the 2007 data using a wavelet-based scheme in \citetalias{MLWH2016} revealed both jet acceleration and significant differences in apparent speeds along the jet edges.
Along the jet between 0.5 and 1~mas there was a higher apparent speed along the north edge relative to the south edge.
Farther downstream between 1 and 4~mas the fast component had higher apparent speed on the north edge relative to the south edge.  
Meanwhile, the slow component had lower apparent speed on the north edge relative to south edge.
The presence of a high speed and a low speed component between 1 and 4~mas led \citetalias{MLWH2016} to infer the presence of a velocity shearing sheath.

\subsubsection{Counter-Jet/Jet Intensity Sidedness Ratios}
The counter-jet structure was studied with two recent, high-sensitivity epochs that were processed for higher resolution and also with an average of 49 epochs (all but the last) that allows the counter-jet to be followed to about 1.7 mas.
In general, the intensity sidedness ratio increases as a function of core distance indicating that the jet material is accelerating at the core distances where the counter-jet can be seen.  
Beyond that distance, the material speed, combined with the brightness decrease along the jet, drop the counter-jet intensity to below detectable levels.
But the sidedness ratio seen in the 49-epoch average image indicates not as much acceleration as seen in the component motions.
Perhaps the intensity ratio is not as high as expected because of the effects of the slow component seen by \citetalias{MLWH2016} between 1 and 4~mas, which would not be as strongly beamed.
Counter-jet/jet intensity sidedness ratios using the 49-epoch average image along the counter-jet and jet edges symmetric across the core show systematic north/south differences.
At very small scales the two higher resolution image intensity ratios probe the first rapid counter-jet and jet expansion region.  
Again systematic north/south differences are seen.

\subsubsection{Helical Flow}

Apparent sheath motions along the northern jet edge are faster than apparent sheath motions along the southern jet edge. 
These differences in component motions along the north and south jet edges can be explained best as the result of a clockwise helical flow.  
For clockwise helical flow, flow along the northern jet edge is farther from the line-of-sight and closer to the critical angle at which apparent speeds are maximized, whereas flow along the southern jet edge is closer to the line-of-sight and farther from the critical angle.
The helical flow accelerates from $\beta_{\rm h} \approx 0.56$ between 0.5 and 1~mas to $\beta_{\rm h} \approx 0.924$ between 1 and 4~mas with the flow becoming more poloidal at the larger distances.
The systematic north/south differences in the counter-jet/jet intensity sidedness ratios seen in both the 49-epoch average image and the two high resolution images are consistent with, and help confirm, the clockwise helical flow scenario suggested by the jet component motions inside 1~mas.
In the high resolution images, the better quality 2013 image indicates clockwise helical flow with acceleration from $\beta_{\rm h} \approx 0.5$ to $\beta_{\rm h} \approx 0.8$ between 0.4 and 0.6~mas in the rapid counter-jet and jet expansion region consistent with the fast component motions found by \citetalias{MLWH2016} between 1 and 4~mas.
The 2016 image intensity ratios indicate a potential for counter-clockwise helical flow beyond $\unsim 0.6$~mas consistent with the slow component motions found by \citetalias{MLWH2016} between 1 and 4~mas.
 The slow component edge speed differences between 1 and 4~mas could result from a counter-clockwise helical flow with $\beta_{\rm h} \approx 0.55$.
This result suggests a complex-velocity-shear sheath region.

The clockwise helical flow scenario is consistent with the observed tendency toward northward apparent transverse motions across the jet where northwards transverse motions are associated with flow on the nearside of the jet.
Assuming the same helical sense for the filaments would indicate clockwise helical filamentary structure.
Differential Doppler boosting between the nearside and farside of the jet, due to the opening angle, predicts an enhancement of the nearside relative to the farside that can partially compensate for the higher farside intensities nearer to the core along lines-of-sight through the jet axis.
Differential Doppler boosting from rotation would give a brighter southern edge relative to the northern edge as seen beyond about 1 mas.
However, the northern edge is brighter than the southern edge close to the core and this fact along with spatial and temporal variation in relative northern and southern edge intensities at sub-parsec, parsec, and kilo-parsec jet scales all suggest that edge intensity differences are not entirely due to rotation induced differential Doppler boosting.

\subsection{Long-Term Transverse Motions}
The long-term, roughly annual observations show that the position angle of the jet has been changing, with both a systematic increase and with an approximately  sinusoidal oscillation with a period of roughly 8 to 10 years.
A recent 3D GRMHD simulation by \citet{TNM2011} indicated that quasi-periodic variation in jet direction is a natural consequence of the jet acceleration and collimation process.
The observed change in jet angle propagates outward at about $3.2$ \masr\ ($\beta_{\rm p}^{\rm app}\approx 0.85$) which is less than the high speed components ($\beta^{\rm app} \gtrsim 2$) seen by \citetalias{MLWH2016} and by our component analysis between 1 and 4~mas.
This non-ballistic propagation speed is the expected result for a Kelvin-Helmholtz instability in a jet with density comparable to or less dense than the cocooning environment \citep[see][]{Hardee2007}.
We note however that there is no evidence for bulk sideways displacement of the jet at arcsecond scales until knot A.
Damping of a bulk displacement at larger scales will occur naturally in the presence of strong magnetic fields \citep[see][]{Mizuno2007} or if the quasi-periodicity frequency exceeds a local resonant frequency that scales inversely with the jet radius and declines along an expanding jet.

\subsection{Magnetic Fields}
While most of the polarization data have yet to be reduced, two epochs show similar systematic polarization structure.
Near to the core the polarization is up to 1.5\% on the jet side where the magnetic field vectors are perpendicular to the jet axis, suggesting a toroidal configuration.
Just beyond the core, the magnetic field vectors on the edge-brightened jet appear parallel to the jet axis.
Polarization on the counter-jet side of the core is less than 1\% and the magnetic field vectors appear parallel to the counter-jet axis.
Flow aligned magnetic field vectors might be indicative of velocity shear in the sheath.
On the other hand, the emission apparently on the counter-jet side could actually be from the optically-thick portion of the inner main jet but spread toward the counter-jet by resolution effects, as a $90\arcdeg$ polarization angle change between optically-thick and thin parts of the jet is seen in many VLBI sources.

\subsection{Future Work}
The data sets reported on here constitute a rich source of information about the M\,87 jet on gravitational scales that cannot be reached in most systems.  
There is much left to be done.  
We have only barely begun to utilize the polarization information that should shed light on the magnetic field structure and its dynamics along the inner jet.  
Also, the observing sessions since 2009 included enough time at 24~GHz to make images at that frequency of similar quality to those at 43 GHz.  
This additional frequency will allow delineation of the spectral index in the inner regions and also, because the jet is somewhat brighter at the lower frequency and the phase stability is a better, should allow extension of some of the studies reported here to larger scales.  
The observations also included astrometric observations targeting the relative positions of M\,87 and another large, Virgo cluster elliptical, M\,84.  
The astrometry will shed light on any possible core wander and on the relative transverse motions of the two galaxies.  
The astrometric study is the close to completion with most of the data reduction done (Davies et al., in preparation).

\begin{acknowledgements}
P. Hardee acknowledges support from NSF awards AST-0506666 and AST-0908010, and NASA/ATP  awards NNX08AG83G and NNX12AH06G to the University of Alabama.
This work made use of the Swinburne University of Technology software correlator, developed as part of the Australian Major National Research Facilities Programme and operated under license.
Earth Orientation and Ionospheric data hosted on the Crustal Dynamics Data Information System (CDDIS) at the NASA Goddard Space Flight Center was used in this project.
This study makes use of 43 GHz VLBA data from the VLBA-BU Blazar Monitoring Program (VLBA-BU-BLAZAR; http://www.bu.edu/blazars/VLBAproject.html), funded by NASA through the {\it Fermi} Guest Investigator Program.   C. Ly and F. Davies were summer students at the NRAO when they became involved in this project.

\end{acknowledgements}

\appendix
\onecolumngrid
\setcounter{table}{0}
\renewcommand{\thetable}{A\arabic{table}}

\section{SYMBOL TABLE}
\label{AppSymbols}

%  If twocolumn mode gets used, the following forces the table to span both columns.
%  Some deleted items are last in svn revision 813.
\onecolumngrid

\startlongtable
\begin{deluxetable}{ll}
\tablecaption{Symbol Definitions
 \label{SymDef}
}
\tablehead{
  \colhead{Symbol} &
  \colhead{Definition}
}
\startdata
$a_1$            & Linear change in $R_c$ with distance \\
$a_2$            & Linear change in $R_c$ with time. \\
$a_3$            & Sine wave amplitude at $z = 0$. \\
$a_4$            & Power law index of sine wave amplitude dependence on $z$. \\
$a$              & Sound speed.\\
$a_{\rm j,e}$    & Jet, external medium sound speed.\\
$B$              &  Magnetic field strength.\\
$B_{\rm j,e}$    &  Magnetic field strength in jet, external medium.\\
$c$              & Speed of light.\\
$f$              & Sine wave frequency (cycles mas$^{-1}$). \\
$G$              & Gravitational constant. \\
$I_{\rm j,cj}$   &  Intensity on the jet, counter-jet side.\\
$I^{\rm N,S}_{\rm j,cj}$ &  Intensity on the north, south side in the jet, counter-jet direction.\\
$I^{\rm Max}_{\rm j,cj}$ &  Maximum intensity on jet, counter-jet side.\\
$I_{\rm cen}$    &  Intensity along projected jet center.\\
$I_{\rm edge}$   &  Intensity along the edge of the jet.\\
$\ell_{\rm edge}$ &  Path length through sheath, cylindrical jet.\\
$\ell_{\rm cen}$ &  Path length through sheath along projected jet centerline.\\
$M$              & Mass of black hole.\\
$M_{\odot}$      & Solar mass. \\
$P$              &  Pressure. \\
$R_{\rm S}$      & Schwarzschild radius $(2GM/c^2)$.\\
$R_{\rm c}$      & Offset from the nominal jet centerline. \\
$R_0$            & $R_c$ at date 2000.0.  \\
$R_{\rm NS}$     &  $\left({\rm I}^{\rm N}_{\rm cj}/I^{\rm S}_{\rm j}\right)^{1/(2 + \alpha)}~$.\\
$R_{\rm SN}$     &  $\left({\rm I}^{\rm S}_{\rm cj}/I^{\rm N}_{\rm j}\right)^{1/(2 + \alpha)}~$.\\
$R_j$            &  Radius of jet. \\
$R_{\rm sp}$     &  Spine radius. \\
$t_{yr}$         & Date in fractional years.\\
$t$              & Date - 2000.0 (for fits). \\
$t_{\rm d}$      & Time minus pattern propagation time from core $(t - z / v_{\rm p})$. \\
$u_{\rm e,j}$    &  External medium, internal jet flow speed. \\
$v^{\rm app}$    & Observed (apparent) speed.\\
$v_{\rm h}$      & Helical mode wave speed at low frequency. \\
$v_{\rm h}^{\ast}$ & Helical mode wave speed at resonance. \\
$v_{\rm p}$      & Pattern propagation speed (\masr). \\
$v_{\rm A}$      &  Alfv\'en speed.\\
$v_{\rm Aj,Ae}$  &  Jet, external medium Alfv\'en speed.\\
$v_{\rm w}$      &  Helical wave speed below resonance.\\
$v_{\rm wj,we}$  &  $\equiv \left( a_{\rm j},v_{\rm Aj}\right) $ in sonic or Alfv\'enic limits for jet, external medium.\\
$V_{\rm A}$      &  $\equiv B^2/(4\pi W)$ \\
$V_{\rm As}$     &  ``surface'' Alfv\'{e}n speed.\\
$W$              &  Enthalpy.\\
$W_{\rm j,e}$    &  Enthalpy in jet, external fluid.\\
$z$              & Core distance in mas.  An increment: $\Delta z$. \\
$\alpha$         &  Spectral index $I \propto \nu^{-\alpha}$.\\
$\beta$          & Intrinsic speed as a fraction of the speed of light. \\
$\beta^{\rm app}_{\rm edge}$ & Apparent speed in projection on edge of jet.\\
$\beta^{\rm app}$              & Apparent speed in projection. \\
$\beta^{\rm app}_{\rm N,S}$    & Apparent speed on north, south edge of the jet.\\
$\beta^{\rm app}_{\rm j,cj}$   & Apparent speed on jet, counter-jet side.\\
$\beta_{\rm p}$                & Intrinsic pattern speed. \\
$\beta_{\rm p}^{\rm app}$      & Apparent pattern speed. \\
$\beta_{\rm edge}$    & $\beta$ on edge of jet.\\
$\beta_{\rm sp,sh}$   & Intrinsic $\beta$ in the spine, sheath.\\
$\beta_{\rm h}$       & Intrinsic $\beta$ along a helical path.\\
$\beta_{\rm e}$       & External sheath medium flow speed.\\
$\gamma$              & Lorentz factor.\\
$\gamma_{\rm sp,sh}$  &  Spine, sheath Lorentz factor.\\
$\gamma_{\rm e,j}$    &  External medium, jet Lorentz factor.\\
$\gamma_{\rm w,wj,we}$  &  Sonic or Alfv\'enic Lorentz factor in jet, external medium.\\
$\gamma_{\rm aj,ae}$  &  Sonic Lorentz factor, jet or external medium.\\
$\gamma_{\rm A}$  &  Alfv\'enic Lorentz factor.\\
$\gamma_{\rm Aj,Ae}$  &  Alfv\'enic Lorentz factor, jet or external medium.\\
$\Gamma$              &  Adiabatic index.\\
$\Delta W$            & Change in jet width. \\
$\Delta R$            &  Sheath thickness. \\
$\delta_{\rm sp,sh}$  &  Spine, sheath Doppler factor.\\
$\delta_{\rm ns,fs}$  &  Doppler factor on near, far side, of the jet.\\
$\delta_{\rm S,N}$    & Doppler boost factor of northern and southern edges of jet.\\
$\epsilon$            &  Deboosting of spine emission.\\
$\eta$                &  $\equiv \gamma _{\rm j}^{2}W_{\rm j}\left/ \gamma _{\rm e}^{2}W_{\rm e}\right. $.\\
$\theta$              & Viewing angle between jet axis and line-of-sight.\\
$\theta_{\rm edge}$   & Viewing angle between jet edges and line-of-sight.\\
$\theta_{\rm fs,ns}$  & Viewing angle of jet edges along line-of-sight through jet axis.\\
$\theta_{\rm crit}$   & Critical angle to line-of-sight for maximum $\beta^{\rm app}$.\\
$\Theta$              & Jet intrinsic full opening angle in a jet segment.\\
$\Theta^{\rm app}$    & Full width half-maximum apparent opening angle.\\
$\nu$                 &  Frequency.\\
$\rho$                &  Density.  \\
$\phi_0$              & Sine wave phase (deg) at $t_d = 0$. \\
$\Delta \phi_h$       & Flow angle relative to pure poloidal flow along jet edges.\\
$\psi$                & Intrinsic half opening angle. \\
$\omega^{\ast}$       &  Helical mode resonant frequency. \\
$\omega$              &  Angular wave frequency.\\
\enddata
\end{deluxetable}

% Try to get the section heading to center, not go to column 2.
\onecolumngrid
\section{LIST OF OBSERVATIONS}
\label{AppObs}

%  The table is more than a page long so needs to split.  The \startlongtable
%  allows the float to be split.  Note all deluxetables are floats.

\startlongtable

%  If twocolumn mode gets used, the following forces the table to span both columns.
\onecolumngrid

\begin{deluxetable}{llccccc}
\tablecaption{VLBA observation sessions that contributed data to this paper.  
  \label{ObsTable}
}
\tablehead{
   \colhead{Observation} &
   \colhead{Date} &
   \colhead{MJD} &
   \colhead{Bit Rate\tablenotemark{a}} &
   \colhead{RMS \tablenotemark{b}}&
   \colhead{Integrated \tablenotemark{c}} &
   \colhead{Peak \tablenotemark{d}}
 \\
   \colhead{Code} &
   \colhead{} &
   \colhead{} &
   \colhead{(Mbps)} &
   \colhead{(\ujb)} &
   \colhead{Flux Density} &
   \colhead{Flux Density}
 \\
   \colhead{} &
   \colhead{} &
   \colhead{} &
   \colhead{} &
   \colhead{} &
   \colhead{(Jy)} &
   \colhead{(\jb)}
}
\startdata
  GJ009\tablenotemark{e}   % Global
 & 1999mar04               % AIPS says 1999mar03 but mid point after midnight.
 & 51241
 & 256
 & 204
 & 0.52
 & 0.28
\\
  BJ031B\tablenotemark{f}  % Plus VLA27
 & 2000apr09               % AIPS says 2000apr08 but mid point on second day
 & 51643
 & 256 
 & 387
 & 1.23
 & 0.60
\\
  BW054C\tablenotemark{g}
 & 2001oct12
 & 52194
 & 256 
 & 709
 & 1.53
 & 0.53
\\
  BU021\tablenotemark{h}
 & 2002jun02               % AIPS says 2000jun01 but mid point on second day
 & 52427
 & 256
 & 540
 & 1.49
 & 0.58
\\
  BW070\tablenotemark{i}   % Had GB and EB, but both had poor weather.
 & 2004apr05               % AIPS says 2004apr04 despite there being on hour on apr03.
 & 53100
 & 256
 & 773
 & 1.26
 & 0.61
\\
  BW082A
 & 2006apr08
 & 53833
 & 128
 & 247
 & 1.59
 & 0.80
\\
  BW082B
 & 2006apr11
 & 53836
 & 128
 & 235
 & 1.60
 & 0.80
 \\
  BW082D
 & 2006apr18
 & 53843
 & 128
 & 273
 & 1.69
 & 0.82
\\
  BW082E
 & 2006apr21
 & 53846
 & 128
 & 268
 & 1.46
 & 0.81
\\
  BW082F
 & 2006may14
 & 53869
 & 128
 & 307
 & 1.67
 & 0.79
\\
  BW082G
 & 2006jul14
 & 53930
 & 128
 & 276
 & 1.35
 & 0.57
\\
  BW088A
 & 2007jan27
 & 54127
 & 256
 & 184
 & 1.75
 & 0.64
\\
  BW088B
 & 2007feb17
 & 54148
 & 256
 & 209
 & 1.60
 & 0.60
\\
  BW088D
 & 2007mar13
 & 54172
 & 256
 & 205
 & 1.80
 & 0.70
\\
  BW088E
 & 2007apr03
 & 54193
 & 256
 & 201
 & 1.70
 & 0.68
\\
  BW088F
 & 2007apr23
 & 54213
 & 256
 & 198
 & 1.86
 & 0.67
\\
  BW088G
 & 2007may10
 & 54230
 & 256
 & 175
 & 1.82
 & 0.70
\\
  BW088H
 & 2007jun03
 & 54254
 & 256
 & 199
 & 1.63
 & 0.60
\\
  BW088I
 & 2007jun22
 & 54273
 & 256
 & 185
 & 1.43
 & 0.51
\\
  BW088J
 & 2007jul16
 & 54297
 & 256
 & 225
 & 1.58
 & 0.58
\\
  BW088K
 & 2007aug04
 & 54316
 & 256
 & 216
 & 1.65
 & 0.60
\\
  BW088L
 & 2007aug26
 & 54338
 & 256
 & 223
 & 1.80
 & 0.68
\\
%  BW088M
% & 2007sep20
% & 54363
% & 256
% & .
% & .
% & .
%\\
%  BW088N
% & 2007oct07
% & 54380
% & 256
% & .
% & .
% & .
%\\
  BW088P
 & 2007nov02
 & 54406
 & 256
 & 192
 & 1.53
 & 0.58
\\
%  BW088Q
% & 2007nov17
% & 54421
% & 256
% & .
% & .
% & .
%\\
%  BW088S
% & 2007dec15
% & 54449
% & 256
% & .
% & .
% & .
%\\
%  BW088T
% & 2007dec30
% & 54464
% & 256
% & 395
% & .
% & .
%\\
  BW088U
 & 2008jan21
 & 54486
 & 256
 & 202
 & 1.86
 & 0.81
\\
  BW090A
 & 2008jan26
 & 54491
 & 256
 & 197
 & 1.88
 & 0.83
\\
  BW090B
 & 2008jan31
 & 54496
 & 256
 & 189
 & 1.97
 & 0.82
\\
  BW090C
 & 2008feb04
 & 54500
 & 256
 & 291
 & 1.97
 & 0.82
\\
%  BW090D
% & 2008feb10
% & 54506
% & 256
% & .
% & 
% & .
%\\
  BW090E
 & 2008feb15
 & 54511
 & 256
 & 193
 & 2.10
 & 0.94
\\
  BW090F
 & 2008feb20
 & 54516
 & 256
 & 249
 & 2.19
 & 1.01
\\
  BW090G
 & 2008feb25
 & 54521
 & 256
 & 169
 & 2.23
 & 1.00
\\
  BW090H
 & 2008mar02
 & 54527
 & 256
 & 191
 & 2.24
 & 1.04
\\
%   BW090I
%  & 2008mar06
%  & 54531
%  & 256
%  & 484 (orig. res)
%  & .
%  & .
%\\
  BW090J
 & 2008mar12
 & 54537
 & 256
 & 158
 & 2.36
 & 1.10
\\
  BW090K
 & 2008mar19
 & 54544
 & 256
 & 170
 & 2.38
 & 1.10
\\
  BW090L
 & 2008mar24
 & 54549
 & 256
 & 159
 & 2.37
 & 1.09
\\
%   BW090M
%  & 2008mar29
%  & 54554
%  & 256
%  & 437 (orig res)
%  & 2.41
%  & 1.11
%\\
  BW090N
 & 2008apr05
 & 54561
 & 256
 & 167
 & 2.44
 & 1.13
\\
  BW092A
 & 2009jan19
 & 54850
 & 256
 & 268
 & 1.46
 & 0.66
\\
%  BW092B
% & 2009mar13
% & 54903
% & 256
% & .
% & 2.44
% & 1.12
%\\
  BW093A
 & 2010jan18
 & 55214
 & 256
 & 271
 & 1.41
 & 0.66
\\
  BW093B
 & 2010apr04
 & 55290
 & 256
 & 210
 & 1.55
 & 0.68
\\
  BW093C
 & 2010apr19
 & 55305
 & 256
 & 231
 & 1.58
 & 0.65
\\
  BW093D
 & 2010may01
 & 55317
 & 256
 & 227
 & 1.48
 & 0.66
\\
  BW093E
 & 2010may16
 & 55332
 & 256
 & 240
 & 1.42
 & 0.63
\\
  BW093F
 & 2010may31
 & 55347
 & 256
 & 247
 & 1.31
 & 0.62
\\
  BW095B
 & 2011feb13
 & 55605
 & 512
 & 182
 & 1.71
 & 0.86
\\
  BW097A
 & 2012jan06
 & 55932
 & 512
 & 185
 & 1.43
 & 0.64
\\
  BW098A\tablenotemark{j}
 & 2013jan12
 & 56304
 & 2048
 & 67
 & 1.51
 & 0.62
\\
  BW106\tablenotemark{k}
 & 2013jun04
 & 56447
 & 2048
 & 80
 & 1.40
 & 0.56
\\
  BW107\tablenotemark{l}
 & 2013dec27
 & 56653
 & 2048
 & 149
 & 1.44
 & 0.61
\\
  BW115A
 & 2015Jan26
 & 57048
 & 2048
 & 92
 & 1.66
 & 0.73
\\
  BW115D
 & 2016mar14
 & 57461
 & 2048
 & 72
 & 1.78
 & 0.75
\\
  BW115H\tablenotemark{m}
 & 2016apr10
 & 57488 
 & 2048
 & 84   
 & 1.65 
 & 0.70 
\\
\enddata

\tablenotetext{a}{The total recording rate in bits per second.  For these projects, 
there are 2 bits per sample, 2 samples per Hz of bandwidth, and 2 polarizations so 
the recording rate is 8 times the spanned bandwidth.}

\tablenotetext{b}{The noise level was measured using a histogram fit to the final image prior to adjusting the convolving beam used for the point clean components to match the common resolution.  When the beam size adjustment is made (AIPS task CCRES), the residual image, including any low resolution components from the multi-resolution clean, is scaled by the ratio of beam areas.  This is correct for the regions with emission, but is somewhat problematic for the noise.  The RMS of the unscaled image should provide a better comparison of the quality of the images.  Note that the scaling of the residual image is by less than 20\% for all but 4 images, but there are outliers.}

\tablenotetext{c}{The integrated flux density is the sum of all clean components of all resolutions from the very deep cleans used in the imaging.}

\tablenotetext{d}{The peak flux density is measured on the images for which the point clean components are convolved with a common beam size of $0.43 \times 0.21$~mas, elongated along position angle $-16\arcdeg$}

\tablenotetext{e}{Global observations involving VLBA and European VLBI Network antennas.  Single polarization.  The flux density calibration is highly suspect, but has not been revisited.  See \citet{Junor1999}}

\tablenotetext{f}{Included the phased VLA. See \citet{Ly2007}.}

\tablenotetext{g}{\citet{Ly2004}}

\tablenotetext{h}{Archival data.  M\,87 was a phase reference calibrator.  See \citet{Ly2007}}

\tablenotetext{i}{See \citet{Ly2007}}

\tablenotetext{j}{First epoch to use the Roach Digital Back End (RDBE) that provides
wide bandwidths for all following observations.}

\tablenotetext{k}{Used Digital Down Converter personality of the RDBE backend.  Other RDBE observations used the Polyphase Filter Bank personality.}

\tablenotetext{l}{Image quality poor.  The image was used in the stacked image of Figure~\ref{Stack49} but not in the long term variation analysis or in Figure~\ref{LTE3}.}

\tablenotetext{m}{This image was used in Figure~\ref{LTVfit} but not in the image montage.  The quality is not quite as good as BW115D.  The structure is similar.}

\tablecomments{Additional observations were made and are in the
archive, but were not used in this paper for various reasons including
of poor data quality and processing difficulties.}

\end{deluxetable}

%  If twocolumn mode gets used, the following forces the table to span both columns.
\onecolumngrid

\begin{deluxetable}{ccccc}[h]
\tablecaption{Averages of multiple images used in the long term image sequence.
  \label{StackTable}
}
\tablehead{
   \colhead{Observation\tablenotemark{a}} &
   \colhead{Average Date} &
   \colhead{RMS\tablenotemark{b}} &
   \colhead{Integrated\tablenotemark{c}} &
   \colhead{Peak}
 \\
   \colhead{Code} &
   \colhead{} &
   \colhead{(\ujb)} &
   \colhead{Flux Density} &
   \colhead{Flux Density}
 \\
   \colhead{} &
   \colhead{} &
   \colhead{} &
   \colhead{(Jy)} &
   \colhead{(\jb)}
}
\startdata
  BW082
 & 2006.3
 & 137
 & 1.54
 & 0.77
\\
  BW088
 & 2007.4
 & 74
 & 1.64
 & 0.63
\\
  BW090
 & 2008.1
 & 75
 & 2.13
 & 0.98
\\
  BW093
 & 2010.3
 & 127
 & 1.25
 & 0.65
\\
\enddata

\tablenotetext{a}{The stack is a noise-weighted average of all observations in
Table~\ref{ObsTable} using this base observation code.  An exception is BW088U
which is included in the 2008 stack, not the 2007 stack.  The images used are those with the common resolution of $0.43 \times 0.21$~mas, elongated along position angle $-16\arcdeg$}

\tablenotetext{b}{The RMS is measured with a histogram fit to the full stacked image.  This will be subject to the scaling issue noted in note b of Table~\ref{ObsTable}, but
the CCRES scale factors will average toward 1.0 so the effect should be small}

\tablenotetext{c}{Flux density is measured by integrating in a box from $1$ to $-8$ mas RA offset and $-1$ to $4$ mas Declination offset from the core.  At larger distances, low 
surface brightness and poor short spacing UV coverage make image plane integration
unreliable.}

\end{deluxetable}

\newpage

\section{HELICAL MODE RESONANT FREQUENCY \& PROPAGATION}
\label{HMtheory}

In this Appendix we briefly review the important analytical limiting equations giving the resonant frequency and describing the motion of the helical mode at and below the resonant frequency used to constrain the jet external medium interface properties that would lead to the observed long-term pattern motion.

An approximate lower limit to the helical mode resonant  frequency,  $\omega^{\ast}$, of 
\begin{equation}
\label{hmrfreq}
\omega^{\ast} R_{\rm j}/v_{\rm we} \approx \frac{3\pi /4}{ \left[1-
\left(v_{\rm we}^{2}/u_{\rm j}^{2}+2 \frac{\gamma_{\rm wj}}{\gamma_{\rm we}}v_{\rm we}
v_{\rm wj}/\gamma_{\rm j} u_{\rm j}^2+\frac{\gamma
_{\rm wj}^2}{\gamma_{\rm we}^2}v_{\rm wj}^2/\gamma_{\rm j}^2u_{\rm j}^2\right)\right]^{1/2}}~,
\end{equation}
is found in the sonic (fluid) or  Alfv\'enic (magnetic) limits \citep[see eq.(16) in][]{Hardee2007} in the limit of no flow, $u_{\rm e} = 0$, in the external (e) medium.
In Equation~(\ref{hmrfreq})  $\gamma_{\rm j}$ and $u_{\rm j}$ refer to the jet (j) Lorentz factor and speed, with $v_{\rm wj}\equiv \left( a_{\rm j}~{\rm or}~v_{\rm Aj}\right)$ and $v_{\rm we}\equiv \left(
a_{\rm e}~{\rm or}~v_{\rm Ae}\right)$ in the sonic or Alfv\'enic limits, respectively.
The sound speed $a$ and the Alfv\'en wave speed $v_{\rm A}$ are defined by
$$
a\equiv \left[ \frac{\Gamma P}{ W}\right] ^{1/2}~{\rm and}~v_{\rm A}\equiv \left[ \frac{V_{\rm A}^{2}}{1+V_{\rm A}^{2}/c^{2}}\right] ^{1/2}~,
$$
where the adiabatic index $4/3\leq \Gamma \leq 5/3$, enthalpy $W\equiv \rho +\left[ \Gamma /\left( \Gamma -1\right) \right] P/c^{2} $ where $\rho$ and P are the density and pressure, and
$V^2_{\rm A} \equiv B^2/(4\pi W)$ where $B$ is the magnetic field strength with
$\gamma _{\rm w}\equiv (1-v_{\rm w}^{2}/c^{2})^{-1/2}$ the sonic or
Alfv\'{e}nic Lorentz factor accompanying $v_{\rm wj}\equiv \left(
a_{\rm j}~{\rm or}~v_{\rm Aj}\right)$ and $v_{\rm we}\equiv \left( a_{\rm e}~{\rm or}~v_{\rm Ae}\right)$.
Note that the sonic Lorentz factor $\gamma_{\rm aj,e} = (1-a_{\rm j,e}^{2}/c^{2})^{-1/2} \le 1.225$ whereas the Alfv\'enic Lorentz factor $\gamma_{\rm Aj,e}\equiv (1+V_{\rm Aj,e}^{2}/c^{2})^{1/2} = (1-v_{\rm Aj,e}^{2}/c^{2})^{-1/2}$ can be large.
The resonant frequency is a function of the velocity shear, $u_{\rm j} - u_{\rm e}$, as well as the sound and Alfv\'en wave speeds in the jet and external medium where $\omega^{\ast }\longrightarrow \infty $ as  
$$
\frac{u_{\rm j}-u_{\rm e}}{1-u_{\rm j}u_{\rm e}/c^{2}}\longrightarrow \frac{v_{\rm wj}+v_{\rm we}}{
1+v_{\rm wj}v_{\rm we}/c^{2}}~,
$$
where the left hand side approaches the right hand side from above.

An estimate for the conditions required for the lower limit to the resonant frequency to exceed this angular frequency can be obtained by assuming that the sound or Alfv\'en speed in the jet and external medium are the same, i.e., $a_{\rm j} = a_{\rm e}$ or $v_{\rm Aj} = v_{\rm Ae}$ in the sonic or Alfv\'enic limits.
In this case we would find that 
\begin{equation}
\label{resfreq}
\omega^{\ast} \approx \frac{3 \pi}{4} \frac{u_{\rm j}/R_{\rm j}}{(u_{\rm j}/v_{\rm w} - 1)}  \approx \frac{3 \pi}{4} \frac{c/R_{\rm j}}{(c/v_{\rm w} - 1)} \ge \frac{3 \pi}{4} \frac{v_{\rm w}}{R_{\rm j}}~,
\end{equation}
where $v_w =$ ($a$ or $v_{\rm A}$) in the sonic or Alfv\'enic limits.
Below the resonant frequency the helical mode wave speed, $v_{\rm h}$, is described by the real part of Equation~(8) in \citet{Hardee2007}
\begin{equation}
v_{\rm h} = \frac{\eta u_{\rm j}+u_{\rm e}}{(1+V_{\rm Ae}^{2}/\gamma
_{\rm e}^{2}c^{2})+\eta (1+V_{\rm Aj}^{2}/\gamma _{\rm j}^{2}c^{2})} ~,
\label{H8}
\end{equation}
where $\eta \equiv \gamma _{\rm j}^{2}W_{\rm j}\left/ \gamma _{\rm e}^{2}W_{\rm e}\right. $.
The helical mode wave speed at resonance, $v_{\rm h}^{\ast}$, can be obtained in the sonic or Alfv\'enic limits and is given by  (Equation~(15) in \citet{Hardee2007}) 
\begin{equation}
v_{\rm h}^{\ast }\equiv \frac{\gamma _{\rm j}(\gamma
_{\rm we}v_{\rm we})u_{\rm j}+\gamma _{\rm e}(\gamma _{\rm wj}v_{\rm wj})u_{\rm e}}{\gamma
_{\rm j}(\gamma _{\rm we}v_{\rm we})+\gamma _{\rm e}(\gamma _{\rm wj}v_{\rm wj})}~.
\label{H15}
\end{equation}
In both these expressions we now allow for an external medium flow $u_{\rm e} > 0$.

If pattern motion is governed by helical mode propagation below the resonant frequency,  a little manipulation allows us to rewrite Equation~\ref{H8} for $W_{\rm j}/W_{\rm e}$ as
\begin{equation}
\label{Wlowfreq}
W_{\rm j}/W_{\rm e} = \left[ \frac{\gamma^2_{\rm e} (v_{\rm p} - u_{\rm e}) + v_{\rm p}(B_{\rm e}^2 + B_{\rm j}^2)/ 4 \pi W_{\rm e} c^2}{\gamma^2_{\rm j} (u_{\rm j} - v_{\rm p})}\right] ~,
\end{equation}
where we now define $v_{\rm p} \equiv v_{\rm h}$.

If pattern motion is more accurately described by helical mode propagation near the resonant frequency, a little manipulation allows us to write Equation~\ref{H15} as
\begin{equation}
\label{Wresfreq}
W_{\rm j}/W_{\rm e} = \left[\left(\frac{\gamma_{\rm e}}{\gamma_{\rm j}} \right) \left(\frac{\gamma_{\rm aj}}{\gamma_{\rm ae}} \right) \frac{(v_{\rm p} - u_{\rm e})}{(u_{\rm j} - v_{\rm p})}\right]^2 ~{\rm or} ~ W_{\rm j}/W_{\rm e} = \left[\left(\frac{\gamma_{\rm e}}{\gamma_{\rm j}} \right) \left(\frac{B_{\rm e}}{B_{\rm j}} \right)\frac{(v_{\rm p} - u_{\rm e})}{(u_{\rm j} - v_{\rm p})}\right]^2~,
\end{equation}
in the sonic or Alfv\'enic limits, respectively, and where we now define $v_{\rm p} \equiv v^*_{\rm h}$. Here we see that the sonic and Alfv\'enic limiting expressions are comparable with $\gamma_{\rm aj} / \gamma_{\rm ae} \approx 1$ in the sonic case and with $B_{\rm e}/B_{\rm j} \approx 1$ in the Alfv\'enic case when there is magnetic pressure balance between the jet and external medium.

Theory and numerical simulation show that damping of a displacement will occur in the presence of strong magnetic fields (see \citet{Mizuno2007})  when 
\begin{equation}
\label{damping}
( u_{\rm j}-u_{\rm e})^{2}-V_{\rm As}^{2}/\gamma_{\rm j}^{2}\gamma _{\rm e}^{2} < 0~, 
\end{equation}
where $V_{\rm As}$ is a ``surface'' Alfv\'{e}n speed defined by
$$
V_{\rm As}^{2}\equiv \left( \gamma _{\rm Aj}^{2}W_{\rm j}+\gamma _{\rm Ae}^{2}W_{\rm e}\right) 
\frac{B_{\rm j}^{2}+B_{\rm e}^{2}}{4\pi W_{\rm j}W_{\rm e}}~.
$$ 
Damping of a bulk displacement will also occur if the wave frequency is less than the resonant frequency. This can occur on an expanding jet  as a wave frequency $\omega$ that is initially less than the local resonant frequency $\omega^{\ast}$  can propagate down the expanding jet to where $\omega$ is above the local resonant frequency because $\omega^{\ast} \propto R^{-1}$ declines along an expanding jet.

\end{document}